\documentclass[aps,prx,twocolumn,groupedaddress,superscriptaddress,nofootinbib,notitlepage,floatfix]{revtex4-1}
\usepackage{times,bm,bbm,bbold,amssymb,amsmath,amsfonts,dsfont,graphics,graphicx,braket,color,xcolor,hyperref}
\usepackage[normalem]{ulem}

\hypersetup{colorlinks,linkcolor={purple},citecolor={purple},urlcolor= {purple}}
\newcommand{\ignore}[1]{}
\newcommand{\ave}[1]{\langle{#1}\rangle}
\newcommand{\lbrac}[0]{[\hskip-0.9mm[}
\newcommand{\rbrac}[0]{]\hskip-0.9mm]}

\newtheorem{remark}{Remark}
\DeclareFontFamily{OT1}{pzc}{}
\DeclareFontShape{OT1}{pzc}{m}{it}
              {<-> s * [1.25] pzcmi7t}{}
\DeclareMathAlphabet{\mathpzc}{OT1}{pzc}
                                 {m}{it}
\usepackage[T1]{fontenc}
\usepackage{newtxtext,newtxmath}

\begin{document}

\title{Correlation Picture Approach to Open-Quantum-System Dynamics}

\author{S. Alipour}
\affiliation{QTF Center of Excellence, Department of Applied Physics, Aalto University, P. O. Box 11000, FI-00076 Aalto, Espoo, Finland}
\email{sahar.alipour@aalto.fi}

\author{A. T. Rezakhani}
\affiliation{Department of Physics, Sharif University of Technology, Tehran 14588, Iran}
\email{rezakhani@sharif.edu}

\author{A. P. Babu}
\affiliation{QTF Center of Excellence, Department of Applied Physics, Aalto University, P. O. Box 11000, FI-00076 Aalto, Espoo, Finland}

\author{K. M{\o}lmer}
\affiliation{Department of Physics and Astronomy, Aarhus University, Ny Munkegade 120, DK-8000, Aarhus C, Denmark}

\author{M. M\"{o}tt\"{o}nen}
\affiliation{QCD Labs, QTF Center of Excellence, Department of Applied Physics, Aalto University, P. O. Box 13500, FI-00076 Aalto, Espoo, Finland}
\date{\today}

\author{T. Ala-Nissila}
\affiliation{QTF Center of Excellence, Department of Applied Physics, Aalto University, P. O. Box 11000, FI-00076 Aalto, Espoo, Finland}
\affiliation{Interdisciplinary Centre for Mathematical Modelling and Department of Mathematical Sciences, Loughborough University, Loughborough, Leicestershire LE11 3TU, UK}
\email{tapio.ala-nissila@aalto.fi}

\begin{abstract}  
We introduce a new dynamical picture, referred to as correlation picture,' which connects a correlated state to its uncorrelated counterpart. Using this picture allows us to derive an exact dynamical equation for a general open-system dynamics with system--environment correlations included. This exact dynamics is in the form of a Lindblad-like equation even in the presence of initial system-environment correlations. For explicit calculations, we also develop a \textit{weak-correlation} expansion formalism that allows us to perform systematic perturbative approximations. This expansion provides approximate master equations which can feature advantages over existing weak-coupling techniques. As a special case, we derive a Markovian master equation, which is different from existing approaches. We compare our equations with corresponding standard weak-coupling equations by two examples, where our correlation picture formalism is more accurate, or at least as accurate as weak-coupling equations.
\end{abstract}
\maketitle

\section{introduction}
\label{sec:intro}

The recent rise in high-fidelity quantum technological devices has necessitated detailed understanding of open quantum systems and how their environment influences their performance. The theory of open quantum systems has spurred numerous possibilities to harness the power of the environments in various physical tasks \cite{eng-dis,light-harvesting}. Although in this framework quantum correlations play a key role, it has remained unresolved how to explicitly keep track of correlations between a system and its environment in the dynamical equation.

The description of the joint evolution of a quantum system and its environment (or bath) through the unitary evolution given by the Schr\"{o}dinger equation is hampered by the large dimensionality of the Hilbert space of the bath. Although there already exist a plethora of approximate methods \cite{book:Breuer-Petruccione, book:Rivas-Huelga, Nakajima, Zwanzig, Breuer-proj, devega, Maniscalco, Breuer-Kappler} for obtaining dynamics of an open quantum system by a closed set of equations, these approaches in general do not incorporate correlations between the system and its environment (see, e.g., Refs. \cite{Nazir-etal, pollock-1,paz-silva} for attempts to take correlations into account). Despite previous attempts to prove the existence of universally valid time-local Lindblad-like master equations for general dynamics \cite{Lindblad,GKS,Andersson,Lidar-Whaley}, a microscopic derivation which incorporates correlations within the dynamics has been missing. Whether a time-local master equation exists that can generally describe dynamics of systems with correlations is still an open question \cite{PRL-Chruscinski}. 

Here we resolve this issue by introducing the \textit{correlation picture}, a new technique through which we relate any correlated state of a composite system in the Schr\"{o}dinger picture to an uncorrelated description of that system. Using the dynamical picture allows us to assign a \textit{correlation parent operator} to the instantaneous system-bath correlation. This correlation parent operator is used as a key element in a microscopic derivation of the dynamical equation of the system. This introduced framework for studying open quantum dynamics enables us to derive a universal Lindblad-like (ULL) equation which is time-local and valid for any quantum dynamics. The ULL form is derived without approximations and it is valid also when the system is initially correlated with the bath. A Lindblad-like form for general dynamical equations has also been noted in recent attempts to generalize quantum mechanics \cite{Weinberg-Lindblad,Bassi-Lindblad,Diosi-Lindblad}.

We study the behavior of the ULL equation under chosen approximations and are able to derive conveniently solvable master equations which to a good degree of accurately reproduce the exact dynamics in the corresponding parameter regimes. In particular, in the vicinity of time instants where the correlations become negligible, the ULL equation reduces to a Markovian Lindblad-like (MLL) master equation, in which the jump rates are positive. We prove that this equation correctly characterizes the universal quadratic short-time behavior of the system dynamics \cite{Braun}, in contrast to the standard Lindblad equation which gives a purely linear behavior for short times \cite{Rivas-short-time,DelCampo}. In addition, we demonstrate that our MLL equation, which does not utilize the secular approximation, may in some cases faithfully describe the long-time behavior of the system. This MLL equation thus constitutes a useful framework for studying open-quantum-system dynamics beyond the weak-coupling regime. Based on the MLL methodology, for explicit or practical calculations we also develop a systematic perturbative weak-correlation expansion of higher orders which include non-Markovian effects. Interestingly, we show that even the lowest order of the constructed ULL equation (ULL$2$) can outperform corresponding weak-coupling master equations. This shows that giving the principal role to correlations rather than coupling offers a strong alternative to existing weak-coupling techniques.

\section{Correlating transformation}
\label{sec:corr-transform}

Any given system-bath state at an arbitrary instant of time $\varrho_{\mathsf{SB}}(\tau)$ can be decomposed in terms of an uncorrelated part, given by the tensor product of the instantaneous reduced states of the subsystems $\varrho_{\mathsf{S}}(\tau)=\mathrm{Tr}_{\mathsf{B}}[\varrho_{\mathsf{SB}}(\tau)]$ and $\varrho_{\mathsf{B}}(\tau) = \mathrm{Tr}_{\mathsf{S}}[\varrho_{\mathsf{SB}}(\tau)]$, and the remainder $\chi(\tau)$ which carries all correlations within the total state,
\begin{equation}
\varrho_{\mathsf{SB}}(\tau)=\varrho_{\mathsf{S}}(\tau) \otimes \varrho_{\mathsf{B}}(\tau)+\chi(\tau),
\label{rho-decomposed}
\end{equation}
where $\mathrm{Tr}_{\mathsf{S}}[\chi]$ and $ \mathrm{Tr}_{\mathsf{B}}[\chi]$ are null operators on the bath and system Hilbert spaces, respectively. Below, we refer to $\chi(\tau)$ as the correlation operator or simply the correlation. It includes all kinds of correlations, whether classical or quantum mechanical. The latter, in the form of entanglement, discord, or other more complex types, have a rich and resourceful nature in physics, e.g., in energy fluctuations of thermodynamical systems \cite{Sampaio} and in quantum information tasks \cite{book:Nielsen,vedral-rmp,modi}.

\begin{figure}[tp]
\includegraphics[width=8cm]{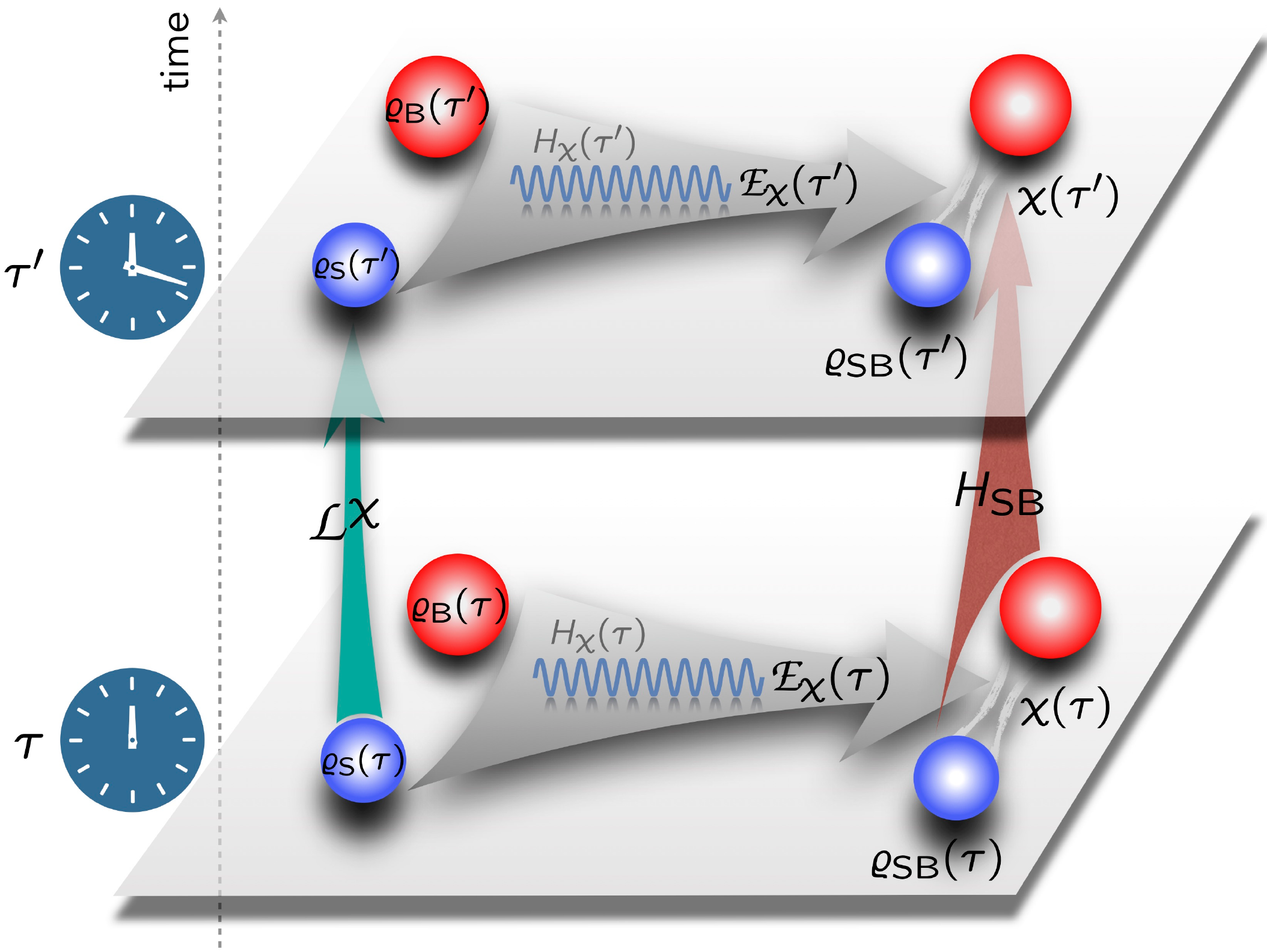}
\caption{Description of the correlation picture. At any time $\tau$ (or $\tau'$) a correlating transformation $\mathpzc{E}_{\chi}$ transforms an uncorrelated state $\varrho_{\mathsf{S}}\otimes \varrho_{\mathsf{B}}$ to a correlated state $\varrho_{\mathsf{SB}}=\varrho_{\mathsf{S}}\otimes \varrho_{\mathsf{B}}+\chi$, at the same instant of time, due to an abstract correlation-dependent parent operator given by $H_{\chi}$. Using this transformation we obtain the temporal evolution of the uncorrelated system with a universal Lindblad-like generator $\mathpzc{L}^{\chi}$ [see Eq. \eqref{LL}] constructed from $H_\mathsf{SB}$, the generator of the total system dynamics in the Schr\"{o}dinger picture.}
\label{fig:correlation-picture}
\end{figure}

To define our correlation picture, we introduce an operation $\mathpzc{E}_{\chi}$ which transforms the uncorrelated state $\varrho_{\mathsf{S}}(\tau)\otimes \varrho_{\mathsf{B}}(\tau)$ to the correlated state $\varrho_{\mathsf{SB}}(\tau)$. We refer to the opposite operation relating the correlated state to the uncorrelated one as \textit{decorrelating}. These are interesting operations, which also appear in the context of quantum statistical physics, where they are dubbed the quantum Boltzmann map and relate to the \textit{stosszahlansatz} \cite{Wolf-lectures}. Decorrelating transformations have already been investigated in the literature \cite{D'Ariano-Demkowicz:PRL}, and it is known that a universal decorrelating machine would violate linearity of quantum mechanics \cite{Terno}. Our correlating transformation, indeed, is not universal, i.e., $\mathpzc{E}_{\chi}$ depends on the states. To gain insight on how to find $\mathpzc{E}_{\chi}$, it is helpful to start with the case of creating a weakly correlated state. We emphasize that this case study will serve merely to set the scene, whereas later we do not assume any weak correlation in our general analysis.

\textit{Correlating transformation for a weakly correlated state.---}Let us aim to transform an uncorrelated state $\varrho_{\mathsf{SB}}=\varrho_{\mathsf{S}}\otimes\varrho_{\mathsf{B}}$ to a \textit{weakly} correlated state $\varrho_{\mathsf{SB}}=\varrho_{\mathsf{S}}\otimes\varrho_{\mathsf{B}}+\chi$ with $\Vert \chi\Vert\ll1$ ($\Vert\cdot\Vert$ being the operator norm). A natural way to do so is to apply an entangling gate on the uncorrelated state. Consider, e.g., an entangling gate \cite{entangler3} described by a unitary transformation $U_{\chi}=e^{-i H_{\chi}}$, with $\Vert H_{\chi}\Vert\ll1$ \textcolor{blue}{\cite{nb}}. That is, here the transformation ($U_{\chi}$) is assumed to be given and we look for the corresponding generated correlation operator $\chi$. This gate results in a weakly correlated state $U_{\chi}\varrho_{\mathsf{S}}\otimes \varrho_{\mathsf{B}} U_{\chi}^{\dag} \approx \varrho_{\mathsf{S}}\otimes \varrho_{\mathsf{B}}-i [H_{\chi},\varrho_{\mathsf{S}}\otimes \varrho_{\mathsf{B}}]$. Comparing this state with $\varrho_{\mathsf{SB}}=\varrho_{\mathsf{S}}\otimes\varrho_{\mathsf{B}}+\chi$, we observe that up to $O(\|H_{\chi}\|^2)$ the correlation $\chi$ obeys the following equation:
\begin{align}
\chi = -i[H_{\chi},\varrho_{\mathsf{S}}\otimes \varrho_{\mathsf{B}}].
\label{correlate}
\end{align}
We refer to the dimensionless operator $H_{\chi}$ as the correlation parent operator. The equivalent correlating transformation in this regime is obtained by inserting Eq. \eqref{correlate} into Eq. \eqref{rho-decomposed} as $\varrho_{\mathsf{SB}}=\mathpzc{E}_{\chi}[\varrho_{\mathsf{S}}\otimes\varrho_{\mathsf{B}}]:=\varrho_{\mathsf{S}}\otimes\varrho_{\mathsf{B}}-i [H_{\chi},\varrho_{\mathsf{S}}\otimes\varrho_{\mathsf{B}}]$.

\textit{Correlating transformation for a general state.}---In the above scenario, we applied a \textit{known} unitary transformation to create a correlated state and obtained the correlation $\chi$ in terms of the input uncorrelated state $\varrho_{\mathsf{S}}\otimes \varrho_{\mathsf{B}}$. Now, we return to our question of interest: to find the operation which transforms a given $\varrho_{\mathsf{S}}\otimes\varrho_{\mathsf{B}}$ to its associated $\varrho_{\mathsf{SB}}$ with a definite correlation operator $\chi$ which is not necessarily small.

Although for any given pair of quantum states $\varrho_{1}$ and $\varrho_{2}$, it is possible to find a quantum map or channel $\mathpzc{E}$ such that $\mathpzc{E}[\varrho_{1}]=\varrho_{2}$ \cite{Rabitz}, such a map does not necessarily have a unitary representation. Hence if we postulate the form given in Eq. \eqref{correlate} as the equation connecting $\varrho_{\mathsf{S}}\otimes \varrho_{\mathsf{B}}$ and $\chi$, for some $H_{\chi}$, it does not generally have a Hermitian solution for $H_{\chi}$. However, with the insight gained from the weakly correlated case, we still choose $\chi$ as in Eq. \eqref{correlate} but relax the Hermiticity constraint on $H_{\chi}$. Since the left side of Eq. \eqref{correlate} is Hermitian, to ensure the Hermiticity of the right side, with a non-Hermitian $H_{\chi}$, we introduce a generalized commutator $\lbrac A,B \rbrac=A B-B^{\dag} A^{\dag}$  \cite{MR} and replace Eq. \eqref{correlate} with
\begin{align}
\chi=-i\lbrac H_{\chi},\varrho_{\mathsf{S}}\otimes \varrho_{\mathsf{B}}\rbrac.
\label{chi-H_{chi}}
\end{align}
We consider this relation as the general definition of $H_{\chi}$, valid for any $\chi$. Fortunately, this equation has always a solution for $H_{\chi}$ provided that $P_{0}(\tau)\chi(\tau) P_{0}(\tau)=0$, where $P_{0}(\tau)$ is the projector onto the null-space of $\varrho_{\mathsf{S}}(\tau)\otimes \varrho_{\mathsf{B}}(\tau)$ \cite{Djordjevic}. We prove that this condition is always satisfied \cite{SM}---and we also provide the solution for $H_{\chi}$. Using Eq. \eqref{chi-H_{chi}}, we define our correlating transformation $\mathpzc{E}_{\chi}$ as
\begin{align}
\varrho_{\mathsf{SB}}=\mathpzc{E}_{\chi}[\varrho_{\mathsf{S}}\otimes\varrho_{\mathsf{B}}]:=\varrho_{\mathsf{S}}\otimes\varrho_{\mathsf{B}}-i\lbrac H_{\chi},\varrho_{\mathsf{S}}\otimes\varrho_{\mathsf{B}}\rbrac.
\label{correlating-trans}
\end{align}

We note that neither the solution of Eq. \eqref{chi-H_{chi}} for $H_{\chi}$ nor the formal connection of $\chi$ to $\varrho_{\mathsf{S}}\otimes \varrho_{\mathsf{B}}$ is unique (one may suggest other forms). This means that there is a kind of \textit{gauge freedom} in choosing the correlating transformation. However, this nonuniqueness does not affect our formalism, because any solution of Eq. \eqref{chi-H_{chi}} and any properly chosen form for $\mathpzc{E}_{\chi}$ must generate the same given $\chi$---which is the physical quantity which should remain unchanged under these gauge transformations.

Note that the uncorrelated state $\varrho_\mathsf{S}\otimes \varrho_\mathsf{B}$ is not the state of the total system (because in general $\chi\neq 0$); rather, we take this state as the description of the total system in a pertaining correlation picture. In order to keep the dynamics of the state in this picture faithful to the Schr\"{o}dinger equation, we need to devise an appropriate formulation. Figure \ref{fig:correlation-picture} depicts the correlating transformation and the correspondingly emerging correlation picture---which is explained in detail below.

\section{Correlation picture dynamics}

We aim to apply our correlation picture transformation between the correlated and uncorrelated states, $\varrho_{\mathsf{SB}}$ and $\varrho_{\mathsf{S}}\otimes \varrho_{\mathsf{B}}$, respectively, to obtain a dynamical equation for $\varrho_{\mathsf{S}}$. Our approach can be considered in the spirit of the derivation of the Nakajima--Zwanzig (NZ) equation \cite{Nakajima,Zwanzig}. However, rather than applying a decorrelating projector to omit system--bath correlations (while implicitly carrying correlations into another complementary equation), we employ our correlating transformation within the Schr\"{o}dinger equation of the total system. Hence we shall explicitly retain contributions to the system dynamics from the correlations in the total system.

Let us assume that the total Hamiltonian of the system and the bath is given by $H_{\mathsf{SB}}=H_{\mathsf{S}}+H_{\mathsf{B}}+H_{\mathrm{I}}$, where the last term denotes the system--bath interaction. We employ the correlating transformation \eqref{correlating-trans} to obtain a counterpart for the Schr\"{o}dinger picture generator $\mathpzc{D}_{\mathpzc{s}}[\circ]=-i[H_{\mathsf{SB}},\circ]$ (throughout this paper we have assumed the natural units $\hbar \equiv k_{\mathrm{B}}\equiv 1$). More precisely, we define this operator $\mathpzc{D}_{\mathpzc{c}}$ such that
\begin{align}
\mathpzc{D}_{\mathpzc{s}}[\varrho_{\mathsf{SB}}(\tau)]
=&\mathpzc{D}_{\mathpzc{c}}[\varrho_{\mathsf{S}}(\tau)\otimes\varrho_{\mathsf{B}}(\tau)].
\label{sch-cor-pic}
\end{align}
By inserting the correlating transformation \eqref{correlating-trans} in the Schr\"{o}dinger equation as $\mathpzc{D}_{\mathpzc{s}}[\varrho_{\mathsf{SB}}(\tau)]=\mathpzc{D}_{\mathpzc{s}}\big[\mathpzc{E}_{\chi}[\varrho_{\mathsf{S}}(\tau)\otimes \varrho_{\mathsf{B}}(\tau)]\big]$ we obtain the correlation-picture generator as
\begin{align}
\mathpzc{D}_{\mathpzc{c}}[\circ]=-i[H_{\mathsf{SB}},\circ]-\big[H_{\mathsf{SB}},\lbrac H_{\chi},\circ \rbrac \big].
\label{cor-gen}
\end{align}
Although the dynamics described by $\mathpzc{D}_{\mathpzc{c}}$ is fully equivalent to the Schr\"{o}dinger picture dynamics governed by $\mathpzc{D}_{\mathpzc{s}}$, working in the correlation picture offers clear advantages.

\section{Universality of the Lindblad-like form for open system dynamics} 
\label{sec:ULLeq}

We show in the following that working in the correlation picture leads to a universal Lindblad-like (ULL) equation. In the next sections, we shall discuss applications and further properties of the ULL formalism.

\subsection{General theory}
\label{sec:gentheory}

From Eq. \eqref{cor-gen} we can readily obtain the dynamics of the subsystem by tracing over the bath degrees of freedom in $\dot{\varrho}_{\mathsf{SB}}(\tau)=\mathpzc{D}_{\mathpzc{c}}[\varrho_{\mathsf{S}}(\tau)\otimes \varrho_{\mathsf{B}}(\tau)]$. To show that the subsystem dynamics has a Lindblad-like form, we use the expansions of $H_{\mathrm{I}}=\sum_{i=1}^{d_{\mathsf{S}}^2-1}\mathpzc{S}_{i}\otimes \mathpzc{B}_{i}$ and $H_{\chi}(\tau) =\sum_{j=0}^{d_{\mathsf{S}}^2-1} \mathpzc{S}_j \otimes \mathpzc{B}_j^{\chi}(\tau)$, where $\{\mathpzc{S}_i\}_{i=1}^{d_{\mathsf{S}}^2-1}$ is the basis of Hermitian operators defined on the system Hilbert space. Here, $d_{\mathsf{S}}$ is the dimension of the Hilbert space of the system and $\mathpzc{S}_{0}=\mathbbmss{I}$. We emphasize the difference between the operators $\mathpzc{B}_i$ and $\mathpzc{B}_i^{\chi}(\tau)$, which are related to $H_{\mathrm{I}}$ and $H_{\chi}$, respectively. Inserting these expansions into $\dot{\varrho}_{\mathsf{S}}=\mathrm{Tr}_{\mathsf{B}}\big[-i[H_{\mathsf{SB}}, \varrho_{\mathsf{S}}\otimes \varrho_{\mathsf{B}}]-\big[H_{\mathsf{SB}},\lbrac H_{\chi},\varrho_{\mathsf{S}} \otimes \varrho_{\mathsf{B}} \rbrac \big]\big]$ yields
\begin{align}
\dot{\varrho}_{\mathsf{S}}=&-i\big[H_{\mathsf{S}} + \mathrm{Tr}_{\mathsf{B}}[H_{\mathrm{I}}\,\varrho_{\mathsf{B}}],\varrho_{\mathsf{S}}\big]\nonumber\\
&- \textstyle{\sum_{i\neq 0,j}} \big[\mathpzc{S}_i,c_{ij}\mathpzc{S}_j \varrho_{\mathsf{S}} -c_{ij}^{\ast}\varrho_{\mathsf{S}} \mathpzc{S}_j \big],
\label{pre-LL}
\end{align}
where
\begin{align}
c_{ij}(\tau)=\ave{ \mathpzc{B}_i\mathpzc{B}_j^{\chi}(\tau)}_{\mathsf{B}}
\label{c_{ij}0}
\end{align}
are the elements of the covariance matrix $C$ of the bath operators (with $\ave{\circ}_{\mathsf{B}}=\mathrm{Tr}_{\mathsf{B}}[\varrho_{\mathsf{B}}(\tau) \circ]$). Here, unlike the standard Lindblad equation, these bath operators are obtained not only from the interaction Hamiltonian but also from the correlation parent operator $H_{\chi}$ defined in Eq. \eqref{chi-H_{chi}}. We rewrite $C(\tau)$ in Eq. \eqref{pre-LL} in terms of its Hermitian and anti-Hermitian parts as $C(\tau)=A(\tau)+i B(\tau)$, with Hermitian matrices $A(\tau)=[a_{ij}(\tau)]$ defined by $a_{ij}(\tau)=(1/2)[C(\tau)+C^{\dag}(\tau)]_{ij}$ and  $B(\tau)=[b_{ij}(\tau)]$ defined by $b_{ij}(\tau)=(-i/2)[C(\tau)-C^{\dag}(\tau)]_{ij}$, for $i,j\geqslant 1$. This leads to an exact Lindblad-like master equation for the system,
\begin{align}
\dot{\varrho}_{\mathsf{S}}&=\mathpzc{L}^{\chi}[\varrho_{\mathsf{S}}]\nonumber\\
&=-i[H_{\mathsf{S}}+{\mathbbmss{h}}_{{\mathsf{L}}}^{\chi},\varrho_{\mathsf{S}}]+\sum_{m} \gamma_m^{\chi} \big(2 L^{\chi}_m\varrho_{\mathsf{S}} L^{\chi \dag}_m-\{L^{\chi \dag}_m L^{\chi}_m,\varrho_{\mathsf{S}}\}),
\label{LL}
\end{align}
where $\{X,Y\}\equiv XY+YX$ denotes the anticommutator. In this equation, the quasi-rates $\gamma_m^{\chi}$ are the eigenvalues of the matrix $A(\tau)$. The jump operators are given by $L_m^{\chi}=\sum_{j\neq 0} V_{mj} \mathpzc{S}_j$ where \{$V_{mj}\}_j$ are the elements of the eigenvector corresponding to the eigenvalue $\gamma^{\chi}_m$. Furthermore, the Lamb-shift-like Hamiltonian ${\mathbbmss{h}}_{{\mathsf{L}}}^{\chi}(\tau)=\ave{H_{\mathrm{I}}}_{\mathsf{B}}+2 \textstyle{\sum_{i\neq0}} \mathrm{Im}\big(c_{i0}(\tau)\big)\mathpzc{S}_i+\textstyle{\sum_{i \neq 0, j \neq 0 }} b_{ij}(\tau) \mathpzc{S}_i\mathpzc{S}_j$, where $b_{ij}$ are the elements of the matrix $B(\tau)$. For more details on the derivation see Ref. \cite{SM}. 

Several remarks are in order here:

(i) Although by combining known results in the literature one may infer time-local Lindblad-like forms for the dynamics under the assumptions of linearity or absence of initial system-environment correlations \cite{Andersson, GKS, book:Breuer-Petruccione}, the ULL equation  (\ref{LL}) is completely general; we have made no assumptions on the initial system-environment correlations or the strength of the system-environment interaction. In addition, the ULL equation is built explicitly on a microscopic theory of correlations in the total system.

(ii) Unlike the well-known Markovian embedding, where a non-Lindblad (non-Markovian) evolution is mapped to a Lindblad (Markovian) evolution for a specific \textit{larger} system employing an ancillary system \cite{garrahan}, we have proven here that the Lindblad-like form is derived for the open-system dynamics itself.

(iii) We note that the coefficients in the ULL master equation (\ref{c_{ij}0}) refer to the correlation functions $\ave{ \mathpzc{B}_i\mathpzc{B}_j^{\chi}(\tau)}_{\mathsf{B}}$, defined by the correlation parent operator and are thus different from the conventional correlation functions of bath operators $\ave{ \mathpzc{B}_i(\tau) \mathpzc{B}_j(\tau)}_{\mathsf{B}}$ appearing in the standard Markovian Lindblad equation \cite{book:Breuer-Petruccione}, where the average is taken on a constant bath state and $\mathpzc{B}_{i}(\tau)=U^{\dag}_{\mathsf{B}}(\tau) \mathpzc{B}_{i}U_{\mathsf{B}}(\tau)$ [with $U_{\mathsf{B}}(\tau)=e^{-i\tau H_{\mathsf{B}}}$].

(iv) Since $\mathpzc{L}^{\chi}$ depends on the state of the system, Eq. \eqref{LL} is formally a nonlinear equation. Indeed, the linearity constraint on the full dynamics of quantum systems does not imply a similar restriction on the dynamics of a subsystem, and this nonlinearity is naturally expected for a general dynamical equation. Nevertheless, we show in Ref. \cite{SM} that our ULL master Eq. \eqref{LL} is linear in two important cases: (a) if there is no initial correlation, i.e., $\chi(0)=0$, where we show $\chi(\tau)$ can be explicitly expressed in terms of the system-bath product state, and (b) if the domain of $\mathpzc{L}^{\chi}$ is restricted to a set of states $\{\varrho_{\mathsf{S}}^{(i)}\}$ forming a convex decomposition of the state of the system, i.e., $\varrho_{\mathsf{S}}=\sum_{i} p_{i} \varrho_{\mathsf{S}}^{(i)}$, but here the initial total state may be correlated. 

\subsection{Example I: Jaynes--Cummings model with initial correlation}
\label{sec:JCexample}

To illustrate universality of the dynamical Eq. \eqref{LL}, even in the presence of initial system--bath correlations, we begin with a proof-of-principle example, the well-known, exactly solvable Jaynes--Cummings model \cite{Book: Gerry-Knight}, and show that the dynamics of the two-level system is described by the ULL equation even when the system is correlated with a bosonic mode.

Consider a two-level system interacting with a single bosonic mode under the Jaynes--Cummings Hamiltonian. The system Hamiltonian is $H_{\mathsf{S}}=(\omega_{0}/2) \sigma_{z}$, where $\sigma_{\pm}=(\sigma_{x}\pm i \sigma_{y})/2$ and $\sigma_{x}$, $\sigma_{y}$, and $\sigma_{z}$ are the $x$, $y$, and $z$ Pauli operators, respectively, the bosonic bath Hamiltonian is $H_{\mathsf{B}}=\omega \hat{a}^\dagger \hat{a} $, where $\hat{a}^\dagger$ ($\hat{a}$) is the creation (annihilation) operator of the bosonic mode, and $H_{\mathrm{I}}=\lambda(\sigma_{+}\otimes \hat{a} +\sigma_{-}\otimes \hat{a}^{\dag})$ describes the system--bath interaction. For simplicity, we assume that $\omega_{0}=\omega$ and that the initial state of the total system is in a correlated state $|\psi(0)\rangle=r_{1}|\mathrm{e},0\rangle+r_{2}|\mathrm{g},1\rangle$, where both $r_{1}$ and $r_{2}$ are real numbers. Choosing $\mathpzc{S}_{0}=\mathbbmss{I}/\sqrt{2}$, $\mathpzc{S}_{1}=\sigma_{x}/\sqrt{2}$, $\mathpzc{S}_{2}=\sigma_{y}/\sqrt{2}$, and $\mathpzc{S}_3=\sigma_{z}/\sqrt{2}$ as the basis operators, we can find $\{\mathpzc{B}^{\chi}_{i}\}_i$ (see Ref. \cite{SM}). Thus, the bath covariances are obtained as $c_{10}=c_{20}=0$, $c_{11}=c_{22}=\lambda (-2 i r_{1} r_{2}+\alpha_{1} \alpha_{2})/(2\alpha_{1}^2-2)$, and $c_{12}=-c_{21} =\lambda (2 r_{1} r_{2} \alpha_{1} +i \alpha_{2})/\left(2(1- \alpha_{1}^2)\right)$, where $\alpha_{1}=(1-2r_{1}^2)\cos(2\lambda \tau)$ and $\alpha_{2}=(1-2r_{1}^2)\sin(2\lambda \tau)$. Following the steps of the derivation of Eq. \eqref{LL}, we obtain
\begin{align}
\dot{\varrho}_{\mathsf{S}}=&-i [H_{\mathsf{S}}+\widetilde{\omega}_{0} \sigma_z, \varrho_{\mathsf{S}}] +\gamma_{1}^{\chi} (2 \sigma_{-} \varrho_{\mathsf{S}} \sigma_{+} - \{\sigma_{+} \sigma_{-}, \varrho_{\mathsf{S}}\} )\nonumber\\
&-\gamma_{2}^{\chi} (2 \sigma_{+} \varrho_{\mathsf{S}} \sigma_{-} - \{\sigma_{-} \sigma_{+}, \varrho_{\mathsf{S}} \}),
\end{align}
where
$\widetilde{\omega}_{0}=4 \lambda r_{1} r_{2} \alpha_{1}/\big(1+4 r_{1}^2 -4 r_{1}^4 - (\alpha_{1}^2-\alpha_{2}^2)\big)$, $\gamma_{1}^{\chi}=-\lambda\alpha_{2}/\big(2(1-\alpha_{1})\big)$, and $\gamma_{2}^{\chi}=\lambda\alpha_{2}/\big(2(1+\alpha_{1})\big)$ \cite{SM}. We emphasize that this equation is in the ULL form and is valid even with initial system--bath correlations.

\section{Reduction to a Markovian equation}
\label{sec:MLLeq}

Based on our general dynamical equation where system--bath correlations are fully incorporated, we can obtain simpler expressions for the case where the correlations are small. This approach is valid, e.g., in the vicinity of time instants at which the correlation vanishes or becomes negligible. In other words, we introduce a \textit{weak-correlation} approximation. In such cases, we can simplify our ULL master equation into a Markovian Lindblad-like (MLL) master equation, in which jump rates are positive---as expected from Markovian dynamics \cite{Maniscalco}. Below, we show that this equation correctly characterizes the universal quadratic short-time behavior of the system dynamics where the standard Lindblad master equation may fail \cite{Rivas-short-time,DelCampo}. 

We assume that at $\tau_{0}$ the correlation vanishes. Without loss of generality we take $\tau_{0}=0$, thus $\chi(0)=0$. We allow the correlations to accumulate in the subsequent time steps due to the dynamics. To first order in the time argument $\tau$, we find that the correlation satisfies Eq. \eqref{chi-H_{chi}} with $H_{\chi}(\tau)=\tau \widetilde{H}_{\mathrm{I}}(\tau)$, where $\widetilde{H}_{\mathrm{I}}(\tau)= \textstyle{\sum_{i\neq 0}} \mathpzc{S}_i\otimes (\mathpzc{B}_i-\ave{\mathpzc{B}_i}_{\mathsf{B}})- \textstyle{\sum_{i\neq 0}} \ave{\mathpzc{S_i}}_{\mathsf{S}}\mathpzc{B_i}$ and $\ave{\circ}_{\mathsf{S}}=\mathrm{Tr}_{\mathsf{S}}[\varrho_{\mathsf{S}}(\tau) \circ]$. Thus, from the knowledge of $H_{\chi}$, we can read $\mathpzc{B}_j^{\chi}(\tau)=\tau(\mathpzc{B}_j-\langle \mathpzc{B}_j\rangle_{\mathsf{B}})$, where $j\geqslant 1$. Substituting these expressions into Eq. \eqref{c_{ij}0} the bath covariance matrix becomes
\begin{align}
c_{ij}(\tau)&=\tau\, \mathrm{Cov}_{\mathsf{B}}(\mathpzc{B}_i,\mathpzc{B}_j),~~~i,j\geqslant 1,
\label{Markov-cov0}
\end{align}
where $\mathrm{Cov}_{\mathsf{B}}(O_{1},O_{2})=\ave{O_{1} O_{2}}_{\mathsf{B}}-\ave{O_{1}}_{\mathsf{B}}\ave{O_{2}}_{\mathsf{B}}$. Since the covariance matrix $\bm{c}(\tau)$ is positive-semidefinite, $a_{ij}(\tau)=c_{ij}(\tau)$ and $b_{ij}(\tau)=0$. The positivity of $A$ implies positivity of the rates $\gamma_m^{\chi}\geqslant  0$, which is a necessary feature of a Markovian dynamical evolution. To obtain an equation with no dependence on the state of the bath (recall that $\mathbbmss{h}_{{\mathsf{L}}}^{\chi}(\tau)$ and $c_{ij}(\tau)$ depend on $\varrho_{\mathsf{B}}(\tau)$), we also expand $\varrho_{\mathsf{B}}(\tau)$ around $\tau_{0}=0$ and keep relevant terms up to the first order in $\tau$. Thus, we obtain
\begin{align}
a_{ij}(\tau) \approx & \tau\, \mathrm{Cov}_{\mathsf{B}_{0}}(\mathpzc{B}_i,\mathpzc{B}_j);~~~~~~ b_{ij}(\tau) =0;\nonumber\\
 {{\mathbbmss{h}}_{{\mathsf{L}}}^{\chi}}(\tau) \approx &\ave{H_{\mathrm{I}}}_{\mathsf{B}_{0}}-i \tau \ave{[H_{\mathrm{I}},\widetilde{H}_{\mathsf{B}}]}_{\mathsf{B}_{0}}\nonumber\\
 &-2 \tau \textstyle{\sum_{(i,j) \neq (0,0)}} \ave{\mathpzc{S}_j}_{\mathsf{S}_{0}} \,\mathrm{Im}\ave{\mathpzc{B}_i \mathpzc{B}_j}_{\mathsf{B}_{0}} \mathpzc{S}_i,
\label{aij-cij-markovian}
\end{align}
where subscripts $\mathsf{B}_{0}$ and $\mathsf{S}_{0}$ indicate that the averages or covariances are taken with respect to $\varrho_{\mathsf{B}}(0)$ and $\varrho_{\mathsf{S}}(0)$ rather than $\varrho_{\mathsf{B}}(\tau)$ and $\varrho_{\mathsf{S}}(\tau)$. In Eq. \eqref{aij-cij-markovian}, we have defined $\widetilde{H}_{\mathsf{B}}=H_{\mathsf{B}}+\ave{H_{\mathrm{I}}}_{\mathsf{S}_{0}}$ (see Ref. \cite{SM} for more details). Equation \eqref{LL}---bearing in mind Eq. \eqref{aij-cij-markovian}---describes the short-time Markovian dynamics around a point of vanishing correlation. 

We emphasize that our weak-correlation assumption is {\textit{exact}} up to the first order in $\tau$. If we extend this Markovian dynamical equation to longer times, it may still work as an approximation for the exact dynamics, e.g., when the correlation becomes repeatedly zero \cite{Lidar-Whaley,SM}. Although at first sight expanding around a point of  vanishing correlation may seem equivalent to the standard Born approximation, we will illustrate in the next example that the MLL equation can be different from the Redfield equation.  
In addition, unlike the Redfield equation \cite{book:Breuer-Petruccione, book:Rivas-Huelga}, the MLL equation always keeps the state positive, hence avoiding the so-called slippage issue which afflicts the Redfield equation \cite{slippage,Whitney}.

\begin{figure*}[tp]
\includegraphics[width=8cm]{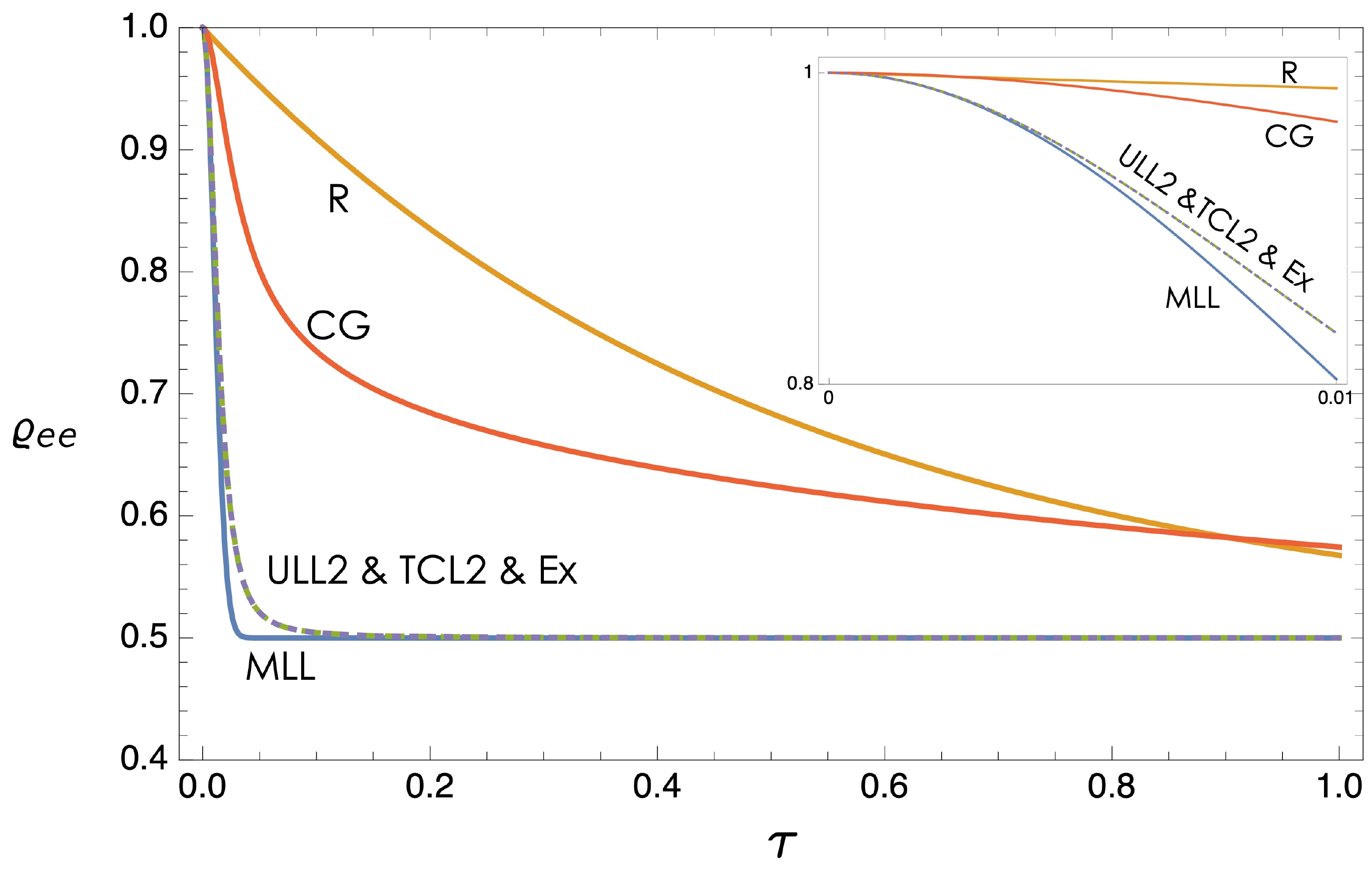}\hskip5mm\includegraphics[width=8cm]{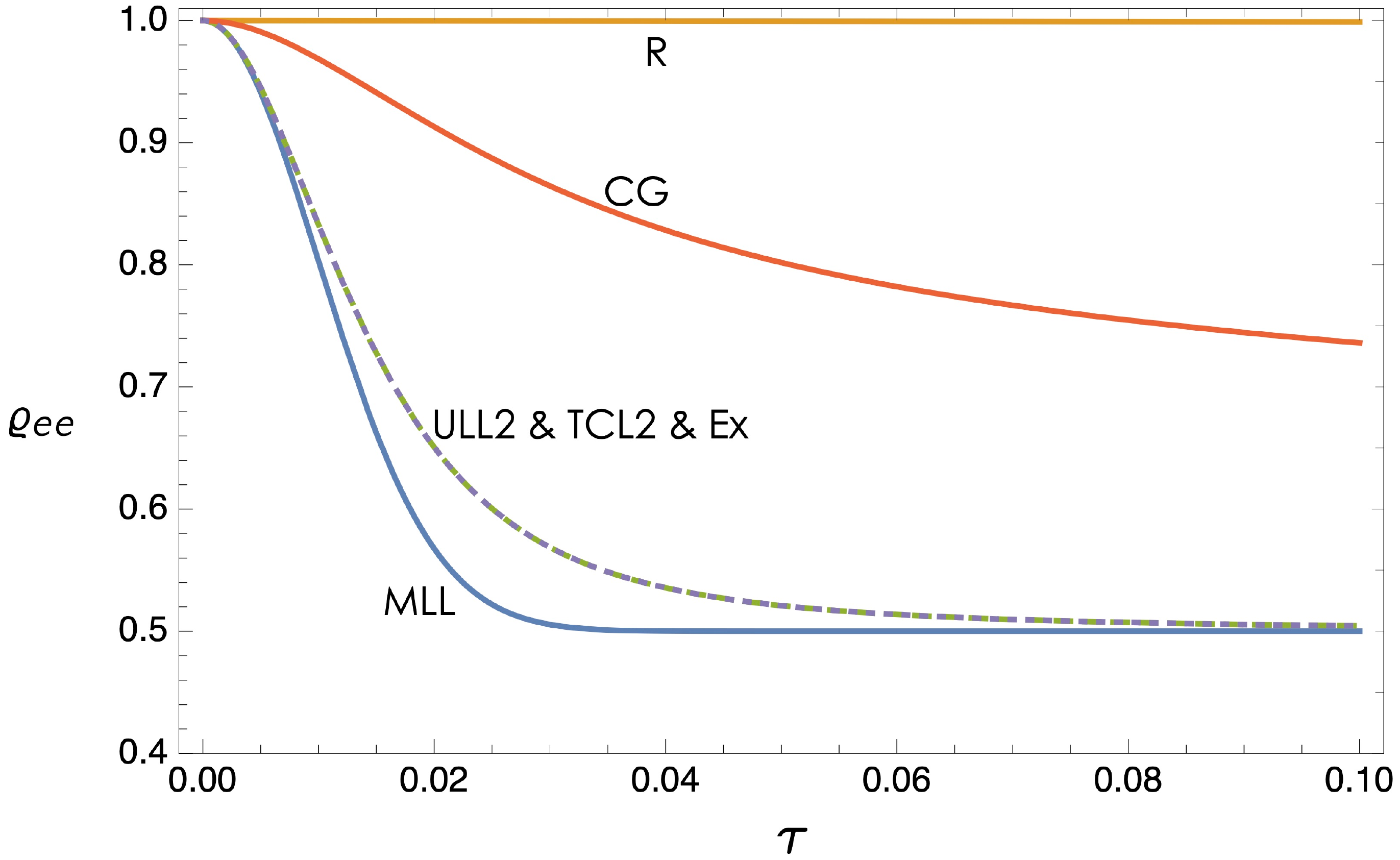}
\caption{An atom in a bosonic bath. (Left) Population of the excited state of the atom as a function of time for $\beta=1$, $\eta=0.5$, $\omega_c=100$, and $\omega_{0}=0$ (all in natural units), where the atom is initially in the excited state. The labels TCL$2$ \& Ex, MLL, ULL$2$, CG, and R denote, respectively, the data from the second-order time-convolutionless (= Exact), MLL, time-local ULL$2$ (= Exact), coarse-graining, and the Redfield (= Markovian Lindblad) solutions. The inset plot is a magnified depiction of the population vs. time for short times. (Right) The same as Left but for $\beta=100$.}
\label{fig:spin-boson-1}
\end{figure*}

A final remark regarding the applicability of the MLL approximation in other regimes is in order. If the system has an asymptotic state, a candidate for such a state can be $\varrho^{\star}=\sum_{n} \langle E_{n}|\varrho_{\mathsf{SB}}(0) |E_n\rangle  |E_n\rangle\langle E_n|$, where $\{|E_n\rangle\}$ are the eigenvectors of $H_{\mathsf{SB}}$ \cite{SM}. Now assume that (i) the system is strongly interacting with the bath, i.e., $H_{\mathrm{I}}$ is the dominant term in the total Hamiltonian such that $H_{\mathsf{SB}}\approx H_{\mathrm{I}}$, (ii) the interaction Hamiltonian contains only one term, $H_{\mathrm{I}}=\mathpzc{S}\otimes \mathpzc{B}$, and (iii) the initial state of the system and the bath is uncorrelated, $\chi(0)=0$. It is straightforward to show that under these conditions $\varrho^{\star}$ is an uncorrelated state, and thus, also in this case an MLL equation describes the asymptotic dynamics \cite{SM}. In the next section we provide another case where the MLL approximation also holds, and later we illustrate these behaviors with two examples.

\section{Dynamics of the correlation: systematic weak-correlation expansion}
\label{sec:systematic}

The above MLL approximation shows that the correlation picture and the ULL equation can offer far-reaching practical implications beyond their fundamental appeal. We identify that the \textit{weak-correlation approximation} is the basic ingredient of the MLL equation. Expanding upon this is desirable as it can make the ULL methodology more amenable to practical investigations of a diverse set of systems. 

\subsection{General theory}
\label{sec:UUL-formalism}

To go further and demonstrate that the ULL equation systematically enables such a rich approximative structure, in the following we develop a perturbative weak-correlation expansion for the ULL equation. In particular, by using the MLL toolbox as a starting point for a perturbative expansion of the correlation matrix in the interaction picture, we find an exact dynamical equation for the correlation $\bm{\chi}$ (boldface denotes the interaction picture) and expand it in terms of the interaction Hamiltonian $\bm{H}_{\mathrm{I}}$ as 
\begin{equation}
\bm{\chi}(\tau)=\textstyle{\sum_{\ell=0}^{\infty}} \bm{f}_{\ell}(\tau),
\label{wcorr-expansion}
\end{equation}
 where $\bm{f}_{0}(\tau)\equiv \bm{\chi}(0)$ and $\bm{f}_{\ell}(\tau)$ is of the order of $\bm{H}^{\ell}_{\mathrm{I}}$, for $\ell\geqslant1$. We incorporate the first $\ell+1$ terms within this expansion in the ULL dynamical equations for the system and the bath and derive the associated $\ell$th-order approximate ULL equations, referred to as the ``ULL$\ell$'' equations. 

Let us provide an outline of the weak-correlation expansion---for details see Ref. \cite{SM}. We start by the decomposition of the state of the total system as in Eq. (\ref{rho-decomposed}). The total system at a later time $\tau+\xi$ is given by
\begin{equation}
\varrho_{\mathsf{SB}}(\tau+\xi)=U_{\xi}\varrho_{\mathsf{S}}(\tau)\otimes\varrho_{\mathsf{B}}(\tau) U_{\xi}^{\dag}+U_{\xi}\chi(\tau) U_{\xi}^{\dag},
\label{eq1}
\end{equation}
where $U_{\xi}=e^{-i H_{\mathsf{SB}}\xi}$. The first term of the above equation represents the evolution of an uncorrelated state $\varrho_{\mathsf{S}}(\tau)\otimes\varrho_{\mathsf{B}}(\tau)$ which up to the first order in $\xi$ is given by the MLL dynamical equation. Using Eq. (\ref{eq1}) and the definition of the correlation opertaor \eqref{rho-decomposed} at time $\tau+\xi$ one obtains after some algebra \cite{SM}  
\begin{align}
\dot{\bm{\chi}}(\tau)=-i [\widetilde{\bm{H}}_{\mathrm{I}}(\tau),\bm{\varrho}_{\mathsf{S}}(\tau)\otimes \bm{\varrho}_{\mathsf{B}}(\tau)]+\bm{\mathcal{Y}}_{\tau}[\bm{\chi}(\tau)],
\label{chi-dot-intpic}
\end{align}
where we have defined 
\begin{align}
\bm{\mathcal{Y}}_{\tau}[\circ]:=&-i [\bm{H}_{\mathrm{I}}(\tau),\circ]+i \mathrm{Tr}_{\mathsf{B}}[\bm{H}_{\mathrm{I}}(\tau) \, ,\circ]\otimes \bm{\varrho}_{\mathsf{B}}(\tau)\nonumber\\
&+i \bm{\varrho}_{\mathsf{S}}(\tau)\otimes\mathrm{Tr}_{\mathsf{S}} [\bm{H}_{\mathrm{I}}(\tau) \, ,\circ].
\label{def-of-F-intpic}
\end{align}
Using integration and iterations a solution to Eq. \eqref{chi-dot-intpic} is obtained as 
\begin{widetext}
\begin{align}
\bm{\chi}(\tau)=& \textstyle{\sum_{k=0}^{\infty}} \int_{0}^{\tau} ds_{1} \int_{0}^{s_{1}} ds_{2}\, \cdots \int_{0}^{s_{k-1}} ds_{k}\, \bm{\mathcal{Y}}_{s_{1}} \Big[\bm{\mathcal{Y}}_{s_{2}}\cdots \big[\bm{\mathcal{Y}}_{s_{k}} [\bm{\chi}(0)]\big]\ldots\Big]\nonumber\\
&+ \textstyle{\sum_{k=0}^{\infty}} \int_{0}^{\tau} ds_{1} \int_{0}^{s_{1}} ds_{2}\, \cdots \int_{0}^{s_{k-1}} ds_{k}\, \bm{\mathcal{Y}}_{s_{1}} \left[\bm{\mathcal{Y}}_{s_{2}}\cdots \Big[\bm{\mathcal{Y}}_{s_{k}}\big[-i\int_{0}^{s_k} ds \, [\widetilde{\bm{H}}_{\mathrm{I}}(s),\bm{\varrho}_{\mathsf{S}}(s)\otimes \bm{\varrho}_{\mathsf{B}}(s)]\big]\Big]\ldots\right],
\label{chi-exact-intPic}
\end{align}
\end{widetext}
which is symbolically in the form of Eq. (\ref{wcorr-expansion}). Starting from Eq. \eqref{chi-exact-intPic}, we can systematically approximate the correlation and hence the ULL dynamical equation. If the initial system-bath correlation vanishes, the first-order approximation (with respect to $\widetilde{\bm{H}}_{\mathrm{I}}$) gives 
\begin{align}
\bm{\chi}^{(1)}(\tau)=-i \textstyle{\int_{0}^{\tau}} ds \, [\widetilde{\bm{H}}_{\mathrm{I}}(s),\bm{\varrho}_{\mathsf{S}}(s)\otimes \bm{\varrho}_{\mathsf{B}}(s)].
\label{chi-1-uncorrelated-initial}
\end{align}
Using this relation to derive the correlation parent operator \cite{SM} one can obtain a weak-correlation approximation for the ULL dynamical equation, which is of second order with respect to $\widetilde{\bm{H}}_{\mathrm{I}}$---hence it is referred to as the ULL$2$ equation. We need to solve the coupled differential equations for $\dot{\varrho}_{\mathsf{S}}(\tau)$ and $\dot{\varrho}_{\mathsf{B}}(\tau)$ in a self-consistent fashion, to obtain approximate states of the system and the bath---see Ref. \cite{SM} for details and further elaborations. 

For sufficiently short times Eq. (\ref{chi-1-uncorrelated-initial}) reduces to 
\begin{equation}
\bm{\chi}^{(1)}_{\mathrm{MLL}}(\tau)\approx-i \tau [\widetilde{\bm{H}}_{\mathrm{I}}(\tau),\bm{\varrho}_{\mathsf{S}}(\tau)\otimes \bm{\varrho}_{\mathsf{B}}(\tau)],
\end{equation}
which is identical to the correlation operator obtained in the derivation of the MLL dynamical equation \cite{SM}. Hence although in both  ULL$2$ and the MLL equations $\chi$ is of the first order with respect to $\widetilde{\bm{H}}_{\mathrm{I}}$, the ULL$2$ equation can be expected to lead to an improvement over the performance of the MLL approximation \cite{SM}. 

Let us discuss another asymptotic-time case where the MLL approximation holds. This underlines that the utility of the MLL approximation is not necessarily limited to the short-time dynamics. If (i) the interaction is sufficiently weak (weak-coupling or weak-correlation regime), and (ii) the subsystem dynamics reaches a steady state in a finite time, one can approximate $\bm{\varrho}_{\mathsf{S}}(s)\otimes \bm{\varrho}_{\mathsf{B}}(s)$ in Eq. (\ref{chi-1-uncorrelated-initial}) with the tensor product of subsystem steady states $\bm{\varrho}_{\mathsf{S}}^{\star}\otimes \bm{\varrho}_{\mathsf{B}}^{\star}$ and hence $\widetilde{\bm{H}}_{\mathrm{I}}(s)\approx \widetilde{\bm{H}}_{\mathrm{I}}^{\star}$ at most times, which yields the MLL approximation $\bm{\chi}^{(1),\star}_{\mathrm{MLL}}(\tau) \approx -i \tau [\widetilde{\bm{H}}_{\mathrm{I}}^{\star},\bm{\varrho}_{\mathsf{S}}^{\star}\otimes \bm{\varrho}_{\mathsf{B}}^{\star}]$. Since Eq. \eqref{chi-1-uncorrelated-initial} becomes exact in the weak-coupling limit and that typical quantum systems reach their steady state in finite times \cite{rapid-equilibration1,rapid-equilibration2,rapid-equilibration3,rapid-equilibration4}, one can conclude that the MLL equation modified by the subsystem steady states may hold at long times for typical systems \cite{SM}. 

Applying additional approximations on $\bm{\chi}^{(1)}$ by imposing time-locality and also assuming that the bath state remains constant in time, i.e., approximating $\bm{\chi}^{(1)}\approx -i\int_{0}^{\tau} ds \, [\widetilde{\bm{H}}_{\mathrm{I}}(s),\bm{\varrho}_{\mathsf{S}}(\tau)\otimes \bm{\varrho}_{\mathsf{B}}(0)]$, yields a time-local ULL$2$ equation. 

Systematic derivation of higher-order correlation terms up to $\bm{\chi}^{(\ell-1)}$ yielding the ULL$\ell$ approximations is straightforward but algebraically heavy, and so we leave this for future work. For details see Ref. \cite{SM}. However, similar to other approximate techniques, it usually suffices to consider the lowest order ULL$2$ or at most a few lowest orders. For a comparison with other techniques see also Ref. \cite{SM}.

We emphasize that although the validity of the weak-correlation expansion hinges on the strength of the interaction Hamiltonian, it shows clear differences with standard weak-coupling approximations \cite{book:Breuer-Petruccione, book:Rivas-Huelga}. In particular, we note that our expansion uses the correlation in a direct fashion as the key ingredient.

In the following, we illustrate our weak-correlation ULL method through two examples. In example II we show that the MLL equation captures the exact dynamics with a good accuracy, and outperforms the standard Markovian Lindblad equation. In addition, the time-local ULL$2$ equation becomes tantamount to the second-order time-convolutionless (TCL$2$) dynamical equation for the system, which gives the exact dynamics for this example. We also compare our solutions with a coarse-graining (CG) method \cite{book:Rivas-Huelga, Lidar-Whaley, Rivas-short-time, Schallerbook, Schaller-Brandes, Schaller-Brandes-2,Majenz}, which fails to exceed the performance of the ULL$2$ solution. In example III, we show that $\bm{\chi}^{(1)}$ of Eq. \eqref{chi-1-uncorrelated-initial} leads to a significant improvement in predicting the dynamics of the system whereas there the TCL$2$ and other approximating techniques such as the CG method do not provide such accuracy.

\subsection{Example II: Atom in a bosonic bath}
\label{example-atom}

Let us consider a two-level system (atom) interacting with a manymode bosonic bath initially in the thermal state $\varrho^{\beta}_{\mathsf{B}}=e^{-\beta \sum_n \omega_n \hat{a}^{\dag}_n \hat{a}_n }/\mathrm{Tr}[e^{-\beta \sum_n \omega_n \hat{a}^{\dag}_n \hat{a}_n }]$ at temperature $T=1/\beta$. Here $\hat{a}_l$ is the annihilation operator for mode $l$. The total Hamiltonian reads
\begin{align}
H_{\mathsf{SB}}=\omega_{0} \sigma_{+} \sigma_{-}+ \textstyle{\sum_n} \omega_n \hat{a}^{\dag}_n \hat{a}_n -\sigma_{x} \otimes \mathpzc{O}_{\mathsf{B}},
\end{align}
where $\mathpzc{O}_{\mathsf{B}}=\sum_n \kappa_n (\hat{a}_n+\hat{a}^{\dag}_n)$. Assuming that the atom at all times retains only a small correlation with the bath, we conclude that Eq. \eqref{aij-cij-markovian} applies and we obtain the following master equation:
\begin{equation}
\dot{\varrho}_{\mathsf{S}}(\tau)=-i[H_{\mathsf{S}},\varrho_{\mathsf{S}}(\tau)]+\gamma(\tau) \big(\sigma_{x} \varrho_{\mathsf{S}}(\tau)\sigma_{x}- \varrho_{\mathsf{S}}(\tau)\big),
\label{examp1}
\end{equation}
where $\gamma(\tau)=2\tau\,\mathrm{Cov}_{\mathsf{B}_{0}}(\mathpzc{O}_{\mathsf{B}},\mathpzc{O}_{\mathsf{B}})$, and $\mathrm{Cov}_{\mathsf{B}_{0}}(\mathpzc{O}_{\mathsf{B}},\mathpzc{O}_{\mathsf{B}})=\textstyle{\int_{0}^{\infty}} d\omega\, J(\omega) \big(2 n(\beta,\omega)+1\big)$ is given in terms of a spectral density function $J(\omega)$ and the bosonic occupation number $n(\beta,\omega)=(e^{\beta \omega}-1)^{-1}$. Equation \eqref{examp1} describes pure dephasing in the eigenbasis of $\sigma_x$ and gives the population of the excited state of the atom as
\begin{align}
\varrho_{\mathrm{ee}}(\tau)=1/2 + \big(\varrho_{\mathrm{ee}}(0) - 1/2\big) e^{-2 \tau^2\, \mathrm{Cov}_{\mathsf{B}_{0}}(\mathpzc{O}_{\mathsf{B}},\mathpzc{O}_{\mathsf{B}})}.
\label{exp2-DynamicalEq}
\end{align}
The solution of the exact dynamics for this example has been provided in Ref. \cite{Braun} for $\omega_{0}=0$ and under the assumption of an initial thermal state for the bath and an Ohmic spectral density for the couplings of the interaction Hamiltonian, $J(\omega)=\eta \omega(1+\omega^2/\omega^2_c)^{-2}$, where $\omega_c$ is the cutoff frequency and $\eta$ denotes the coupling strength between the system and the bath. This provides a convenient means of studying the accuracy of Eq. \eqref{exp2-DynamicalEq}. As we argued in Sec. \ref{sec:MLLeq}, under the mentioned conditions even in the highly strong-coupling regime, the MLL equation is exact in the asymptotic time. In this example, $H_{\mathrm{I}}$ has only one term, $H_{\mathsf{S}}$ is set to zero by $\omega_{0}=0$, and for the chosen spectral density and the uncorrelated initial state, all the conditions are satisfied. Hence the MLL equation gives an exact prediction for the asymptotic state, see Fig. \ref{fig:spin-boson-1}.

Figure \ref{fig:spin-boson-1} shows the evolution of the excited-state population and compares our MLL and ULL$2$ methods with the Redfield equation (an equation obtained by applying only the weak-coupling and time-locality approximations on the exact dynamics \cite{book:Rivas-Huelga}), the TCL$2$ master equation \cite{book:Breuer-Petruccione,book:Rivas-Huelga}, and the exact solution. In addition, to make a comparison with a CG dynamical equation we used the results of Ref. \cite{Schaller-Brandes-2}; see Fig. \ref{fig:spin-boson-1}. For this particular example the time-local ULL$2$ dynamical equation is identical to the TCL$2$ dynamical equation \cite{SM}, and both coincide with the exact dynamics. From this figure and the explicit form of the Redfield equation \cite{SM}, it is clear that the Redfield equation is less accurate in the low-temperature limit. The MLL equation follows the exact solution relatively well, whereas the Redfield equation exhibits a relatively slower decay. Note that when $\omega_{0}=0$ the standard Lindblad equation is equivalent to the Redfield equation. For details of the derivation of the Redfield and the TCL$2$ equations and for the analysis of the short-time dynamics using the Lindblad-like model and the exact evolution, see Ref. \cite{SM}. 

In the following example, we illustrate that our Markovian approximation works well to describe the short-time dynamics in a system with non-Markovian features. We then show that the ULL$2$ equation outperforms other methods for later times and captures the long-time dynamics more accurately.

\begin{figure}[tp]
\includegraphics[width=6.3cm]{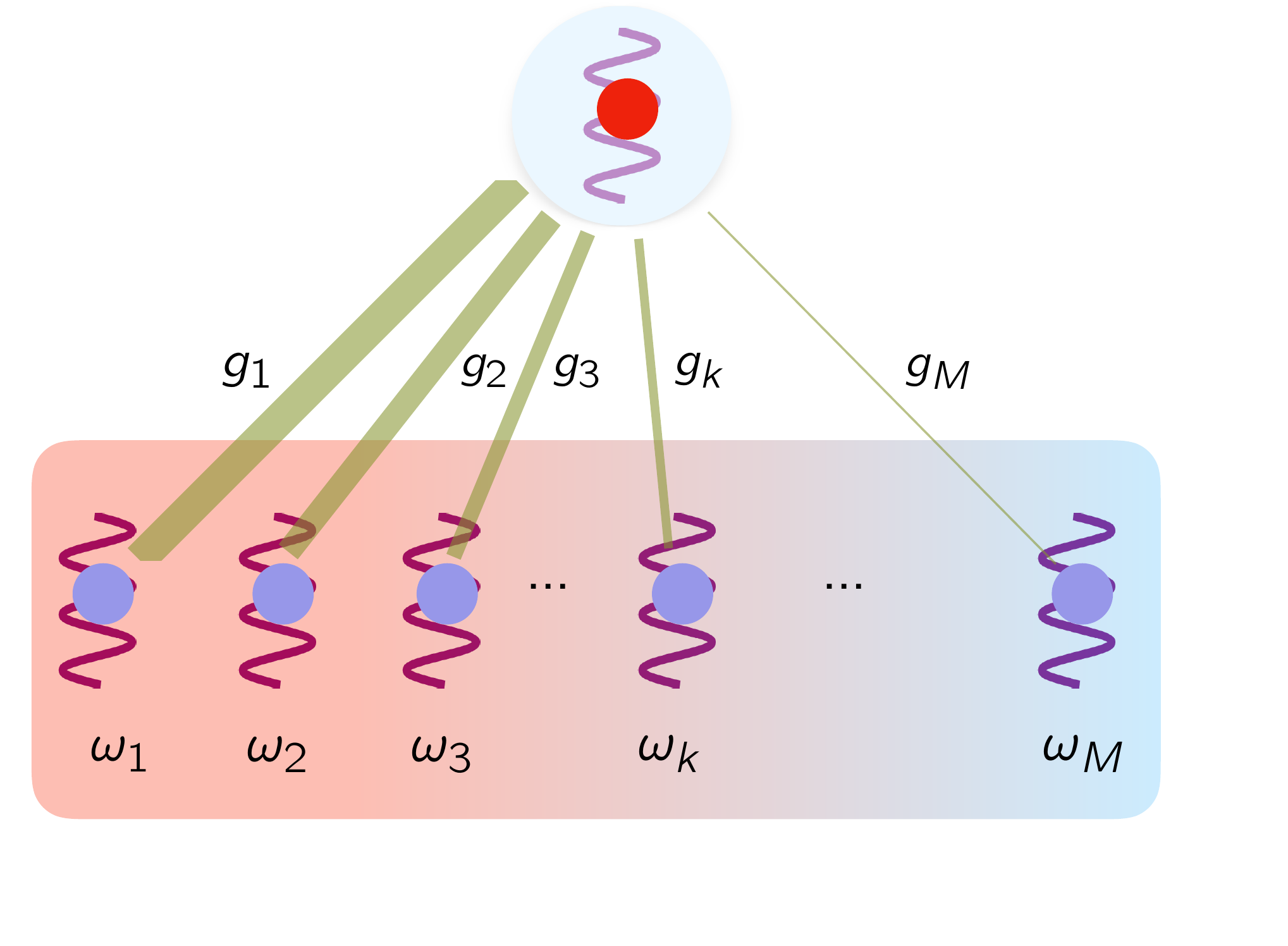}
\caption{Schematic model of a damped harmonic oscillator. See the text for details.}
\label{fig:harmonic0-schematic}
\end{figure}

\begin{figure*}[tp]
\includegraphics[width=8cm]{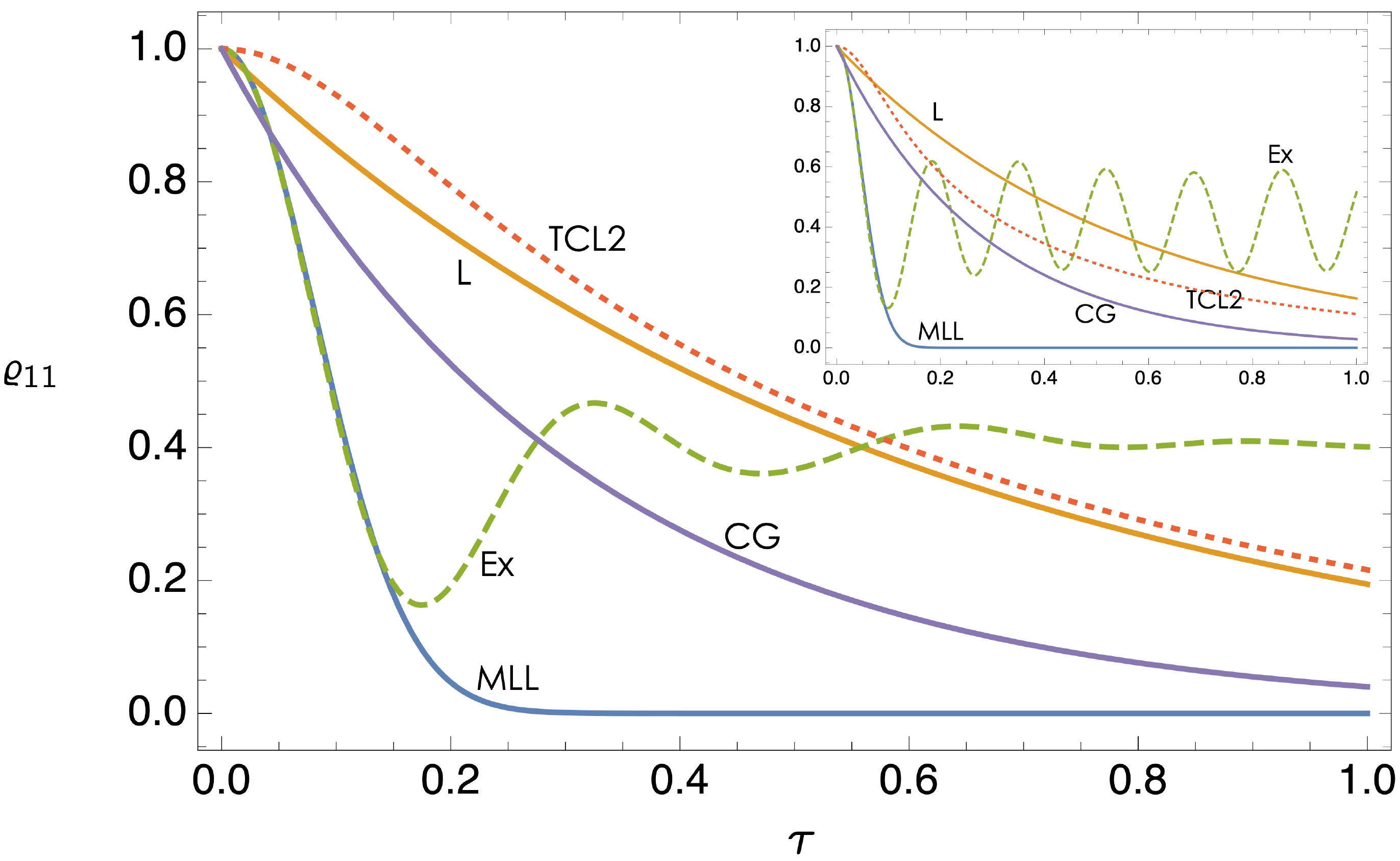}
\hskip5mm
\includegraphics[width=8cm]{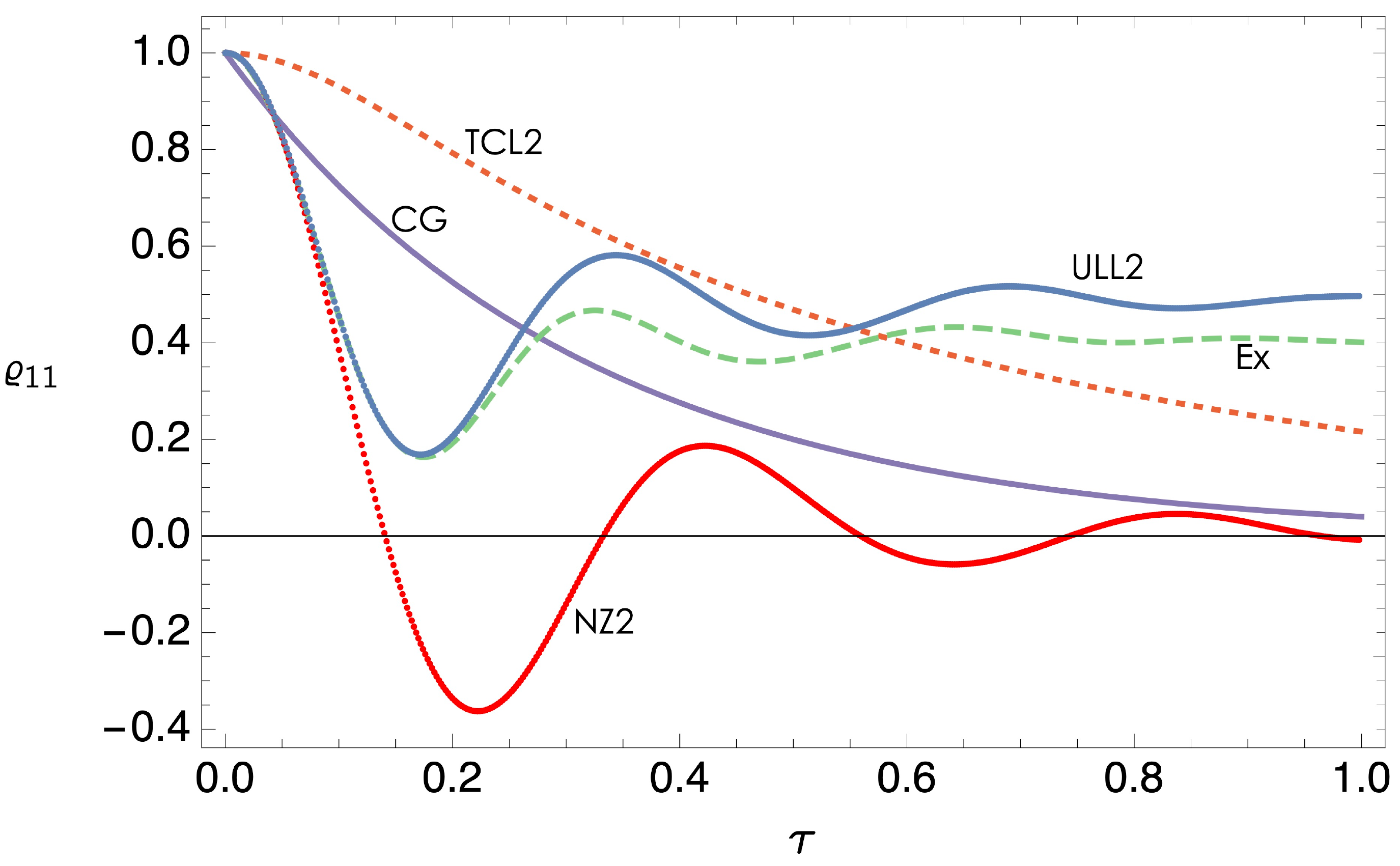}
\caption{Damped harmonic oscillator. 
(Left) Population of the first excited state of the system harmonic oscillator vs. time for $\omega_{0}=1$ and $\omega_{c}=5$ (all in natural units) when the system is initially in the state $|\psi(0) \rangle_{\mathsf{S}} = |1\rangle_{\mathsf{S}}$ and the bath has infinite oscillators ($M\to\infty$) at zero temperature. For the numerical solution of the exact dynamics, we have assumed $M=255$, $g_k=\sqrt{J(\omega_k)}$, and $\omega_k=0.1 k$, where $1\leqslant k\leqslant M$. The labels CG, Ex, L, MLL, and TCL$2$ denote, respectively, data from the CG, exact, Lindblad, MLL, and TCL$2$ solutions. The inset shows an identical case except that $\omega_c=10$. (Right) The same as Left, but only for the ULL$2$, NZ$2$, CG, and TCL$2$, and exact solutions.}
\label{fig:harmonic0}
\end{figure*}

\subsection{Example III: Damped harmonic oscillator within a bath of oscillators}
\label{example-oscillator}

Consider a quantum harmonic oscillator interacting with a bath of oscillators with the total Hamiltonian given by
\begin{align}
H_{\mathsf{SB}}=\omega_{0} \hat{a}^{\dag} \hat{a}+ \textstyle{\sum_{k=1}^{M}}\omega_{k} \hat{b}_k^{\dag} \hat{b}_k+ \textstyle{\sum_{k=1}^{M}} g_k (\hat{a}^{\dag} \hat{b}_k+ \hat{a} \hat{b}_k^{\dag}),
\end{align}
where $M$ is the number of the bath oscillators---see Fig. \ref{fig:harmonic0-schematic}. 
For simplicity of the analysis, we assume the initial system-bath state $(c_{0}|0\rangle+c_{1} |1\rangle)_{\mathsf{S}}\otimes |0\rangle^{\otimes M}_{\mathsf{B}}$, where $|i\rangle$ denotes the eigenstate of the corresponding number operator with eigenvalue $i$.

To obtain the MLL equation, we choose $\mathpzc{S}_{1}=\hat{a}+\hat{a}^{\dag}$ and $\mathpzc{S}_{2}=i(\hat{a}-\hat{a}^{\dag})$, and hence $\mathpzc{B}_{1}=\sum_k g_k (\hat{b}_k+\hat{b}_k^{\dag})/2$ and $\mathpzc{B}_{2}=\sum_k i g_k (\hat{b}_k-\hat{b}_k^{\dag})/2$. Inserting these into Eq. \eqref{aij-cij-markovian} yields $a_{11}=a_{22}=\tau G/4$, $a_{21}=a_{12}^{\ast}=i \tau G/4$, $\mathrm{Im} \ave{\mathpzc{B}_{1}\mathpzc{B}_{1}}_{\mathsf{B}_{0}}=\mathrm{Im} \ave{\mathpzc{B}_{2}\mathpzc{B}_{2}}_{\mathsf{B}_{0}}=0$, and $\mathrm{Im} \ave{\mathpzc{B}_{2}\mathpzc{B}_{1}}_{\mathsf{B}_{0}}=-\mathrm{Im} \ave{\mathpzc{B}_{1}\mathpzc{B}_{2}}_{\mathsf{B}_{0}}=G/4$, where $G=\sum_{k}|g_{k}|^{2}$; hence $\mathbbmss{h}_{\mathsf{L}}^{\chi}=0$. The MLL equation thus reads 
\begin{align}
\dot{\varrho}_{\mathsf{S}}(\tau)=&-i[\omega_{0} \hat{a}^{\dag} \hat{a}, \varrho_{\mathsf{S}}(\tau)]+ G \tau \big(2\hat{a} \varrho_{\mathsf{S}}(\tau) \hat{a}^{\dag}- \hat{a}^{\dag} \hat{a} \varrho_{\mathsf{S}}(\tau)\nonumber\\
&- \varrho_{\mathsf{S}}(\tau) \hat{a}^{\dag} \hat{a}\big).
\label{MLL-harmonic}
\end{align} 
The standard Lindblad equation for this model has been given in Ref. \cite{Rivas-damped} (see also Refs. \cite{Isar,grabert} for more general Lindblad forms for this model and their solutions). For the special case of the chosen initial state the Lindblad equation becomes 
\begin{align}
\dot{\varrho}_{\mathsf{S}}(\tau)=&-i [(\omega_{0}+\delta) \hat{a}^{\dag} \hat{a},\varrho_{\mathsf{S}}(\tau)]\nonumber\\
&+\gamma \big(2\hat{a} \varrho_{\mathsf{S}}(\tau) \hat{a}^{\dag}- \hat{a}^{\dag} \hat{a} \varrho_{\mathsf{S}}(\tau) - \varrho_{\mathsf{S}}(\tau) \hat{a}^{\dag} \hat{a}\big),
\end{align} 
where $\delta=\mathbbmss{P}[\int_{0}^{\infty} d\omega\, J(\omega)\,(\omega_{0} - \omega)^{-1}]$, $\gamma=\pi J(\omega_{0})$, $J(\omega)=\sum_{k=1}^M g_k^2 \delta(\omega -\omega_k)$ (the spectral density function), and $\mathbbmss{P}$ denotes the Cauchy principal value. 

To compare the MLL equation with the standard Lindblad equation, the ULL$2$ equation, and the exact dynamics, we choose an Ohmic spectral density as $J(\omega)=(\omega/\pi) e^{-\omega/\omega_c}$ (similar to Ref. \cite{Rivas-damped}). For numerical simulation of this model, we take $\omega_k=0.1 k$ and $M=255$. In order for the MLL equation and the simulations to be comparable with the standard Lindblad equation we choose the coupling strength as $g_k=\sqrt{J(\omega_k)}$. It is shown in Fig. \ref{fig:harmonic0} (left) that the short-time dynamics is well captured by the MLL equation, for all chosen cutoff values, whereas the Lindblad equation fails to capture this and the TCL$2$ equation seems to capture only the initial moments of the dynamics. When $\omega_c=10$ the exact dynamics shows nondecaying oscillations, which may be due to non-Markovianity. It is interesting to note that even in this non-Markovian case the MLL equation---which is Markovian---can capture the short-time dynamics well. Considering the dynamics at long times, we observe in Fig. \ref{fig:harmonic0} (right) that the ULL$2$ equation can capture the exact dynamics with good accuracy, while the second-order NZ (NZ$2$) equation \cite{book:Breuer-Petruccione} gives unphysical results. It is interesting that although the complete ULL and the NZ dynamical equations are exact, their second-order approximations, i.e., ULL$2$ and NZ$2$, approximate the dynamics differently. This is due to the different underlying approaches in applying the second-order approximation.

In addition, by calculating the population of the first excited state using the suggested asymptotic state $\varrho^{\star}$ \cite{SM}, we obtain that $\varrho^{\star}_{11}=0.3963$ equals the asymptotic value in Fig. \ref{fig:harmonic0}.

\section{Summary and conclusions}

We have introduced the correlation picture as a new dynamical picture. By using the Schr\"{o}dinger equation of the total system, using a correlating transformation, and tracing over the environment degrees of freedom, we have found the dynamical equation of the subsystem without invoking any approximations. We have shown that this exact dynamical equation is in the Lindblad form, even if the system is initially correlated or is in the strong-coupling regime. Hence the Lindblad form for the dynamical equation is general, and the obtained master equation is a universal Lindblad-like (ULL) equation. We have provided a way to derive a Markovian master equation (MLL) from the ULL equation. In particular, we have shown that Markovianity can emerge if we apply a weak-correlation approximation, and the MLL equation becomes exact at instants with vanishing correlations. We have demonstrated that, not only at the initial time (if the system and the bath are prepared in a product state) but also in the asymptotic time, this weak-correlation approximation can be valid under certain conditions. 

The correlation picture has also enabled us to formulate a systematic weak-correlation perturbative expansion, from which we have introduced approximate second- and higher-order tractable master equations. The MLL methodology plays an important role in this construction of the master equations which can feature non-Markovian effects. We have shown that existing and widely used weak-coupling-based equations emerge as special cases of our perturbative constructions, and thus our weak-correlation master equations are expected to outperform or perform as accurately as corresponding weak-coupling solutions. In particular, we have illustrated through three examples our results for the existence of the ULL equation, validity of the MLL equation around weak-correlation points for initial and asymptotic times, and have compared the MLL and ULL$2$ equations as approximate solutions with other Markovian and non-Markovian equations. We have shown that in these examples our equations describe the dynamics more accurately. We expect that introducing the correlation picture can pave the way for developing new techniques for controlling and harnessing system-environment correlations. We also anticipate a wide range of applications of our theory from quantum thermodynamics to quantum computation. In particular, our approach may help to understand whether and how quantum systems thermalize, and it may shed light on the role of correlations in quantum algorithms and the robustness of quantum error correction against correlated noise mechanisms. 

\begin{acknowledgements}

Discussions with J. Anders, E. Aurell, H.-P. Breuer, L. A. Correa, S. F. Huelga, A. Isar, S. J. Kazemi, J. Piilo, \'{A}. Rivas, R. Sampaio, J. Tuorila, and S. Vinjanampathy are acknowledged. This work was supported by the Academy of Finland's Center of Excellence program QTF Project 312298 and Sharif University of Technology's Office of Vice President for Research and Technology. A.T.R. also acknowledges hospitality and support of the QTF Center of Excellence at Aalto University.

\end{acknowledgements}


\twocolumngrid

\begin{widetext}
\newpage

\appendix

\begin{center}
\textbf{Supplementary Material: Correlation Picture Approach to Open-Quantum-System Dynamics}
\end{center}

\section{Solution of the equation $\chi=-i\lbrac H_{\chi},\varrho_{\mathsf{S}}\otimes \varrho_{\mathsf{B}}\rbrac$ for $H_{\chi}$}
\label{sec:sol-hchi}

Equation (3) of the main text can be rewritten as
\begin{equation}
\varrho^{\dag}_{\mathsf{S}}(\tau)\otimes\varrho^{\dag}_{\mathsf{B}}(\tau) \big[i H^{\dag}_{\chi}(\tau)\big] + \big[i H^{\dag}_{\chi}(\tau)\big]^{\dag} \varrho_{\mathsf{S}}(\tau)\otimes\varrho_{\mathsf{B}}(\tau)=\chi(\tau),
\label{chi-exp}
\end{equation}
which is of the general form $E^{\dag}X+ X^{\dag}E=F$ (with unknown $X$), where we take $E=\varrho_{\mathsf{S}}(\tau)\otimes\varrho_{\mathsf{B}}(\tau)$, $X=i H^{\dag}_{\chi}(\tau)$, and $F=\chi(\tau)$. We define the instantaneous projection operator onto the null-space of $\varrho_{\mathsf{S}}(\tau)\otimes \varrho_{\mathsf{B}}(\tau)$ by $P_0(\tau)$. From the result of Ref. [36] of the main text, if $P_0(\tau) \chi(\tau) P_0(\tau)=0$, then
\begin{align}
-i H_{\chi}(\tau)=&\frac{1}{2}\big(\mathbbmss{I}+P_0(\tau)\big)\chi(\tau) \varrho_{\mathsf{S}}^{-1}(\tau)\otimes \varrho_{\mathsf{B}}^{-1}(\tau),
\label{Gamma-solution}
\end{align}
is a solution of Eq. \eqref{chi-exp}, where $\varrho_{\mathsf{S}}^{-1}(\tau)$ and $\varrho_{\mathsf{B}}^{-1}(\tau)$ are the pseudo-inverses \cite{pseudo-invers} of $\varrho_{\mathsf{S}}(\tau)$ and $\varrho_{\mathsf{B}}(\tau)$, respectively. The solution is not unique and may include further terms, $\varrho_{\mathsf{S}}(\tau)\otimes\varrho_{\mathsf{B}}(\tau)Z(\tau)\Big(\mathbbmss{I}-P_0(\tau)\Big) +Y(\tau) P_0(\tau)$, where $Y(\tau)$ is an arbitrary operator and $Z(\tau)$ is an operator satisfying $\varrho_{\mathsf{S}}(\tau) \otimes \varrho_{\mathsf{B}}(\tau)\big(Z(\tau)+Z^{\dag}(\tau)\big)\varrho_{\mathsf{S}}(\tau)\otimes \varrho_{\mathsf{B}}(\tau)=0$. For our purposes in this paper, we take $Y(\tau)$ and $Z(\tau)$ equal to zero---with no impact on the state $\varrho_{\mathsf{SB}}$.

To show that Eq. \eqref{chi-exp} always has a solution, we need to ensure that the condition $P_0(\tau) \chi(\tau) P_0(\tau)=0$ is always satisfied, or equivalently, $P_0(\tau)\varrho_{\mathsf{SB}}(\tau)P_0(\tau)=0$, since by definition $P_0(\tau)\varrho_{\mathsf{S}}(\tau)\otimes \varrho_{\mathsf{B}}(\tau)P_0(\tau)=0$. To this end we write $\varrho_{\mathsf{SB}}$ in terms of its spectral decomposition $\varrho_{\mathsf{SB}}=\sum_i \xi_i |\xi_i \rangle \langle\xi_i|$, and use the Schmidt decomposition (see Ref. [26] of the main text) for each eigenstate $\ket{\xi_i}= \textstyle{\sum_j} \sqrt{\lambda_j^{(i)}} \ket{e_{j}^{(i)}}_{\mathsf{S}}\ket{f_{j}^{(i)}}_{\mathsf{B}}$ to reach
\begin{align}
\varrho_{\mathsf{SB}}= \textstyle{\sum_{ijl}} \xi_i \sqrt{\lambda_j^{(i)}\lambda_l^{(i)}} \ket{e_{j}^{(i)}}_{\mathsf{S}}\bra{e_{l}^{(i)}} \otimes \ket{f_{j}^{(i)}}_{\mathsf{B}}\bra{f_{l}^{(i)}}.
\label{rho_sb}
\end{align}
From $\mathrm{Tr}[P_0(\tau)\varrho_{\mathsf{S}}(\tau)\otimes \varrho_{\mathsf{B}}(\tau)]=0$ one concludes that
\begin{align}
\textstyle{\sum_{ijkl}} \xi_i \xi_k  \lambda_j^{(i)}  \lambda_l^{(k)} \bra{e_{j}^{(i)}, f_{l}^{(k)}} P_0\ket{e_{j}^{(i)}, f_{l}^{(k)}}=0.
\label{eq.a}
\end{align}
Since $\xi_i$ and $\lambda_j^{(i)}$ are positive numbers and $P_0$ is a positive--semidefinite matrix, for Eq. \eqref{eq.a} to hold it is needed that $P_0(\tau)\ket{e_{j}^{(i)},f_{l}^{(k)}}_{\mathsf{SB}}=0$. Hence it is seen that applying $P_0(\tau)$ on Eq. \eqref{rho_sb} leads to $P_0(\tau)\varrho_{\mathsf{SB}}(\tau)P_0(\tau)=0$, which accordingly gives $P_0(\tau)\chi(\tau) P_0(\tau)=0$ .

\section{Derivation of the ULL equation}
\label{ULL-der}

By tracing out over the bath from both sides of the dynamical equation in the correlation picture, we reach
\begin{align}
\dot{\varrho}_{\mathsf{S}}(\tau)=&\mathrm{Tr}_{\mathsf{B}}\big[\mathpzc{D}_{\mathpzc{c}}[\varrho_{\mathsf{S}}(\tau)\otimes \varrho_{\mathsf{B}}(\tau)]\big]\nonumber\\
=&-i \mathrm{Tr}_{\mathsf{B}}[H_{\mathsf{SB}},\varrho_{\mathsf{S}}(\tau)\otimes \varrho_{\mathsf{B}}(\tau)]- \mathrm{Tr}_{\mathsf{B}}\big[H_{\mathsf{SB}},\lbrac H_{\chi},\varrho_{\mathsf{S}}(\tau)\otimes \varrho_{\mathsf{B}}(\tau) \rbrac \big].
\label{methods1}
\end{align}
Inserting $H_{\mathsf{SB}}=H_{\mathsf{S}}+H_{\mathsf{B}}+H_{\mathrm{I}}$ and the identities (cyclicity of partial trace)
\begin{align}
\mathrm{Tr}_{\mathsf{B}}\big[H_{\mathsf{S}},\lbrac H_{\chi}, \varrho_{\mathsf{S}}\otimes \varrho_{\mathsf{B}} \rbrac\big]
&=i \big[H_{\mathsf{S}},\mathrm{Tr}_{\mathsf{B}} [\chi]\big]=0,\nonumber\\
\mathrm{Tr}_{\mathsf{B}}\big[H_{\mathsf{B}},\lbrac H_{\chi}, \varrho_{\mathsf{S}}\otimes \varrho_{\mathsf{B}} \rbrac\big] &\equiv 0,
\end{align}
into Eq. \eqref{methods1} yields
\begin{align}
\dot{\varrho}_{\mathsf{S}}=&-i \mathrm{Tr}_{\mathsf{B}}[\widetilde{H}_{\mathsf{S}},\varrho_{\mathsf{S}}]- \mathrm{Tr}_{\mathsf{B}}[H_{\mathrm{I}},\lbrac H_{\chi},\varrho_{\mathsf{S}}\otimes \varrho_{\mathsf{B}} \rbrac ],
\label{method3}
\end{align}
where
\begin{equation}
\widetilde{H}_{\mathsf{S}}(\tau)=H_{\mathsf{S}}+\mathrm{Tr}_{\mathsf{B}}[H_{\mathrm{I}}\,\varrho_{\mathsf{B}}(\tau)].
\label{H-tilde-S}
\end{equation}
For future reference, we can also define $\widetilde{H}_{\mathsf{B}}(\tau)$ in a similar fashion. Expanding $H_{\mathrm{I}}$ in terms of a Hermitian operator basis $\{\mathpzc{S}_{i}\}_{i=1}^{d_{\mathsf{S}}^2-1}$ (with $\mathpzc{S}_0=\mathbbmss{I}$) as
\begin{align}
H_{\mathrm{I}}&=\textstyle{\sum_{i=1}^{d_{\mathsf{S}}^2-1}}\mathpzc{S}_{i}\otimes \mathpzc{B}_{i},
\label{H_{int}-expanded}
\end{align}
and replacing it into Eq. \eqref{method3} yields
\begin{align}
\dot{\varrho}_{\mathsf{S}}(\tau)=&-i\big[\widetilde{H}_{\mathsf{S}}(\tau),\varrho_{\mathsf{S}}(\tau)\big]-\textstyle{\sum_{i=1}^{d_{\mathsf{S}}^2-1}}\Big[\mathpzc{S}_{i},\mathrm{Tr}_{\mathsf{B}}\left[\mathpzc{B}_i \lbrac H_{\chi} , \varrho_{\mathsf{S}}(\tau) \otimes \varrho_{\mathsf{B}}(\tau) \rbrac \right]\Big].
\label{method2}
\end{align}

We now expand $H_{\chi}(\tau)$ in the same basis
\begin{align}
H_{\chi}(\tau) &=\textstyle{\sum_{j=0}^{d_{\mathsf{S}}^2-1}} \mathpzc{S}_{j} \otimes \mathpzc{B}_j^{\chi}(\tau),
\label{H_{chi}-expanded0}
\end{align}
and insert it into Eq. \eqref{method2}, which gives
\begin{align}
\dot{\varrho}_{\mathsf{S}}(\tau)=&-i\big[ \widetilde{H}_{\mathsf{S}}(\tau),\varrho_{\mathsf{S}}(\tau)\big]-\textstyle{\sum_{i\neq 0}}\Big[\mathpzc{S}_{i},\mathrm{Tr}_{\mathsf{B}} \big[\left( H_{\chi}(\tau) \varrho_{\mathsf{S}}(\tau)\otimes \varrho_{\mathsf{B}}(\tau)- \varrho_{\mathsf{S}}(\tau) \otimes \varrho_{\mathsf{B}}(\tau)  H_{\chi}^{\dag}(\tau)\right)\mathpzc{B}_i\big]\Big],\nonumber\\
=& -i\big[\widetilde{H}_{\mathsf{S}}(\tau),\varrho_{\mathsf{S}}(\tau)\big]
-\textstyle{\sum_{i\neq 0,j}}\Big[\mathpzc{S}_{i},\mathpzc{S}_{j} \varrho_{\mathsf{S}}(\tau) \,\mathrm{Tr}_{\mathsf{B}}\big[\mathpzc{B}_j^{\chi}(\tau) \varrho_{\mathsf{B}} (\tau)\mathpzc{B}_i\big]-\varrho_{\mathsf{S}} \mathpzc{S}_{j} \,\mathrm{Tr}_{\mathsf{B}}\big[\varrho_{\mathsf{B}}(\tau) {\mathpzc{B}_j^{\chi}}^{\dag}\mathpzc{B}_i\big]\Big],\nonumber\\
=& -i\big[\widetilde{H}_{\mathsf{S}}(\tau),\varrho_{\mathsf{S}}(\tau)\big]
-\textstyle{\sum_{i\neq 0,j}}\big[\mathpzc{S}_{i},c_{ij}(\tau)\mathpzc{S}_{j} \varrho_{\mathsf{S}}(\tau) -c_{ij}^{\ast}(\tau)\varrho_{\mathsf{S}}(\tau) \mathpzc{S}_{j} \big],\nonumber\\
=& -i\big[\widetilde{H}_{\mathsf{S}}(\tau)-i\textstyle{\sum_{i\neq0}} \big(c_{i0}(\tau)-c_{i0}^{\ast}(\tau)\big)\mathpzc{S}_{i}\mathpzc{S}_0,\varrho_{\mathsf{S}}(\tau)\big] + \textstyle{\sum_{i \neq 0, j \neq 0 }} c_{ij}(\tau)\left(\mathpzc{S}_{j} \varrho_{\mathsf{S}}(\tau) \mathpzc{S}_{i}-\mathpzc{S}_{i} \mathpzc{S}_{j} \varrho_{\mathsf{S}}(\tau) \right)\nonumber\\
&+ \textstyle{\sum_{i \neq 0, j \neq 0 }} c_{ij}^{\ast}(\tau)\big(\mathpzc{S}_{i}\varrho_{\mathsf{S}}(\tau) \mathpzc{S_j}-\varrho_{\mathsf{S}}(\tau) \mathpzc{S}_{j} \mathpzc{S}_{i}\big),\nonumber\\
=& -i\big[\widetilde{H}_{\mathsf{S}}(\tau)+2\textstyle{\sum_{i\neq0}} \mathrm{Im}\big(c_{i0}(\tau)\big)\mathpzc{S}_{i},\varrho_{\mathsf{S}}(\tau)\big]
+ \textstyle{\sum_{i \neq 0, j \neq 0 }} c_{ij}(\tau)\left(\mathpzc{S}_{j} \varrho_{\mathsf{S}}(\tau)\mathpzc{S}_{i}-\mathpzc{S}_{i} \mathpzc{S}_{j} \varrho_{\mathsf{S}}(\tau)\right) \nonumber\\
&+ \textstyle{\sum_{i \neq 0, j \neq 0 }} c_{ji}^{\ast}(\tau)\left(\mathpzc{S}_{j}\varrho_{\mathsf{S}}(\tau) \mathpzc{S_i}-\varrho_{\mathsf{S}}(\tau) \mathpzc{S}_{i} \mathpzc{S}_{j}\right),
\label{non-diagonal-Lindblad}
\end{align}
where it is seen that the elements of the matrix $C$ are given by $c_{ij}(\tau)=\mathrm{Tr}[\varrho_{\mathsf{B}}(\tau) \mathpzc{B}_i\mathpzc{B}_j^{\chi}(\tau)]$. Defining the Hermitian matrices $A=[a_{ij}]_{i\neq 0, j\neq 0}$ and $B=[b_{ij}]_{i\neq 0, j\neq 0}$, see the main text below Eq. (8), we obtain $c_{ij}=a_{ij}+i\,b_{ij}$ and $c^{\ast}_{ji}=a_{ij}-i\,b_{ij}$. Inserting these equations into Eq. \eqref{non-diagonal-Lindblad} leads to the following general form for the dynamical equation:
\begin{align}
\dot{\varrho}_{\mathsf{S}}(\tau)=-i\big[H_{\mathsf{S}}+{\mathbbmss{h}}_{{\mathsf{L}}}^{\chi}(\tau),\varrho_{\mathsf{S}}(\tau)\big] + \textstyle{\sum_{i\neq 0, j\neq 0}} a_{ij}(\tau)\big(2 \mathpzc{S}_{j}\varrho_{\mathsf{S}}(\tau)\mathpzc{S}_{i}-\{\mathpzc{S}_{i}\mathpzc{S}_{j},\varrho_{\mathsf{S}}(\tau)\}\big),
\label{general-dynamics}
\end{align}
where
\begin{align}
{\mathbbmss{h}}_{{\mathsf{L}}}^{\chi}(\tau)=&\ave{H_{\mathrm{I}}}_{\mathsf{B}}+2 \textstyle{\textstyle{\sum_{i\neq0}}} \mathrm{Im}\big(c_{i0}(\tau)\big)\mathpzc{S}_{i}+ \textstyle{\sum_{i\neq 0,j\neq 0}} b_{ij}(\tau) \mathpzc{S}_{i}\mathpzc{S}_{j}.
\label{def:l-h-s}
\end{align}
The only remaining step to obtain the final form of Eq. (9) of the main text is to diagonalize $A=[a_{ij}]_{i\neq0,j\neq0}$, such that $V A V^{\dag}=\Gamma$, where $\Gamma$ is a diagonal matrix with the eigenvalues $\gamma^{\chi}_m$ of $A$ as its diagonal elements.

\section{On the linearity of the ULL equation}
\label{appendix:linearity-zero-corr}

Following the principles of quantum mechanics, the total dynamics should be linear for arbitrary initial states of the \textit{total} system. However, linearity of the subsystem dynamics in general is not required. One can investigate the linearity of the Lindblad-like equation for the subsystem from different perspectives. Consider two reduced system states $\varrho^{(1)}_{\mathsf{S}}$ and $\varrho^{(2)}_{\mathsf{S}}$ obtained, respectively, by the partial trace over the bath of the total states $\varrho^{(1)}_{\mathsf{SB}}=\varrho^{(1)}_{\mathsf{S}}\otimes \varrho^{(1)}_{\mathsf{B}}+\chi^{(1)}$ and $\varrho^{(2)}_{\mathsf{SB}}=\varrho^{(2)}_{\mathsf{S}}\otimes \varrho^{(2)}_{\mathsf{B}}+\chi^{(2)}$. Now consider the convex combination of these states $p_{1}\varrho^{(1)}_{\mathsf{S}} + p_{2}\varrho^{(2)}_{\mathsf{S}}$, with $p_{1},p_{2}\geqslant 0$ and $p_{1}+p_{2}=1$, which is evidently obtained by the partial trace over the total system state $p_{1}\varrho^{(1)}_{\mathsf{SB}} + p_{2}\varrho^{(2)}_{\mathsf{SB}}$. Starting from the Schr\"{o}dinger equation and following the derivation of the subsystem dynamics in Sec. \ref{ULL-der}, it can be seen that in this case the generator of the Lindblad-like equation $\mathpzc{L}^{\chi}$ in general is not linear in the sense that
\begin{equation}
\mathpzc{L}^{\chi}\big[\textstyle{\sum_i} p_i \varrho^{(i)}_{\mathsf{S}}\big]\neq \textstyle{\sum_i} p_i \mathpzc{L}^{\chi^{(i)}} [\varrho^{(i)}_{\mathsf{S}}],
\end{equation}
where $\chi$ denotes the correlation operator when the total system state is $p_{1}\varrho^{(1)}_{\mathsf{SB}} + p_{2}\varrho^{(2)}_{\mathsf{SB}}$. However, below we discuss a restricted case where linearity in this sense can be retrieved.

\subsection{Linearity on a restricted set of states}
\label{Lin-SB}

Consider a total state, defined with a given $\varrho_{\mathsf{B}}$ and $\chi$ such that $\varrho_{\mathsf{SB}}=\varrho_{\mathsf{S}}\otimes \varrho_{\mathsf{B}}+\chi$, and initial subsystem states which are chosen from a restricted set of states forming a convex decomposition $\varrho_{\mathsf{S}}=\sum_i p_i \varrho_{\mathsf{S}}^{(i)}$, with $p_{i}\geqslant 0$ and $\sum_{i}p_{i}=1$. By replacing the convex decomposition of $\varrho_{\mathsf{S}}$ in Eq. (3) of the main text and defining $\chi^{(i)}:=-i\lbrac H_{\chi}, \varrho_{\mathsf{S}}^{(i)}\otimes \varrho_{\mathsf{B}} \rbrac$, we observe that $\chi$ can also be written in the convex combination form of $\chi=\sum_i p_i \chi^{(i)}$. Thus, one can associate a correlation operator $\chi^{(i)}$ to each $\varrho^{(i)}_{\mathsf{S}}$ while $H_{\chi}$ remains identical for each of them. As a result, a single $\mathpzc{L}^{\chi}$ is associated with all of the cases. Following the derivation of the Lindblad-like equation and replacing convex decompositions of $\varrho_{\mathsf{S}}$ and $\chi$, one can obtain from the dynamical equation for $\varrho_{\mathsf{S}}$ the identity
\begin{equation}
\mathpzc{L}^{\chi}\big[\textstyle{\sum_i} p_i \varrho_{\mathsf{S}}^{(i)}\big]=\sum_i p_i \mathpzc{L}^{\chi}[\varrho^{(i)}_{\mathsf{S}}].
\end{equation}

\subsection{Linearity when there is no initial system-bath correlation}

Now we consider another case where linearity holds. If the initial system-bath state is a product state, i.e., $\chi(0)=0$, and the dynamics of the total system is given by the unitary evolution $U(\tau)=e^{-i H_{\mathsf{SB}}}$, we obtain
\begin{align}
\varrho_{\mathsf{SB}}(\tau)&=U(\tau) \varrho_{\mathsf{SB}}(0) U^{\dag}(\tau),\\
\varrho_{\mathsf{S}}(\tau)\otimes \varrho_{\mathsf{B}}(\tau)+\chi(\tau)&=U(\tau) \varrho_{\mathsf{S}}(0)\otimes \varrho_{\mathsf{B}}(0) U^{\dag}(\tau).
\end{align}
Thus, it follows that
\begin{equation}
\chi(\tau)=U(\tau) \varrho_{\mathsf{S}}(0)\otimes \varrho_{\mathsf{B}}(0) U^{\dag}(\tau)-\varrho_{\mathsf{S}}(\tau)\otimes \varrho_{\mathsf{B}}(\tau).
\label{chi-for-chi0}
\end{equation}
Replacing the above $\chi$ with $-i \lbrac H_{\chi},\varrho_{\mathsf{S}}\otimes\varrho_{\mathsf{B}} \rbrac$ in Eq. (\ref{methods1}) yields
\begin{align}
\dot{\varrho}_{\mathsf{S}}(\tau)&=-i [H_{\mathsf{S}}+\mathrm{Tr}_{\mathsf{B}}[H_{\mathrm{I}}\varrho_{\mathsf{B}}(\tau)],\varrho_{\mathsf{S}}(\tau)]-i \mathrm{Tr}_{\mathsf{B}}[H_{\mathrm{I}},U(\tau) \varrho_{\mathsf{S}}(0)\otimes \varrho_{\mathsf{B}}(0) U^{\dag}(\tau)-\varrho_{\mathsf{S}}(\tau)\otimes \varrho_{\mathsf{B}}(\tau)]\nonumber\\
&=-i [H_{\mathsf{S}},\varrho_{\mathsf{S}}(\tau)]-i \mathrm{Tr}_{\mathsf{B}}[H_{\mathrm{I}},U(\tau) \varrho_{\mathsf{S}}(0)\otimes \varrho_{\mathsf{B}}(0) U^{\dag}(\tau)]
\label{Tr_B-schrodinger-1}
\end{align}
It follows from the above equation that replacing $\varrho_{\mathsf{S}}(\tau)$ with $\alpha_{1}\varrho_{\mathsf{S}1}(\tau)+\alpha_{2} \varrho_{\mathsf{S}2}(\tau)$ and keeping $\varrho_{\mathsf{B}}(0)$ unchanged yield
\begin{align}
\dot{\varrho}_{\mathsf{S}}(\tau)=&\alpha_{1}\big(-i [H_{\mathsf{S}},\varrho_{\mathsf{S}1}(\tau)]-i \mathrm{Tr}_{\mathsf{B}}[H_{\mathrm{I}},U(\tau) \varrho_{\mathsf{S}1}(0)\otimes \varrho_{\mathsf{B}}(0) U^{\dag}(\tau)]\big) +\alpha_{2} \big(-i [H_{\mathsf{S}},\varrho_{\mathsf{S}2}(\tau)]-i \mathrm{Tr}_{\mathsf{B}}[H_{\mathrm{I}},U(\tau) \varrho_{\mathsf{S}2}(0)\otimes \varrho_{\mathsf{B}}(0) U^{\dag}(\tau)]\big),
\end{align}
which implies that the reduced dynamics is linear.

\section{Details of example I: the Jaynes-Cummings model}
\label{ex:Jaynes-Cummings}

Choosing $\mathpzc{S}_0=\mathbbmss{I}/\sqrt{2}$, $\mathpzc{S}_1=\sigma_{x}/\sqrt{2}$, $\mathpzc{S}_2=\sigma_{y}/\sqrt{2}$, $\mathpzc{S}_3=\sigma_{z}/\sqrt{2}$ as the system operator basis, we find $\mathpzc{B}_0=0$, $\mathpzc{B}_1=(\lambda/\sqrt{2})(\hat{a}+\hat{a}^{\dag})$, $\mathpzc{B}_2=(i \lambda/\sqrt{2})(\hat{a}-\hat{a}^{\dag})$, and $\mathpzc{B}_3=0$. Using the exact solution of the Jaynes-Cummings model (see Ref. [39] of the main text)
we find $|\psi(\tau)\rangle= e^{-i\tau \omega_0/2} \big\{\big[r_{1}\cos(\lambda \tau)-i r_{2} \sin(\lambda \tau)\big] |\mathrm{e},0\rangle + \big[-i r_{1}\sin(\lambda \tau)+r_{2} \cos(\lambda \tau) |\mathrm{g},1\rangle\big]\big\}$, from which
\begin{align}
\chi(\tau)=&\frac{1}{8} (1+4r_{1}^2-4r_{1}^4 -\alpha_{1}^2+\alpha_{2}^2) \big(|\mathrm{e},0\rangle\langle \mathrm{e},0|+|\mathrm{g},1\rangle\langle \mathrm{g},1|\big)+\frac{1}{4}(\alpha_{1}^2-1) \big(|\mathrm{e},1\rangle\langle \mathrm{e},1|+|\mathrm{g},0\rangle\langle \mathrm{g},0|\big)\nonumber\\
&+r_{1} r_{2} \big(|\mathrm{e},0\rangle\langle \mathrm{g},1|+|\mathrm{g},1\rangle\langle \mathrm{e},0|\big) + i\alpha_{2}/2 \big(|\mathrm{g},1\rangle\langle \mathrm{e},0|-|\mathrm{e},0\rangle\langle \mathrm{g},1|\big),
\end{align}
where $\alpha_{1}=(1-2r_{1}^2)\cos(2\lambda \tau)$ and $\alpha_{2}=(1-2r_{1}^2)\sin(2\lambda \tau)$. Now $H_{\chi}=\sum_{i=0}^{4}\mathpzc{S}_{i}\otimes \mathpzc{B}^{\chi}_i$, with
\begin{align}
\mathpzc{B}^{\chi}_0=&i\alpha_{1}/\big(\sqrt{2}(1-\alpha_{1})\big)|0\rangle\langle 0|-i\alpha_{1}/\big(\sqrt{2}(1+\alpha_{1})\big)|1\rangle\langle 1|,\nonumber\\
\mathpzc{B}^{\chi}_1=&(2 i r_{1} r_{2}+\alpha_{2})/\big[\sqrt{2}(1+\alpha_{1})^2\big]|0\rangle \langle 1|+(2 i r_{1} r_{2}-\alpha_{2})/\big[\sqrt{2}(1-\alpha_{1})^2\big]|1\rangle \langle 0|,\nonumber\\
\mathpzc{B}^{\chi}_2=&\big(-2r_{1} r_{2}+i\alpha_{2}\big)/\big[\sqrt{2}(1+\alpha_{1})^2\big]|0\rangle \langle 1| +\big(2r_{1} r_{2}+i\alpha_{2}\big)/\big[\sqrt{2}(1-\alpha_{1})^2\big]|1\rangle \langle 0|,\nonumber\\
\mathpzc{B}^{\chi}_3=&i/\big[\sqrt{2}(1-\alpha_{1})\big]|0\rangle\langle 0|-i/\big[\sqrt{2}(1+\alpha_{1})\big] |1\rangle\langle 1|.
\end{align}
Thus the bath covariances are obtained as $c_{10}=c_{20}=0$, $c_{11}=c_{22}=\lambda (-2 i r_{1} r_{2}+\alpha_{1} \alpha_{2})/(2\alpha_{1}^2-2)$, and
\begin{align}
 c_{12}=-c_{21} &= \frac{\lambda (2 r_{1} r_{2} \alpha_{1} +i \alpha_{2})}{2(1- \alpha_{1}^2)}.
 \end{align}
After obtaining $A$ and $B$ and diagonalizing $A$, we get
\begin{align}
{\mathbbmss{h}}_{{\mathsf{L}}}^{\chi}= &
-r_{1} r_{2} \lambda/(\alpha_{1}^2-1) \mathbbmss{I}+4 \lambda r_{1} r_{2} \alpha_{1}/\big(1+4 r_{1}^2 -4 r_{1}^4 - (\alpha_{1}^2-\alpha_{2}^2)\big)\sigma_{z},\\
L_1^{\chi}=&
i\sigma_{-},~~~~~~~~~~~~\gamma_1^{\chi}=-\lambda\alpha_{2}/\big(2(1-\alpha_{1})\big),\\
L_2^{\chi}=&
-i\sigma_{+},~~~~~~~~~\gamma_2^{\chi}=\lambda\alpha_{2}/\big(2(1+\alpha_{1})\big),\\
L_3^{\chi}=&\sigma_{z},~~~~~~~~~~~~~~~\gamma_3^{\chi}=0.
\end{align}
Replacing these into Eq. (9) of the main text, the dynamical equation of the system is obtained in the ULL form.

\section{Derivation of the MLL equation}
\label{sec:0corr}

Using the definition of $\chi(\tau)$ from Eq. (1) of the main text and assuming $\chi(\tau_{0})=0$, we obtain
\begin{align}
\dot{\varrho}_{\mathsf{SB}}(\tau)|_{\tau=\tau_{0}}&=-i[H_{\mathsf{SB}},\varrho_{\mathsf{SB}}(\tau_{0})]\nonumber\\
&=-i[H_{\mathsf{SB}},\varrho_{\mathsf{S}}(\tau_{0})\otimes \varrho_{\mathsf{B}}(\tau_{0})+\chi(\tau_{0})]\nonumber\\
&=-i[H_{\mathsf{SB}},\varrho_{\mathsf{S}}(\tau_{0})\otimes \varrho_{\mathsf{B}}(\tau_{0})],
\end{align}
from which, since $\lbrac H_{\chi}(\tau_{0}),\varrho_{\mathsf{S}}(\tau_{0})\otimes \varrho_{\mathsf{B}}(\tau_{0}) \rbrac = i \chi(\tau_{0}) = 0$, we have
\begin{align}
\dot{\varrho}_{\mathsf{S}}(\tau)|_{\tau=\tau_{0}}&=-i[H_{\mathsf{S}}+\mathrm{Tr}_{\mathsf{B}}[H_{\mathrm{I}}\varrho_{\mathsf{B}}(\tau_{0})],\varrho_{\mathsf{S}}(\tau_{0})],
\end{align}
and similarly for the bath,
\begin{align}
\dot{\varrho}_{\mathsf{B}}(\tau)|_{\tau=\tau_{0}}
&=-i[H_{\mathsf{B}}+\mathrm{Tr}_{\mathsf{S}}[H_{\mathrm{I}}\varrho_{\mathsf{S}}(\tau_{0})],\varrho_{\mathsf{B}}(\tau_{0})].
\label{dotrhob}
\end{align}
Using the above equations in $\dot{\chi}(\tau)=\dot{\varrho}_{\mathsf{SB}}(\tau)-\dot{\varrho}_{\mathsf{S}}(\tau)\otimes \varrho_{\mathsf{B}}(\tau)-\varrho_{\mathsf{S}}(\tau)\otimes \dot{\varrho}_{\mathsf{B}}(\tau)$ and expanding $\chi(\tau)$ around $\tau_{0}$ as $\chi(\tau_{0}+\tau)=\chi(\tau_{0})+\tau\dot{\chi}(\tau_{0})+O(\tau^2)$ yields
\begin{align}
\chi(\tau_{0}+\tau)=-i \tau \big[\widetilde{H}_{\mathrm{I}}(\tau_{0}),\varrho_{\mathsf{S}}(\tau_{0})\otimes\varrho_{\mathsf{B}}(\tau_{0})\big]+O(\tau^2),
\label{chi-expanded00}
\end{align}
where (for an arbitrary $s$)
\begin{align}
\widetilde{H}_{\mathrm{I}}(s) &=H_{\mathrm{I}}-\mathrm{Tr}_{\mathsf{B}}[\varrho_{\mathsf{B}}(s)H_{\mathrm{I}}]- \mathrm{Tr}_{\mathsf{S}}[\varrho_{\mathsf{S}}(s)H_{\mathrm{I}}] \label{HI-i}\\
&= \textstyle{\textstyle{\sum_{i\neq 0}}} \mathpzc{S}_{i}\otimes (\mathpzc{B}_i-\ave{\mathpzc{B}_i}_{\mathsf{B}} \mathbbmss{I}_{\mathsf{B}})- \textstyle{\textstyle{\sum_{i\neq 0}}} \ave{\mathpzc{S_i}}_{\mathsf{S}} \mathbbmss{I}_{\mathsf{S}} \otimes \mathpzc{B}_{i}, \label{HI-ii}
\end{align}
where we remind that $\ave{\circ}_{\mathsf{S}}=\mathrm{Tr}[\varrho_{\mathsf{S}}(s) \circ]$ (and similarly for $\ave{\circ}_{\mathsf{B}}$). Equation (\ref{chi-expanded00}) can also be written as
\begin{align}
\chi(\tau_{0}+\tau)=-i \tau \big[\widetilde{H}_{\mathrm{I}}(\tau_{0}+\tau),\varrho_{\mathsf{S}}(\tau_{0}+\tau)\otimes\varrho_{\mathsf{B}}(\tau_{0}+\tau)\big]+O(\tau^2),
\label{chi-expanded0}
\end{align}
which is more convenient for our analysis. Comparing this equation with Eq. (3) of the main text we conclude that
\begin{align}
 H_{\chi}(\tau_{0}+\tau)=\tau \widetilde{H}_{\mathrm{I}}(\tau_{0}+\tau) + O(\tau^2).
\label{markovian-H_{chi}}
\end{align}
Thus from Eq. (\ref{HI-ii}) we conclude that for $j\neq 0$, $\mathpzc{B}_j^{\chi}=\tau (\mathpzc{B}_j-\langle \mathpzc{B}_j\rangle_{\mathsf{B}})$. Hence from Eq. (8) of the main text we obtain for $(i,j)\neq (0,0)$ that
\begin{align}
c_{ij}(\tau)&=\tau(\ave{\mathpzc{B}_i \mathpzc{B}_j}_{\mathsf{B}}-\ave{\mathpzc{B}_i}_{\mathsf{B}}\ave{\mathpzc{B}_j}_{\mathsf{B}})\nonumber\\
&=:\tau\,\mathrm{Cov}_{\mathsf{B}}(\mathpzc{B}_i,\mathpzc{B}_j),
\label{Markov-cov}
\end{align}
which is a positive matrix. Hence $A$ is positive and $B=0$. For $j=0$ we obtain $\mathpzc{B^{\chi}_0}=-\tau \textstyle{\sum_{i\neq0}} \sqrt{d_{\mathsf{S}}} \ave{\mathpzc{S}_{i}}_{\mathsf{S}} \mathpzc{B}_i$, which yields
\begin{equation}
c_{i0}(\tau)=\textstyle{\sum_{j\neq 0}} -\tau\sqrt{d_{\mathsf{S}}} \ave{\mathpzc{S}_{j}}_{\mathsf{S}} \ave{\mathpzc{B}_i \mathpzc{B}_j}_{\mathsf{B}},~~~ i\neq 0,
\end{equation}
and the dynamical equation is obtained as
\begin{align}
\dot{\varrho}_{\mathsf{S}}(\tau)
&=-i\Big[\widetilde{H}_{\mathsf{S}} (\tau)+2\frac{1}{\sqrt{d_{\mathsf{S}}}}\textstyle{\sum_{i\neq0}} \mathrm{Im}\left[c_{i0}(\tau)\right]\mathpzc{S}_{i}, \varrho_{\mathsf{S}}(\tau)\Big]
+  \sum_{i \neq 0, j \neq 0 } c_{ij}(\tau)\Big(2 \mathpzc{S}_{j}\varrho_{\mathsf{S}}(\tau)\mathpzc{S}_{i}-\{\mathpzc{S}_{i}\mathpzc{S}_{j},\varrho_{\mathsf{S}}(\tau)\}\Big)+O(\tau^2).
\label{MarkovianLB}
\end{align}
The above equation depends on the instantaneous state of the bath, which makes it not directly applicable as a system dynamical equation. To write it as an equation which depends only on the state of the system, we note that Eq. \eqref{MarkovianLB} is valid up to the second order in time around zero-correlation points. Thus, we can expand $\varrho_{\mathsf{B}}(\tau)$ around $\tau_{0}$ using Eq. \eqref{dotrhob}, and keep only relevant terms. Replacing $\varrho_{\mathsf{B}}(\tau_{0}+\tau)=\varrho_{\mathsf{B}}(\tau_{0})-i \tau [\widetilde{H}_{\mathsf{B}}(\tau_{0}),\varrho_{\mathsf{B}}(\tau_{0})]+O(\tau^2)$ and $\varrho_{\mathsf{S}}(\tau_{0}+\tau)=\varrho_{\mathsf{S}}(\tau_{0})-i \tau [\widetilde{H}_{\mathsf{S}}(\tau_{0}),\varrho_{\mathsf{S}}(\tau_{0})]+O(\tau^2)$ into Eq. \eqref{Markov-cov} yields
\begin{align}
c_{ij}(\tau) &= \tau\,\mathrm{Cov}_{\mathsf{B}_0}(\mathpzc{B}_i,\mathpzc{B}_j)=\tau \big(\mathrm{Tr}[\mathpzc{B}_i \mathpzc{B}_j \varrho_{\mathsf{B}}(\tau_{0})]-\mathrm{Tr}[\mathpzc{B}_i \varrho_{\mathsf{B}}(\tau_{0})]\, \mathrm{Tr}[\mathpzc{B}_j \varrho_{\mathsf{B}}(\tau_{0})]\big)+O(\tau^2);\nonumber\\
c_{i0}(\tau)&=-\textstyle{\sum_{j\neq 0}} \tau \sqrt{d_{\mathsf{S}}} \ave{\mathpzc{S}_{j}}_{\mathsf{S}_0} \ave{\mathpzc{B}_i \mathpzc{B}_j}_{\mathsf{B}_0};~~~ i\neq 0,
\end{align}
where subscripts $\mathsf{S}_0$ and $\mathsf{B}_0$ mean that the averages are taken with respect to the states of the system and the bath at $\tau=\tau_{0}$. Replacing these into Eq. (\ref{non-diagonal-Lindblad}) leads to the following Markovian dynamical equation:
\begin{align}
\dot{\varrho}_{\mathsf{S}}(\tau_{0}+ \tau)
=&-i\big[\widetilde{H}_{\mathsf{S}}(\tau_{0}) -i \tau\,\mathrm{Tr}_{\mathsf{B}}\big[H_{\mathrm{I}} [\widetilde{H}_{\mathsf{B}}(\tau_{0}) , \varrho_{\mathsf{B}}(\tau_{0}) ] \big]-2 \tau \textstyle{\sum_{(i,j) \neq 0}} \ave{\mathpzc{S}_{j}}_{\mathsf{S}_0} \mathrm{Im}[\ave{\mathpzc{B}_i\mathpzc{B}_j}_{\mathsf{B}_0}] \mathpzc{S}_{i}, \varrho_{\mathsf{S}}(\tau_{0}+\tau)\big] \nonumber\\
&+ \tau \textstyle{\sum_{i \neq 0, j \neq 0 }} \mathrm{Cov}(\mathpzc{B}_i,\mathpzc{B}_j)_{\mathsf{B}_0} \Big(2 \mathpzc{S}_{j}\varrho_{\mathsf{S}}(\tau_{0}+\tau)\mathpzc{S}_{i}-\{\mathpzc{S}_{i}\mathpzc{S}_{j},\varrho_{\mathsf{S}}(\tau_{0}+\tau)\}\Big)+O(\tau^2),
\label{MarkovianLB-const}
\end{align}
where we have used
\begin{equation}
\widetilde{H}_{\mathsf{S}}(\tau_{0}+\tau) = \widetilde{H}_{\mathsf{S}}(\tau_{0}) -i \tau\,\mathrm{Tr}_{\mathsf{B}}\big[H_{\mathrm{I}} [\widetilde{H}_{\mathsf{B}}(\tau_{0}) , \varrho_{\mathsf{B}}(\tau_{0}) ]\big] + O(\tau^2).
\end{equation}
Without loss of generality we can assume $\tau_{0}=0$.

Equation (\ref{MarkovianLB-const}) has been obtained for sufficiently small $\tau$'s. However, if we extend this equation to arbitrary time instants, we shall have the MLL master equation, which is an approximation for the exact dynamics. This extension is a sort of coarse graining in time, yet with clear differences with the standard coarse-graining (CG) methods (see, e.g., Refs. [4,\,17,\,24,\,46--49] of the main text). For example, the MLL coarse-graining and the CG method of Ref. [17] both assume that the correlation can vanish repeatedly during the evolution (a Born approximation), while MLL jump rates are proportional to the evolution time $\tau$---rather than a fixed CG time scale. In addition, a closer inspection shows that the way the Born approximation is implemented in our MLL equation is different from that of the CG methods (and also the Redfield equation). Methodologically, CG techniques often involve short-time, weak-coupling, and the standard Born approximation; whereas our MLL equation hinges on a weak-correlation approximation. Despite evident similarities, the final results appear different. We illustrate this point in the examples.

\section{The MLL equation is exact for short times}
\label{app:der-short-t}

The short-time behavior of the system density matrix around $\tau_{0}$ is obtained by integration of Eq. \eqref{MarkovianLB-const}, which yields
\begin{align}
\varrho_{\mathsf{S}}(\tau)=&\varrho_{\mathsf{S}}(0) -i [\widetilde{H}_{\mathsf{S}}(0),\textstyle{\int_{0}^{\tau}} ds\,\varrho_{\mathsf{S}}(s)]  +i\Big[i \mathrm{Tr}_{\mathsf{B}}\big[H_{\mathrm{I}} [\widetilde{H}_{\mathsf{B}}(0) , \varrho_{\mathsf{B}}(0) ] \big]+2 \textstyle{\sum_{(i,j) \neq (0,0)}} \ave{\mathpzc{S}_{j}}_{\mathsf{S}_0} \ave{\mathpzc{B}_i \mathpzc{B}_j}_{\mathsf{B}_0} \mathpzc{S}_{i}, \textstyle{\int_{0}^{\tau}} s\varrho_{\mathsf{S}}(s) \,ds \Big]\nonumber\\
&+ \textstyle{\sum_{i\neq 0,j\neq0}} \mathrm{Cov}(\mathpzc{B}_i,\mathpzc{B}_j)_{\mathsf{B}_0} \Big(2 \mathpzc{S}_{j}\textstyle{\int_{0}^{\tau}} ds\, s \varrho_{\mathsf{S}}(s)\mathpzc{S}_{i}-\{\mathpzc{S}_{i}\mathpzc{S}_{j},\textstyle{\int_{0}^{\tau}}ds\, s \varrho_{\mathsf{S}}(s)\}\Big) + O(\tau^3).
\label{int-markov-rho}
\end{align}
To calculate $\textstyle{\int_{0}^{\tau}} \varrho(s)\, ds$ and $\textstyle{\int_{0}^{\tau}} s \varrho(s)\, ds$ for short times, we insert Eq. \eqref{int-markov-rho} into the integrals, and thus we obtain
\begin{align}
\textstyle{\int_{0}^{\tau}} \varrho_{\mathsf{S}}(s)\, ds &=\varrho_{\mathsf{S}}(0) \tau-i \big[\widetilde{H}_{\mathsf{S}}(0),\varrho_{\mathsf{S}}(0)\big]\frac{\tau^2}{2}+ O(\tau^{3}),
\label{int1}
\end{align}
and
\begin{align}
\textstyle{\int_{0}^{\tau}} s \varrho_{\mathsf{S}}(s) \,ds &=\varrho_{\mathsf{S}}(0)\frac{\tau^2}{2}+O(\tau^3).
\label{int2}
\end{align}
Inserting Eqs.~\eqref{int1} and \eqref{int2} into Eq. \eqref{int-markov-rho}, the short-time dynamics of the system is obtained up to third order in time as
\begin{align}
\varrho_{\mathsf{S}}(\tau)=&\varrho_{\mathsf{S}}(0)-i \tau [\widetilde{H}_{\mathsf{S}}(0),\varrho_{\mathsf{S}}(0) ]- \tau^2 \big[\widetilde{H}_{\mathsf{S}}(0) , [\widetilde{H}_{\mathsf{S}}(0),\varrho_{\mathsf{S}}(0)]\big]\nonumber\\
&+i \frac{\tau^2}{2} \Big[\ave{[H_{\mathrm{I}},\widetilde{H}_{\mathsf{B}}(0)]}_{\mathsf{B}_{0}}-2i \textstyle{\sum_{(i,j) \neq 0} } \ave{\mathpzc{S}_{j}}_{\mathsf{S}_0} \ave{\mathpzc{B}_i \mathpzc{B}_j}_{\mathsf{B}_0} \mathpzc{S}_{i}, [\widetilde{H}_{\mathsf{S}}(0),\varrho_{\mathsf{S}}(0)]\Big]\nonumber\\
&+ \frac{\tau^2}{2} \Big[\ave{[H_{\mathrm{I}},\widetilde{H}_{\mathsf{B}}(0)]}_{\mathsf{B}_{0}}-2i \textstyle{\sum_{(i,j) \neq 0}} \ave{\mathpzc{S}_{j}}_{\mathsf{S}_0} \ave{\mathpzc{B}_i \mathpzc{B}_j}_{\mathsf{B}_0} \mathpzc{S}_{i}, \big[\widetilde{H}_{\mathsf{S}}(0),[\widetilde{H}_{\mathsf{S}}(0),\varrho_{\mathsf{S}}(0)]\big]\Big]\nonumber\\
&+\frac{\tau^2}{2}\mathbbmss{D}[\varrho_{\mathsf{S}}(0)]+O(\tau^3),
\label{short-time-markov}
\end{align}
where $\widetilde{H}_{\mathsf{S}}=H_{\mathsf{S}}+\ave{H_{\mathrm{I}}}_{\mathsf{B}_{0}}$ and $\mathbbmss{D}[\circ]=\sum_{i \neq 0, j \neq 0 } \mathrm{Cov}(\mathpzc{B}_i,\mathpzc{B}_j)_{\mathsf{B}_0} (2 \mathpzc{S}_{j} \circ \mathpzc{S}_{i}-\{\mathpzc{S}_{i}\mathpzc{S}_{j},\circ \})$. When the system-bath initial state is prepared in a product state (which is often the case), i.e., when $\chi(0)=0$, the short-time behavior of the dynamics, either Markovian or non-Markovian, will be given with the above equation. From this equation it is immediate that if the state of the system at $\tau_{0}$ commutes with $\widetilde{H}_{\mathsf{S}}(0)$, the system dynamics will be proportional to $\tau^{2}$. Otherwise the linear term in $\tau$ can dominate in short-time evolution. In the case where $[\widetilde{H}_{\mathsf{S}}(0),\varrho_{\mathsf{S}}(0)]=0$ and $[\widetilde{H}_{\mathsf{B}}(0),\varrho_{\mathsf{B}}(0)]=0$, Eq. \eqref{short-time-markov} is simplified as
\begin{align}
\varrho_{\mathsf{S}}(\tau)=&\varrho_{\mathsf{S}}(0)+\frac{\tau^2}{2}\mathbbmss{D}[\varrho_{\mathsf{S}}(0)]+O(\tau^3).
\label{short-time-markov-1}
\end{align}
Using the definition of $\mathrm{Cov}(\mathpzc{B}_i,\mathpzc{B}_j)_{\mathsf{B}_{0}}$ as given below Eq. (11) of the main text and the expansion of $H_{\mathrm{I}}$ from Eq. (\ref{H_{int}-expanded}), after some straightforward algebra, the above equation can be recast as
\begin{align}
\varrho_{\mathsf{S}}(\tau)=&\varrho_{\mathsf{S}}(0)-\frac{\tau^2}{2}\mathrm{Tr}_{\mathsf{B}}\big[H_{\mathrm{I}},[\widetilde{H}_{\mathrm{I}(0)}, \varrho_{\mathsf{S}}(0)\otimes \varrho_{\mathsf{B}}(0)]\big]+O(\tau^3).
\label{short-t-mll}
\end{align}

\section{Applicability of the MLL approximation in asymptotic regimes}
\label{asymptotic-state}

 Consider a quantum system which is coupled to an environment and the spectral decomposition of the total Hamiltonian is given by $H_{\mathsf{SB}}=\sum_m E_m |E_m\rangle\langle E_m|$. Let us define the time-averaged state of the total system as
\begin{align}
\bar{\varrho}_{\mathsf{SB}}(\tau):&=\frac{1}{\tau}{\textstyle{\int_{0}^{\tau}}}ds\,\varrho_{\mathsf{SB}}(s)=\frac{1}{\tau}\textstyle{\sum_{mn}}  |E_m\rangle\langle E_n|  \textstyle{\int_{0}^{\tau}}ds\, e^{-i (E_{m}-E_{n})s} \langle E_{m}|\varrho_{\mathsf{SB}}(0)|E_{n}\rangle,
\end{align}
where we have assumed that the total Hamiltonian is time-independent. Since $\lim_{\tau \rightarrow \infty} (1/\tau)\textstyle{\int_{0}^{\tau}}ds~e^{i(\omega-\omega')s}\approx \delta_{\omega,\omega'}$ and with the assumption that there is no degeneracy in the total Hamiltonian, it is seen that the asymptotic limit of $\bar{\varrho}_{\mathsf{SB}}(\tau)$ is obtained as
\begin{align}
\varrho_{\mathsf{SB}}^{\star}&=\textstyle{\sum_{mn}} \langle E_{m}|\varrho_{\mathsf{SB}}(0)|E_{n}\rangle |E_m\rangle\langle E_n|  \left(\lim_{\tau \rightarrow \infty} \frac{1}{\tau}\textstyle{\int_{0}^{\tau}}ds\, e^{-i (E_{m}-E_{n})s}\right)\nonumber\\
&=\textstyle{\sum_{mn}} \langle E_{m}|\varrho_{\mathsf{SB}}(0) |E_{n}\rangle |E_m\rangle\langle E_n| \delta_{E_m,E_n}\nonumber\\
&= \textstyle{\sum_{m}}  \langle E_{m}|\varrho_{\mathsf{SB}}(0) |E_{m}\rangle  |E_m\rangle\langle E_m|,
\end{align}
From this equation it is evident that $[H_{\mathsf{SB}},\varrho_{\mathsf{SB}}^{\star}]=0$, hence $\varrho_{\mathsf{SB}}^{\star}$ is a steady state of the total system. Noting that, for generic system-bath Hamiltonians (with nondegenerate energies and gaps), for almost all observables $A$, we have $\langle A\rangle_{\varrho_{\mathsf{SB}}(\infty)}\approx \langle A\rangle_{\varrho^{\star}_{\mathsf{SB}}}$ \cite{Reimann}, in this sense one may consider the state $\varrho^{\star}_{\mathsf{SB}}$ effectively as a description for the asymptotic state $\varrho_{\mathsf{SB}}(\infty)$. Hence the asymptotic state of the subsystem $\mathsf{S}$ can be read as $\varrho_{\mathsf{S}}^{\star}=\sum_{m} \langle E_{m}|\varrho_{\mathsf{SB}}(0) |E_{m}\rangle \mathrm{Tr}_{\mathsf{B}}{\left[|E_m\rangle\langle E_m|\right]}$.

In the highly strong-coupling regime, the total Hamiltonian can be approximated as $H_{\mathsf{SB}} \approx H_{\mathrm{I}}$. If the interaction Hamiltonian has only one term, i.e., $H_{\mathrm{I}}=\mathpzc{S}\otimes \mathpzc{B}$, the eigenbasis of the total Hamiltonian can be approximated with the eigenbasis of the interaction Hamiltonian. By decomposing $\mathpzc{S}$ and $\mathpzc{B}$ as $\mathpzc{S}=\sum_i \alpha_i |\alpha_i\rangle\langle \alpha_i|$ and $\mathpzc{B}=\sum_j \eta_j |\eta_j\rangle\langle \eta_j|$, the eigenvector of the total Hamiltonian $|E_{m}\rangle$'s can be approximated as $|E_{ij}\rangle=|\alpha_i\rangle \otimes |\eta_j\rangle$. Now by assuming that the initial state is uncorrelated, it is readily seen that $\varrho_{\mathsf{SB}}^{\star}$ is uncorrelated,
\begin{align}
\varrho_{\mathsf{SB}}^{\star} &\approx \textstyle{\sum_{i,j}} \langle E_{ij}|\varrho_{\mathsf{S}}(0)\otimes \varrho_{\mathsf{B}}(0)|E_{ij}\rangle  |E_{ij}\rangle\langle E_{ij}|\nonumber\\
&= \textstyle{\sum_{i}} \langle \alpha_{i}|\varrho_{\mathsf{S}}(0)|\alpha_{i}\rangle  |\alpha_{i}\rangle\langle \alpha_{i}| \otimes  \textstyle{\sum_j} \langle \eta_{j}|\varrho_{\mathsf{B}}(0)|\eta_{j}\rangle |\eta_{j}\rangle\langle \eta_{j}|\nonumber\\
&=\varrho_{\mathsf{S}}^{\star} \otimes \varrho_{\mathsf{B}}^{\star}.
\end{align}
Hence the correlation operator in this long-time limit vanishes.

\begin{remark}
As is clear from the above analysis, the MLL equation for long times can be generally different from the MLL equation obtained by initial conditions. In addition, constructing the MLL equation for long times requires the knowledge of the asymptotic state. However, this information is initially unavailable and its determination is indeed a goal of any dynamical equation in the first place. Despite this difficulty, for some systems the very MLL equation obtained by initial conditions may still provide the asymptotic state $\varrho^{\star}_{\mathsf{S}}$ with good accuracy. Example II in Sec. \ref{example1} provides a case of this type [see also Fig. 2 (left) in the main text]. In Sec. \ref{app:enMLL} we show another time-asymptotic regime in which the MLL approximation still holds, but in the weak-correlation limit. 
\label{rem:MLL}
\end{remark}

\section{Details of the derivation of the exact dynamical equation for the correlation operator and the systematic weak-correlation expansion}
\label{app:ULL-details}

\subsection{Exact dynamical equation for the correlation}

Using the decomposition of the state of the total system at a given time $\tau_{0}$ as $\varrho_{\mathsf{SB}}(\tau_{0})=\varrho_{\mathsf{S}}(\tau_{0})\otimes\varrho_{\mathsf{B}}(\tau_{0})+\chi(\tau_{0})$, the state of the total system at a later time $\tau_{0}+\xi$ is given by
\begin{equation}
\varrho_{\mathsf{SB}}(\tau_{0}+\xi)=U_{\xi}\varrho_{\mathsf{S}}(\tau_{0})\otimes\varrho_{\mathsf{B}}(\tau_{0}) U_{\xi}^{\dag}+U_{\xi}\chi(\tau_{0}) U_{\xi}^{\dag},
\label{eq1}
\end{equation}
where $U_{\xi}$ is the evolution operator of the total system $\mathsf{SB}$ in a time interval $\xi$. In the following we use the notation $\hat\circ$ to indicate that the related operator has been obtained from an uncorrelated state---which is akin to the MLL method.
Let us remind that in our MLL methodology (Sec. \ref{sec:0corr}) we have already shown that the correlation formed between the system and the bath due to the evolution from an \textit{uncorrelated} state (here $\varrho_{\mathsf{S}}(\tau_{0}) \otimes \varrho_{\mathsf{B}}(\tau_{0})$) is given by [Eq. (\ref{chi-expanded00})]
 \begin{align}
 \hat{\chi}(\tau_{0}+\xi)=-i \xi [\widetilde{H}_{\mathrm{I}}(\tau_{0}),\varrho_{\mathsf{S}}(\tau_{0})\otimes \varrho_{\mathsf{B}}(\tau_{0})]+O(\xi^2),
 \label{eq5}	
 \end{align}
which is exact for short times $\xi$'s. Since the \textit{exact} dynamical equation for the system and the bath are given by
 \begin{align}
 \dot{\varrho}_{\mathsf{S}}(\tau)&=-i [\widetilde{H}_{\mathsf{S}}(\tau),\varrho_{\mathsf{S}}(\tau)]-i \mathrm{Tr}_{\mathsf{B}}[H_{\mathrm{I}},
\chi(\tau)],\label{exactS}\\
\dot{\varrho}_{\mathsf{B}}(\tau)&=-i [\widetilde{H}_{\mathsf{B}}(\tau),\varrho_{\mathsf{B}}(\tau)]-i \mathrm{Tr}_{\mathsf{S}}[H_{\mathrm{I}},\chi(\tau)],\label{exactB}
 \end{align}
if we start from a zero-correlation state at $\tau_{0}$, at a later time $\tau_{0}+\xi$ the system and bath states are given up to the first order in time $\xi$ by
\begin{align}	
\hat{\varrho}_{\mathsf{S}}(\tau_{0}+\xi)&=\varrho_{\mathsf{S}}(\tau_{0})-i \xi [\widetilde{H}_{\mathsf{S}}(\tau_{0}),\varrho_{\mathsf{S}}(\tau_{0})]+O(\xi^2),
\label{eq3}\\
\hat{\varrho}_{\mathsf{B}}(\tau_{0}+\xi)&=\varrho_{\mathsf{B}}(\tau_{0})-i \xi [\widetilde{H}_{\mathsf{B}}(\tau_{0}),\varrho_{\mathsf{B}}(\tau_{0})]+O(\xi^2).
\label{eq4}
\end{align}	
Thus
\begin{equation}
\varrho_{\mathsf{SB}}(\tau_{0}+\xi)=\hat{\varrho}_{\mathsf{S}}(\tau_{0}+\xi)\otimes \hat{\varrho}_{\mathsf{B}}(\tau_{0}+\xi)+\hat{\chi}(\tau_{0}+\xi)+U_{\xi}\chi(\tau_{0}) U_{\xi}^{\dag},
\label{Eq1}
\end{equation}
where on the right-hand side (RHS) we should keep terms of the first order in $\xi$. From this relation the state of the system (bath) is obtained by tracing out over the bath (system) as
\begin{align}
\varrho_{\mathsf{S}}(\tau_{0}+\xi)&=\hat{\varrho}_{\mathsf{S}}(\tau_{0}+\xi)+\mathrm{Tr}_{\mathsf{B}}[U_{\xi}\chi(\tau_{0}) U_{\xi}^{\dag}]
\nonumber\\
&= \varrho_{\mathsf{S}}(\tau_{0})-i \xi [\widetilde{H}_{\mathsf{S}}(\tau_{0}),\varrho_{\mathsf{S}}(\tau_{0})]+\mathrm{Tr}_{\mathsf{B}}[U_{\xi}\chi(\tau_{0}) U_{\xi}^{\dag}]+O(\xi^2), \label{Eq5-}\\
\varrho_{\mathsf{B}}(\tau_{0}+\xi)&=\hat{\varrho}_{\mathsf{B}}(\tau_{0}+\xi)+\mathrm{Tr}_{\mathsf{S}}[U_{\xi}\chi(\tau_{0}) U_{\xi}^{\dag}] \nonumber
\\
&= \varrho_{\mathsf{B}}(\tau_{0})-i \xi [\widetilde{H}_{\mathsf{B}}(\tau_{0}),\varrho_{\mathsf{B}}(\tau_{0})]+\mathrm{Tr}_{\mathsf{S}}[U_{\xi}\chi(\tau_{0}) U_{\xi}^{\dag}]+O(\xi^2). \label{Eq5}
\end{align}
From Eqs.~\eqref{Eq1} -- \eqref{Eq5} and using the relation $\chi(\tau+\tau_{0})=\varrho_{\mathsf{SB}}(\tau_{0}+\xi)-\varrho_{\mathsf{S}}(\tau_{0}+\xi)\otimes \varrho_{\mathsf{B}}(\tau_{0}+\xi)$ the system-bath correlation operator at $\tau_{0}+\xi$ (up to the first order in $\xi$) is obtained as follows:
\begin{align}
\chi(\tau_{0}+\xi)=\hat{\chi}(\tau_{0}+\xi)+U_{\xi}\chi(\tau_{0}) U_{\xi}^{\dag}-\mathrm{Tr}_{\mathsf{B}}[U_{\xi}\chi(\tau_{0}) U_{\xi}^{\dag}]\otimes \hat{\varrho}_{\mathsf{B}}(\tau_{0}+\xi)-\hat{\varrho}_{\mathsf{S}}(\tau_{0}+\xi)\otimes\mathrm{Tr}_{\mathsf{S}}[U_{\xi}\chi(\tau_{0}) U_{\xi}^{\dag}],
\label{eq7}
\end{align}
or equivalently,
\begin{align}
\hskip-3mm\chi(\tau_{0}+\xi)=&\chi(\tau_{0})-i \xi [\widetilde{H}_{\mathrm{I}}(\tau_{0}),\varrho_{\mathsf{S}}(\tau_{0})\otimes \varrho_{\mathsf{B}}(\tau_{0})]-i\xi [H_{\mathsf{SB}},\chi(\tau_{0})]+i\xi \mathrm{Tr}_{\mathsf{B}}[H_{\mathrm{I}} \, ,\chi(\tau_{0})]\otimes \varrho_{\mathsf{B}}(\tau_{0})+i\xi \varrho_{\mathsf{S}}(\tau_{0})\otimes\mathrm{Tr}_{\mathsf{S}} [H_{\mathrm{I}} \, ,\chi(\tau_{0})]\nonumber\\
&+O(\xi^2).
\label{eq6}
\end{align}
Equations (\ref{Eq5-}), (\ref{Eq5}), and (\ref{eq6}) indicate that starting at time $\tau_{0}$ with the knowledge of the total state $\varrho_{\mathsf{SB}}(\tau_{0})$ (hence $\varrho_{\mathsf{S}}(\tau_{0})$, $\varrho_{\mathsf{B}}(\tau_{0})$, and $\chi(\tau_{0})$) one can read the corresponding quantities at a close later time $\tau_{0}+\xi$. When $\chi(\tau_{0})=0$, the states $\varrho_{\mathsf{S}}(\tau_{0}+\xi)$ and $\varrho_{\mathsf{B}}(\tau_{0}+\xi)$ only require the knowledge of the system and the bath at time $\tau_{0}$. However, the correlation operator $\chi(\tau_{0}+\xi)$ is required for the estimation of the states $\varrho_{\mathsf{S}}(\tau_{0}+2\xi)$ and $\varrho_{\mathsf{B}}(\tau_{0}+2\xi)$. Figure \ref{fig:flowchart} depicts this hierarchical and algorithmic construction.

One can rewrite the above equation as a \textit{recursive formula} for calculating $\chi$ at a given time $\tau_{n+1}$ through an expression for time $\tau_n$ as
\begin{align}
\chi(\tau_{n+1})=&\chi(\tau_n)-i \xi [\widetilde{H}_{\mathrm{I}}(\tau_n),\varrho_{\mathsf{S}}(\tau_n)\otimes \varrho_{\mathsf{B}}(\tau_n)]-i\xi [H_{\mathsf{SB}},\chi(\tau_n)]+i\xi \mathrm{Tr}_{\mathsf{B}}[H_{\mathrm{I}} \, ,\chi(\tau_n)]\otimes \varrho_{\mathsf{B}}(\tau_{n})+i\xi \varrho_{\mathsf{S}}(\tau_{n})\otimes\mathrm{Tr}_{\mathsf{S}} [H_{\mathrm{I}} \, ,\chi(\tau_n)]\nonumber\\
&+O(\xi^2).
\label{chi-iterative}
\end{align}
From the above equation and by taking the $\xi\rightarrow 0$ limit, one can also derive the \textit{exact} and general formula for $\dot\chi(\tau)$,
\begin{align}
\dot{\chi}(\tau)=-i [\widetilde{H}_{\mathrm{I}}(\tau),\varrho_{\mathsf{S}}(\tau)\otimes \varrho_{\mathsf{B}}(\tau)]-i [H_{\mathsf{SB}},\chi(\tau)]+i \mathrm{Tr}_{\mathsf{B}}[H_{\mathrm{I}} \, ,\chi(\tau)]\otimes \varrho_{\mathsf{B}}(\tau)+i \varrho_{\mathsf{S}}(\tau)\otimes\mathrm{Tr}_{\mathsf{S}} [H_{\mathrm{I}} \, ,\chi(\tau)].
\label{chi-dot}
\end{align}
To simplify the notation, we introduce a linear superoperator $\mathcal{Y}_{\tau}$ as
\begin{align}
\mathcal{Y}_{\tau}[\circ]:=-i [H_{\mathsf{SB}},\circ]+i \mathrm{Tr}_{\mathsf{B}}[H_{\mathrm{I}} \, ,\circ]\otimes \varrho_{\mathsf{B}}(\tau)+i \varrho_{\mathsf{S}}(\tau)\otimes\mathrm{Tr}_{\mathsf{S}} [H_{\mathrm{I}} \, ,\circ],
\label{def-of-F}
\end{align}
using which Eq. \eqref{chi-dot} can be rewritten as
\begin{align}
\dot{\chi}(\tau)=-i [\widetilde{H}_{\mathrm{I}}(\tau),\varrho_{\mathsf{S}}(\tau)\otimes \varrho_{\mathsf{B}}(\tau)]+\mathcal{Y}_{\tau}[\chi(\tau)].
\label{chi-dot-formal}
\end{align}
This exact equation can be solved formally by iteration,
\begin{align}
\chi(\tau)=& \textstyle{\sum_{k=0}^{\infty}} \int_{0}^{s_{0}}  ds_1 \int_{0}^{s_1}  ds_2\, \cdots \int_{0}^{s_{k-1}}  ds_{k}\, \mathcal{Y}_{s_1} \Big[\mathcal{Y}_{s_2}\cdots \Big[\mathcal{Y}_{s_{k}} [\chi(0)]\ldots\Big]\nonumber\\
&+ \textstyle{\sum_{k=0}^{\infty}} \int_{0}^{s_{0}}  ds_1 \int_{0}^{s_1}  ds_2\, \cdots \int_{0}^{s_{k-1}}  ds_{k}\, \mathcal{Y}_{s_1} \Big[\mathcal{Y}_{s_2}\cdots \Big[\mathcal{Y}_{s_{k}}\Big[-i\int_{0}^{s_k}  ds \, [\widetilde{H}_{\mathrm{I}}(s),\varrho_{\mathsf{S}}(s)\otimes \varrho_{\mathsf{B}}(s)]\Big]\ldots\Big],
\label{chi-exact}
\end{align}
where $s_0\equiv \tau$, $k$ counts the number of nested integrals, and the $k=0$ terms in the summations simply denote the identity map. This is an \textit{exact} relation which shows how by having the exact subsystem states $\varrho_{\mathsf{S}}(\tau)$ and $\varrho_{\mathsf{B}}(\tau)$, the interaction Hamiltonian $H_{\mathrm{I}}$, and the initial correlation $\chi(0)$ one can read the development of the correlation operator with time.

We remark that an alternative way to derive Eq. (\ref{chi-dot-formal}) is to start from the very definition of the correlation operator $\chi(\tau)$ [Eq. (1) of the main text], differentiating it with respect to $\tau$ and using Eqs. (\ref{exactS}) and (\ref{exactB}), which yield Eq. (\ref{chi-dot-formal}). However, a clear physical appeal of the recursive-relation approach we developed above is that this way one can inspect how the development of correlation features beyond the Markovian approximation (MLL) play a key role in deriving more accurate dynamical equation.  In addition, this approach better manifests the intricate role of the MLL expansion in obtaining an expansion for $\chi(\tau)$.

Two limiting cases can be discerned: (i) when there is no coupling between the system and the bath ($H_{\mathrm{I}}=\widetilde{H}_{\mathrm{I}}(\tau)=0$), this relation gives $\chi(\tau)=U_{\mathsf{S}}(\tau)\otimes U_{\mathsf{B}}(\tau)\chi(0)U_{\mathsf{S}}^{\dag}(\tau)\otimes U^{\dag}_{\mathsf{B}}(\tau)$, which obviously vanishes for the uncorrelated initial state. (ii) When there is no initial correlation $\chi(0)=0$ (but $H_{\mathrm{I}}\neq0$ ), Eq. (\ref{chi-exact}) reduces to the solution reported in Eq. (\ref{chi-for-chi0}).

As we delineate in the next section, the principal importance of Eq. (\ref{chi-exact}) for our framework is that it also enables a \textit{systematic} approach to approximating $\chi$ perturbatively---and hence the ULL master equation. To rigorously apply a \textit{weak-correlation} expansion, it is important to derive the interaction-picture forms of the above equations.

\begin{figure}[tp]
\includegraphics[scale=0.5]{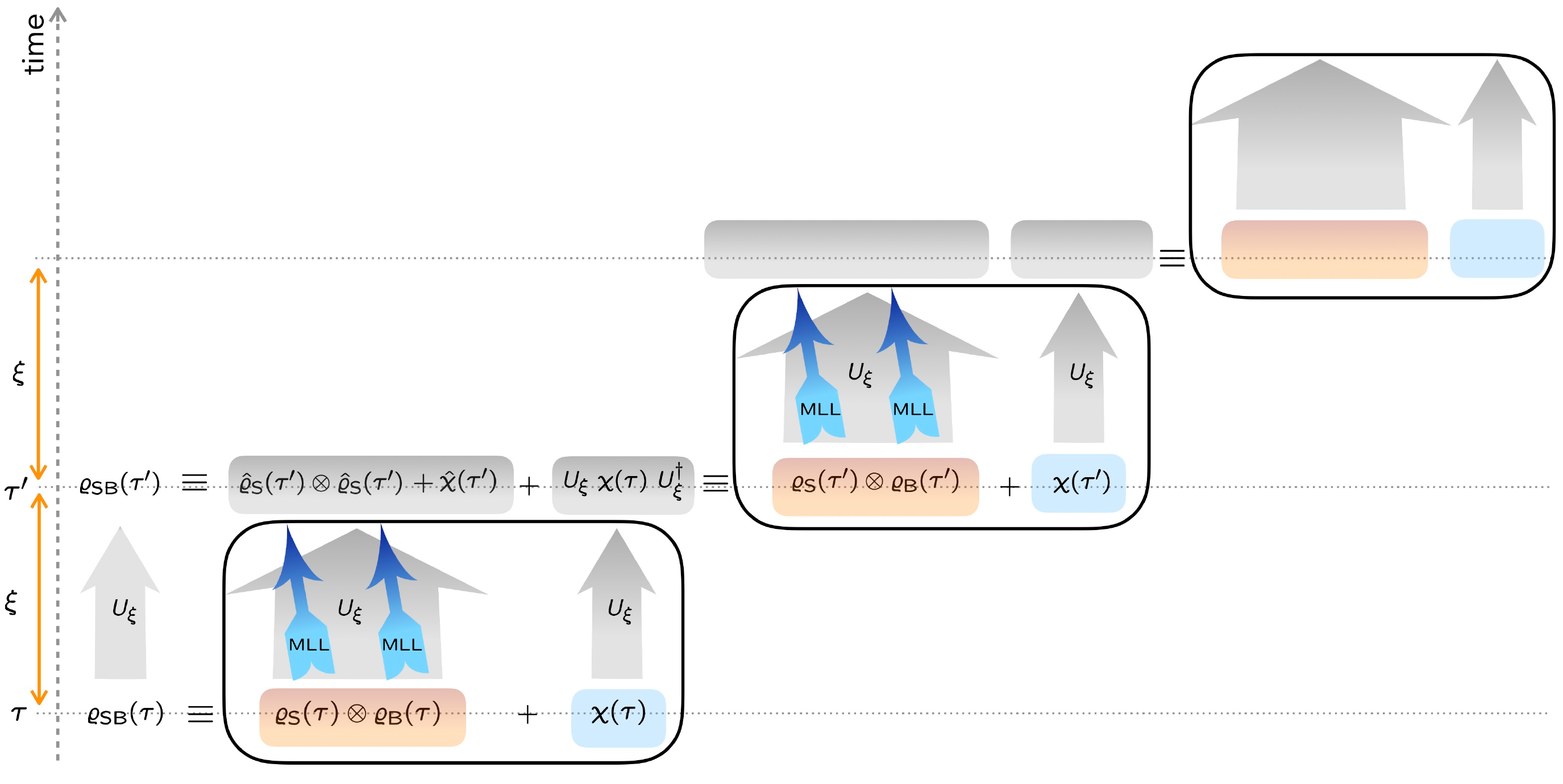}
\caption{Schematic of the systematic perturbative expansion of the ULL equation---constructed from weak-correlation and consecutive MLL expansions [Eq. (\ref{chi-iterative})]. For details see Sec. \ref{app:ULL-details}.}
\label{fig:flowchart}
\end{figure}

\subsection{Systematic weak-correlation perturbative expansion}

The interaction picture forms of Eqs.~\eqref{chi-iterative} -- \eqref{chi-exact} are as follows (we use boldfaced letters for operators in the interaction picture):
\begin{align}
\bm{\chi}(\tau_{n+1})=&\bm{\chi}(\tau_n)-i \xi [\widetilde{\bm{H}}_{\mathrm{I}}(\tau_n),\bm{\varrho}_{\mathsf{S}}(\tau_n)\otimes \bm{\varrho}_{\mathsf{B}}(\tau_n)]-i\xi [\bm{H}_{\mathrm{I}}(\tau_n),\bm{\chi}(\tau_n)]+i\xi \mathrm{Tr}_{\mathsf{B}}[\bm{H}_{\mathrm{I}}(\tau_n) \, ,\bm{\chi}(\tau_n)]\otimes \bm{\varrho}_{\mathsf{B}}(\tau_{n})\nonumber\\
&+i\xi \bm{\varrho}_{\mathsf{S}}(\tau_{n})\otimes \mathrm{Tr}_{\mathsf{S}} [\bm{H}_{\mathrm{I}}(\tau_n) \, ,\bm{\chi}(\tau_n)]+O(\xi^2),
\label{chi-iterative-intpic}\\
\dot{\bm{\chi}}(\tau)=&-i [\widetilde{\bm{H}}_{\mathrm{I}}(\tau),\bm{\varrho}_{\mathsf{S}}(\tau)\otimes \bm{\varrho}_{\mathsf{B}}(\tau)]+\bm{\mathcal{Y}}_{\tau}[\bm{\chi}(\tau)],
\label{chi-dot-intpic}\\
\bm{\mathcal{Y}}_{\tau}[\circ]=&-i [\bm{H}_{\mathrm{I}},\circ]+i \mathrm{Tr}_{\mathsf{B}}[\bm{H}_{\mathrm{I}} \, ,\circ]\otimes \bm{\varrho}_{\mathsf{B}}(\tau)+i \bm{\varrho}_{\mathsf{S}}(\tau)\otimes\mathrm{Tr}_{\mathsf{S}} [\bm{H}_{\mathrm{I}} \, ,\circ],
\label{def-of-F-intpic}\\
\bm{\chi}(\tau)=& \textstyle{\sum_{k=0}^{\infty}} \int_{0}^{\tau}  ds_1 \int_{0}^{s_1}  ds_2\, \cdots \int_{0}^{s_{k-1}}  ds_{k}\, \bm{\mathcal{Y}}_{s_1} \Big[\bm{\mathcal{Y}}_{s_2}\cdots \Big[\bm{\mathcal{Y}}_{s_{k}} [\bm{\chi}(0)]\Big]\nonumber\\
&+ \textstyle{\sum_{k=0}^{\infty}} \int_{0}^{\tau}  ds_1 \int_{0}^{s_1}  ds_2\, \cdots \int_{0}^{s_{k-1}}  ds_{k}\, \bm{\mathcal{Y}}_{s_1} \Big[\bm{\mathcal{Y}}_{s_2}\cdots \Big[\bm{\mathcal{Y}}_{s_{k}}\Big[-i\int_{0}^{s_k}  ds \, [\widetilde{\bm{H}}_{\mathrm{I}}(s),\bm{\varrho}_{\mathsf{S}}(s)\otimes \bm{\varrho}_{\mathsf{B}}(s)]\Big]\Big],
\label{chi-exact-intPic}
\end{align}
where
\begin{gather}
\bm{\varrho}_{\mathsf{S}}(s)= U^{\dag}_{\mathsf{S}}(s) \, \varrho_{\mathsf{S}}(s) \, U_{\mathsf{S}}(s),\\
\bm{\varrho}_{\mathsf{B}}(s)= U^{\dag}_{\mathsf{B}}(s) \, \varrho_{\mathsf{B}}(s) \, U_{\mathsf{B}}(s),\\
\widetilde{\bm{H}}_{\mathrm{I}}(s)= U^{\dag}_{0}(s) \, \widetilde{H}_{\mathrm{I}}(s) \, U_{0}(s),\\
\bm{\chi}(s)= U^{\dag}_{0}(s)\chi(s)U_{0}(s),
\end{gather}
with $U_{0}(s)=e^{-is(H_{\mathsf{S}} + H_{\mathsf{B}})}=U_{\mathsf{S}}(s)\otimes U_{\mathsf{B}}(s)$. We note that in the first summation of Eq. (\ref{chi-exact-intPic}) the $k$th term is of the order of $\bm{H}^{k}_{\mathrm{I}}$, whereas in the second summation the $k$th term is of the order of $\bm{H}^{k+1}_{\mathrm{I}}$. In addition, we can write solution (\ref{chi-exact-intPic}) as
\begin{equation}
\bm{\chi}(\tau)=\textstyle{\sum_{\ell=0}^{\infty}}\bm{f}_{\ell}(\tau),
\label{formal}
\end{equation}
where $\bm{f}_{0}(\tau)\equiv \bm{\chi}(0)$ and $\bm{f}_{\ell}(\tau)$ (for $\ell\geqslant 1$) is the addition of the $\ell$th term of the first summation and the $(\ell-1)$th term of the second summation in Eq. (\ref{chi-exact-intPic}) [hence $\bm{f}_{\ell}$ is of the order of $\bm{H}^{\ell}_{\mathrm{I}}$]. This series solution to Eq. (\ref{chi-dot-intpic}) is convergent as long as the kernel of its corresponding integral equation (the homogenous part) is square integrable \cite{book:Arfken}. This can be readily satisfied for systems with $\Vert \bm{\mathcal{Y}}_{\tau}\Vert <\infty$---or simply $\Vert \bm{H}_{\mathrm{I}}\Vert <\infty$.

\subsubsection{ULL$2$ equation}

From Eq. \eqref{chi-exact-intPic} we can obtain the correlation operator up to the first order in the interaction Hamiltonian $\widetilde{\bm{H}}_{\mathrm{I}}$ as
\begin{align}
\bm{\chi}^{(1)}(\tau)=&\bm{\chi}(0)+ \textstyle{\int_{0}^{\tau}}  ds \, \bm{\mathcal{Y}}_{s} [\bm{\chi}(0)]-i\int_{0}^{\tau}  ds \, [\widetilde{\bm{H}}_{\mathrm{I}}(s),\bm{\varrho}_{\mathsf{S}}(s)\otimes \bm{\varrho}_{\mathsf{B}}(s)].
\label{chi-1}
\end{align}
For an uncorrelated initial state ($\bm{\chi}(0)=0$) the above equation reduces to
\begin{align}
\bm{\chi}^{(1)}(\tau)=-i \textstyle{\int_{0}^{\tau}}  ds \, [\widetilde{\bm{H}}_{\mathrm{I}}(s),\bm{\varrho}_{\mathsf{S}}(s)\otimes \bm{\varrho}_{\mathsf{B}}(s)].
\label{chi-1-uncorrelated-initial}
\end{align}
Inserting this relation in the interaction picture form of the dynamical equation (\ref{exactS}) yields the following second-order equation:
\begin{align}
\dot{\bm{\varrho}}_{\mathsf{S}}(\tau)&=-i\big[\mathrm{Tr}_{\mathsf{B}}[\bm{H}_{\mathrm{I}}(\tau) \bm{\varrho}_{\mathsf{B}}(\tau)],\bm{\varrho}_{\mathsf{S}}(\tau)\big]-\mathrm{Tr}_{\mathsf{B}}\big[\bm{H}_{\mathrm{I}}(\tau), \textstyle{\int_{0}^{\tau}}  ds \, [\widetilde{\bm{H}}_{\mathrm{I}}(s),\bm{\varrho}_{\mathsf{S}}(s)\otimes \bm{\varrho}_{\mathsf{B}}(s)] \big].
\label{rhoS-1-v3}
\end{align}

Returning to the Schr\"{o}dinger picture, we have
\begin{align}
\chi^{(1)}(\tau)=-i \textstyle{\int_{0}^{\tau}}  ds \, U_{0}(\tau-s) [\widetilde{H}_{\mathrm{I}}(s),\varrho_{\mathsf{S}}(s)\otimes \varrho_{\mathsf{B}}(s)] U_{0}(s-\tau).
\label{integchi-1-schrodpic}
\end{align}
Inserting this into the system dynamical equation (\ref{exactS}) gives
\begin{align}
\dot{\varrho}_{\mathsf{S}}(\tau)=-i[\widetilde{H}_{\mathsf{S}}(\tau),\varrho_{\mathsf{S}}(\tau)]-\mathrm{Tr}_{\mathsf{B}}[H_{\mathrm{I}}, \textstyle{\int_{0}^{\tau}}  ds \, U_{0}(\tau-s) [\widetilde{H}_{\mathrm{I}}(s),\varrho_{\mathsf{S}}(s)\otimes \varrho_{\mathsf{B}}(s)] U_{0}(s-\tau)],
\label{rhoS-1}
\end{align}
which can also be written equivalently as
\begin{align}
\dot{\varrho}_{\mathsf{S}}(\tau)&=-i[\widetilde{H}_{\mathsf{S}}(\tau),\varrho_{\mathsf{S}}(\tau)]-\mathrm{Tr}_{\mathsf{B}}\big[U_{\mathsf{B}}(-\tau) H_{\mathrm{I}} U_{\mathsf{B}}(\tau), U_{\mathsf{S}}(\tau) \textstyle{\int_{0}^{\tau}}  ds \, U_{0}(-s) [\widetilde{H}_{\mathrm{I}}(s),\varrho_{\mathsf{S}}(s)\otimes \varrho_{\mathsf{B}}(s)] U_{0}(s) U_{\mathsf{S}}(-\tau) \big].\nonumber\\
&=-i[\widetilde{H}_{\mathsf{S}}(\tau),\varrho_{\mathsf{S}}(\tau)]-\mathrm{Tr}_{\mathsf{B}}\big[U_{\mathsf{B}}(-\tau) H_{\mathrm{I}}U_{\mathsf{B}}(\tau), U_{\mathsf{S}}(\tau) \textstyle{\int_{0}^{\tau}}  ds \, [\widetilde{\bm{H}}_{\mathrm{I}}(s),\bm{\varrho}_{\mathsf{S}}(s)\otimes \bm{\varrho}_{\mathsf{B}}(s)] U_{\mathsf{S}}(-\tau)\big].
\label{rhoS-1-v2}
\end{align}
We need to solve this equation together with the similar equation for $\dot{\varrho}_{\mathsf{B}}(\tau)$,
\begin{equation}
\dot{\varrho}_{\mathsf{B}}(\tau)=-i [\widetilde{H}_{\mathsf{B}}(\tau) \, , \varrho_{\mathsf{B}}(\tau)]- \mathrm{Tr}_{\mathsf{S}}[H_{\mathrm{I}}\, ,\textstyle{\int_0^{\tau}}  ds \, U_{0}(\tau-s) [\widetilde{H}_{\mathrm{I}}(s)\, , \varrho_{\mathsf{S}}(s)\otimes \varrho_{\mathsf{B}}(s)] U_{0}(s-\tau)].\label{eMLL-B}
\end{equation}
We call the dynamical equations (\ref{rhoS-1}) [or Eq. (\ref{rhoS-1-v3})] and (\ref{eMLL-B}) the ``ULL$2$'' equations since they have terms of the order of $\bm{H}^2_{\mathrm{I}}$ (although they have been obtained from the first-order weak-correlation approximation $\bm{\chi}^{(1)}$).

\begin{remark}
In the case of a correlated initial state the analysis is similar to that above with extra terms in the relations.
\label{rem:1}
\end{remark}

\begin{remark}
From the expression for $\bm{\chi}^{(1)}$, one can also obtain an approximate expression for the correlation generator $\bm{H}^{(1)}_{\chi}$ by inserting Eq. (\ref{chi-1-uncorrelated-initial}) into the interaction-picture form of Eq. (\ref{Gamma-solution}).

\label{rem:2}
\end{remark}

\subsubsection{Higher orders}

Higher order approximations to the ULL equation (referred to as ``ULL$k$'') can be obtained by using $\bm{\chi}^{(k-1)}$ (for $k\geqslant 3$) given by Eq. (\ref{chi-exact-intPic}). Deriving the explicit form of the ULL$k$ equation is straightforward, but the expressions become considerably involved. Fortunately, for many problems of interest it may suffice to consider the ULL$2$ or simply a few lowest orders, and we leave a more comprehensive analysis of the higher order ULL$k$ equations for future investigations.

\subsection{Comparing the ULL$2$ equation and corresponding weak-coupling equations}

There exist a number of powerful perturbative techniques in the literature for open systems. Here we discuss how some of these methods can be compared to or regarded as special cases of the ULL$2$ equation. In particular, we discuss the MLL equation, the second-order Nakajima-Zwanzig (NZ$2$) equation, and the second-order time-convolutionless (TCL$2$) equation. In addition, we also compare the ULL$2$ equation with the coarse-grained (CG) master equations \cite{Schaller-CG,Majenz} for the two examples discussed in the main text---and also here in Secs. \ref{example1} and \ref{sec:exIII}. A more comprehensive and detailed comparative analysis with other existing methods is beyond the scope of this paper.

\subsubsection{Comparing the ULL$2$ equation with the MLL equation: ULL$2$ as an enhanced MLL}
\label{app:enMLL}

For short times Eq. (\ref{chi-1-uncorrelated-initial}) reduces to
\begin{equation}
\bm{\chi}^{(1)}_{\mathrm{MLL}}(\tau)\approx-i \tau [\widetilde{\bm{H}}_{\mathrm{I}}(\tau),\bm{\varrho}_{\mathsf{S}}(\tau)\otimes \bm{\varrho}_{\mathsf{B}}(\tau)],
\end{equation}
which is exactly the correlation operator obtained in the derivation of the MLL dynamical equation [cf. Eq. (\ref{chi-expanded0})]. Hence although in both ULL$2$ and the MLL equations $\chi$ is of the first order with respect to $\widetilde{\bm{H}}_{\mathrm{I}}$ [see Eqs. (\ref{chi-1-uncorrelated-initial}) and (\ref{chi-expanded00})], the ULL$2$ equation can expectedly lead to an enhancement over the performance of the MLL approximation and it can also include \textit{non-Markovian} effects. In this sense, the ULL$2$ equation is an \textit{enhanced} MLL dynamical equation.

As is evident from the construction of the MLL approximation, and also recalling Sec. \ref{asymptotic-state}, the utility of the MLL approximation is not necessarily limited to the short-time dynamics. Here we discuss yet another case where the MLL approximation can be applied. If the following conditions hold: (i) the weak-correlation or weak-coupling assumptions hold (in contrast to the strong-coupling case of Sec. \ref{asymptotic-state}), and (ii) the subsystem dynamics reaches a steady state in a finite time, one can approximate $\bm{\varrho}_{\mathsf{S}}(s)\otimes \bm{\varrho}_{\mathsf{B}}(s)$ in Eq. (\ref{chi-1-uncorrelated-initial}) with the tensor product of subsystem steady states $\bm{\varrho}_{\mathsf{S}}^{\star}\otimes \bm{\varrho}_{\mathsf{B}}^{\star}$ and hence $\widetilde{\bm{H}}_{\mathrm{I}}(s)\approx \widetilde{\bm{H}}_{\mathrm{I}}|_{\varrho_{\mathsf{S}}(s)\approx \varrho^{\star}_{\mathsf{S}},\varrho_{\mathsf{B}}(s)\approx \varrho^{\star}_{\mathsf{B}}}=:\widetilde{\bm{H}}^{\star}_{\mathrm{I}}$ at most times, which yields an MLL approximation $\bm{\chi}^{(1)\,\star}_{\mathrm{MLL}}(\tau) \approx -i \tau [\widetilde{\bm{H}}_{\mathrm{I}}^{\star},\bm{\varrho}_{\mathsf{S}}^{\star}\otimes \bm{\varrho}_{\mathsf{B}}^{\star}]$. Since Eq. \eqref{chi-1-uncorrelated-initial} is accurate in the weak-correlation (or weak-coupling) regime, and typical quantum open systems often reach their steady state in finite times (see Refs. [42--45] of the main text), one can conclude that the MLL equation modified by the steady state still holds for long times for such typical systems.

\begin{remark}
Note that the same issue as in Remark \ref{rem:MLL} applies here too; \textit{a priori} knowledge of the subsystem steady state seems necessary to construct the associated MLL equation. However, as shall be discussed later in Sec. \ref{app:dep}, one may alleviate this issue by applying the MLL approximation for \textit{both} system and bath states around the initial time and solving these coupled equations simultaneously. Figure \ref{fig-ex-cons} shows that these initial-time MLL equations can give a significantly more accurate asymptotic state than other techniques (compare also with Fig. 4 of the main text where we report the result of other techniques).
\label{rem:MLL2}
\end{remark}

\subsubsection{Comparing the ULL$2$ equation with the NZ$2$ equation}

We recall that the NZ$2$ equation (see, e.g., Ref. [3] of the main text) reads as
\begin{equation}
\dot{\bm{\varrho}}_{\mathsf{S}}(\tau)=-\mathrm{Tr}_{\mathsf{B}}\big[\bm{H}_{\mathrm{I}}(\tau), \textstyle{\int_{0}^{\tau}}  ds \, [\bm{H}_{\mathrm{I}}(s),\bm{\varrho}_{\mathsf{S}}(s)\otimes \bm{\varrho}_{\mathsf{B}}(0)] \big],
\label{eq:NZ2}
\end{equation}
under the assumption of $\mathrm{Tr}_{\mathsf{B}}[\bm{H}_{\mathrm{I}}(s)\bm{\varrho}_{\mathsf{B}}(0)]=0$. We observe that the ULL$2$ equation [Eq. (\ref{rhoS-1-v3})] is to some extent similar to the NZ$2$ equation. But two key differences can also be discerned by comparing the integrands of Eqs. \eqref{rhoS-1-v3} and (\ref{eq:NZ2}): (i) the state of the bath is time-dependent in the ULL$2$ equation, whereas in the NZ$2$ equation it is a constant state, which is usually assumed to be the initial (or equilibrium) state of the bath, and (ii) in the ULL$2$ equation rather than $\bm{H}_{\mathrm{I}}(s)$ we have its effective form $\widetilde{\bm{H}}_{\mathrm{I}}(s)$. These points suggest that the performance of the ULL$2$ equation is generally different from that of the NZ$2$ equation, and that one may expect that the ULL$2$ equation either performs similarly to the NZ$2$ equation or outperforms it. To illustrate the validity of this prediction, we compare these two equations for example III of the main text (Sec. \ref{sec:exIII} of this supplementary material)---in particular, see Fig. 4 of the main text. 
There we show that the NZ$2$ equation may even give unphysical solutions whereas the ULL$2$ equation gives fairly accurate results.

\subsubsection{Comparing the ULL$2$ equation with the TCL$2$ equations}

We recall that the TCL$2$ equation for a general system in the interaction picture is (see, e.g., Refs. [3,4] of the main text)
\begin{equation}
\dot{\bm{\varrho}}_{\mathsf{S}}(\tau) =-\textstyle{\int_{0}^{\tau}} ds\, \mathrm{Tr}_{\mathsf{B}}\big[\bm{H}_{\mathrm{I}}(\tau),[\bm{H}_{\mathrm{I}}(s),\bm{\varrho}_{\mathsf{S}}(\tau) \otimes \bm{\varrho}_{\mathsf{B}}(0)]\big],
\label{eq:TCL2}
\end{equation}
under the assumption of $\mathrm{Tr}_{\mathsf{B}}[\bm{H}_{\mathrm{I}}(s)\bm{\varrho}_{\mathsf{B}}(0)]=0$. Basically this equation has been obtained by a \textit{time-local} transformation in the steps of the NZ framework---compare Eqs. (\ref{eq:TCL2}) and (\ref{eq:NZ2}). In order to make the ULL$2$ equation comparable (\ref{rhoS-1-v3}) with the TCL$2$ equation, we first need to introduce a time-local version of the ULL$2$ equation.

\subsubsection{Time-local ULL$2$ equation}

Here we argue that by applying appropriate approximations on the ULL$2$ equation (\ref{rhoS-1-v3}), one can obtain a dynamical equation which is comparable to the TCL$2$ equation. Let us start from the correlation operator $\bm{\chi}^{(1)}(\tau)$ [Eq. (\ref{chi-1-uncorrelated-initial})].  If we apply the \textit{time-locality} (or, in some sense, Markovian) approximation $\bm{\varrho}_{\mathsf{S}}(s)\otimes \bm{\varrho}_{\mathsf{B}}(s) \approx \bm{\varrho}_{\mathsf{S}}(\tau)\otimes \bm{\varrho}_{\mathsf{B}}(\tau)$ on the RHS of Eq. (\ref{chi-1-uncorrelated-initial}) and insert this correlation in the ULL$2$ equation, we obtain the following \textit{time-local ULL$2$} equation:
\begin{align}
\dot{\bm{\varrho}}_{\mathsf{S}}(\tau)=-i\big[\mathrm{Tr}_{\mathsf{B}}[\bm{H}_{\mathrm{I}}(\tau) \bm{\varrho}_{\mathsf{B}}(\tau)],\bm{\varrho}_{\mathsf{S}}(\tau)\big]-\textstyle{\int_{0}^{\tau}}  ds\,\mathrm{Tr}_{\mathsf{B}}\big[\bm{H}_{\mathrm{I}}(\tau), [\widetilde{\bm{H}}_{\mathrm{I}}(s)|_{\bm{\varrho}_{\mathsf{S}}(s)\approx \bm{\varrho}_{\mathsf{S}}(\tau),\bm{\varrho}_{\mathsf{B}}(s) \approx \bm{\varrho}_{\mathsf{B}}(\tau)},\bm{\varrho}_{\mathsf{S}}(\tau)\otimes \bm{\varrho}_{\mathsf{B}}(\tau)] \big].
\end{align}
If we further assume (as approximation) that the state of the bath does not change appreciably in time, ${\bm{\varrho}}_{\mathsf{B}}(\tau)\approx {\bm{\varrho}}_{\mathsf{B}}(0)$, and also that $\mathrm{Tr}_{\mathsf{B}}[\bm{H}_{\mathrm{I}}(\tau) \bm{\varrho}_{\mathsf{B}}(0)]=0$ (similar to the derivation of the TCL equation \cite{book:Rivas-Huelga}), the above time-local dynamical equation reduces to
\begin{align}
\dot{\bm{\varrho}}_{\mathsf{S}}(\tau)=-\textstyle{\int_{0}^{\tau}}  ds\,\mathrm{Tr}_{\mathsf{B}}\big[\bm{H}_{\mathrm{I}}(\tau), [\widetilde{\bm{H}}_{\mathrm{I}}(s)|_{\bm{\varrho}_{\mathsf{S}}(s)\approx \bm{\varrho}_{\mathsf{S}}(\tau), \bm{\varrho}_{\mathsf{B}}(s) \approx \bm{\varrho}_{\mathsf{B}}(0)},\bm{\varrho}_{\mathsf{S}}(\tau)\otimes \bm{\varrho}_{\mathsf{B}}(0)] \big].
\label{eq:TL-ULL2}
\end{align}
This equation is to some extent similar to the TCL$2$ equation (\ref{eq:TCL2})---but not fully. Due to the nature of the approximations made to derive this reduced time-local ULL$2$ equation, it is expected that the ULL$2$ equation in general may outperform the TCL$2$ equation, except under particular circumstances where the TCL$2$ equation may become exact. 

\begin{remark}
Let us start from the correlation operator $\bm{\chi}^{(1)}(\tau)$ [Eq. (\ref{chi-1-uncorrelated-initial})] and compare it with the definition of $\bm{\chi}(\tau)=-i [\bm{H}_{\chi}(\tau),\bm{\varrho}_{\mathsf{S}}(\tau)\otimes \bm{\varrho}_{\mathsf{B}}(\tau)]$ [Eq. (3) of the main text in the interaction picture]. If we apply the \textit{time-locality} approximation by replacing $\bm{\varrho}_{\mathsf{S}}(s)\otimes \bm{\varrho}_{\mathsf{B}}(s)$ in the commutator of the integrand of Eq. (\ref{chi-1-uncorrelated-initial}) with its time-local form $\bm{\varrho}_{\mathsf{S}}(\tau)\otimes \bm{\varrho}_{\mathsf{B}}(\tau)$, we can read the correlation generator as
$\bm{H}_{\chi}^{(1)\,\mathrm{TL}}(\tau)= \textstyle{\int_{0}^{\tau}}  ds \, \widetilde{\bm{H}}_{\mathrm{I}}(s)$.
If we also apply the time-locality approximation in $\widetilde{\bm{H}}_{\mathrm{I}}(s)$, the integrand in the above relation changes to $\widetilde{\bm{H}}_{\mathrm{I}}(s)|_{\bm{\varrho}_{\mathsf{S}}(s)\approx \bm{\varrho}_{\mathsf{S}}(\tau),\bm{\varrho}_{\mathsf{B}}(s) \approx \bm{\varrho}_{\mathsf{B}}(\tau)}$ [see Eq. (\ref{HI-i}) in the interaction picture]. Returning to the Schr\"{o}dinger picture, we obtain $H_{\chi}^{(1)\,\mathrm{TL}}(\tau)= \textstyle{\int_{0}^{\tau}}  ds \, U_{0}(\tau-s) \widetilde{H}_{\mathrm{I}}(s)|_{\varrho_{\mathsf{S}}(s)\approx \varrho_{\mathsf{S}}(\tau),\varrho_{\mathsf{B}}(s) \approx \varrho_{\mathsf{B}}(\tau)} U_{0}(s-\tau)$. Consequently, if we insert the above relation in the dynamical equation of the system [Eq. (\ref{method3})] we obtain
\begin{align}
\dot{\varrho}_{\mathsf{S}}(\tau)=-i \big[\widetilde{H}_\mathsf{S}(\tau),\varrho_{\mathsf{S}}(\tau)]- \textstyle{\int_{0}^{\tau}}  ds \,\mathrm{Tr}_{\mathsf{B}}[H_{\mathrm{I}},[U_{0}(\tau-s) \widetilde{H}_{\mathrm{I}}(s)|_{\bm{\varrho}_{\mathsf{S}}(s)\approx \varrho_{\mathsf{S}}(\tau),\varrho_{\mathsf{B}}(s) \approx \varrho_{\mathsf{B}}(\tau)} U_{0}(s-\tau),\varrho_{\mathsf{S}}(\tau)\otimes \varrho_{\mathsf{B}}(\tau)]\big].
\label{IntegChi-TL}
\end{align}

\label{rem:3}
\end{remark}

\subsection{The dependence of the dynamical equation of the system on the instantaneous state of the bath}
\label{app:dep}

It is clear from Eq. \eqref{chi-exact-intPic} that the correlation operator depends on the instantaneous state of the bath. Hence the correlation generator $H_{\chi}$ and $\dot{\varrho}_{\mathsf{S}}(\tau)$ also explicitly depend on $\varrho_{\mathsf{B}}(\tau)$, and the ULL equation is more general but also more complicated than the NZ and TCL equations. Since our weak-correlation methodology hinges on obtaining or approximating the correlation operator (which is a joint property of the system and the bath), once we have an expression for $\chi$, we can employ it to also approximate the state of the bath more accurately. We can then use this approximate state of the bath to obtain a yet more accurate approximation for the dynamical equation of the system. Although this increases the number of the equations to be solved, it enables a more accurate estimation of the system. We will explicitly demonstrate this improvement in the next sections---Secs. \ref{example1} and \ref{sec:exIII}---where we elaborate on examples II and III of the main text.

We also show that it may not always be even approximately adequate to assume the bath to remain unchanged ($\varrho_{\mathsf{B}}(\tau)\approx\varrho_{\mathsf{B}}(0)$) during the dynamics of the system. In particular, in Fig. \ref{fig-ex-cons}, we observe that this assumption leads to unphysical, negative populations in our example III, whereas using our MLL equation for the bath state too provides a considerable improvement over the constant-bath assumption. The situation can be improved further by using the ULL$2$ equations for both the system and the bath---Fig. \ref{fig:eMLL-cutoff10} (and also Fig. 4 of the main text). We remark that other approximate methods may also provide some improvements over the constant-bath assumption. 

\begin{figure}[tp]
\includegraphics[width=0.4\linewidth]{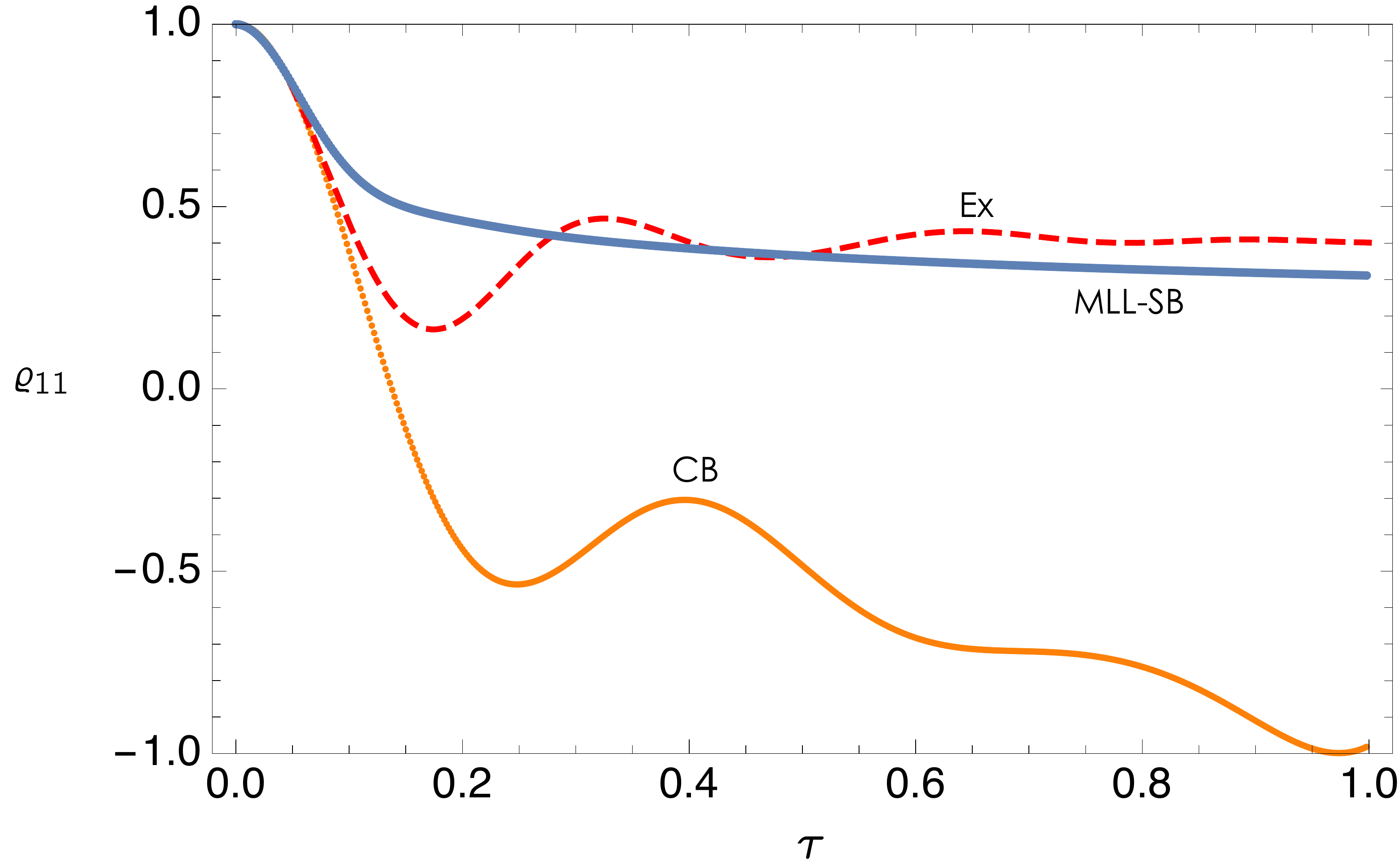}
\caption{Population of the first excited state of the system in example III of the main text. The dashed-red curve shows the exact dynamics (``Ex''), the solid-blue curve is obtained with a constant state of the bath,  $\varrho_{\mathsf{B}}(0)$ (``C$\mathsf{B}$'') and explores unphysical, negative values for the population. The solid-orange curve shows that the MLL dynamical equations for both the system and the bath (``MLL-$\mathsf{SB}$'') provide a clear improvement over the equation assuming a constant bath.}
\label{fig-ex-cons}
\end{figure}

\section{Details of example II: Two-level system in a bosonic bath}
\label{example1}

Here we provide the details related to example II introduced in Sec. VI A of the main text.

\subsection{Redfield equation}

Following the standard steps for derivation of the Redfield equation (e.g., see Ref. [3] of the main text) 
we obtain for the dynamics of a two-level system (atom)
\begin{align}
\dot{\varrho}_{\mathsf{S}}=&-i[H_{\mathsf{S}},\varrho_{\mathsf{S}}]+\frac{1}{2} \Big( [S(\beta,-\omega_0)+S(\beta,\omega_0)](\sigma_{+} \varrho_{\mathsf{S}}\sigma_{+} + \sigma_{-} \varrho_{\mathsf{S}}\sigma_{-})+2S(\beta,-\omega_0)\sigma_{+}\varrho_{\mathsf{S}}\sigma_{-}+S(\beta,-\omega_0)\lbrace\varrho_{\mathsf{S}},\sigma_{-}\sigma_{+}\rbrace
\nonumber\\
&+2S(\beta,\omega_0)\sigma_{-}\varrho_{\mathsf{S}}\sigma_{+} +S(\beta,\omega_0)\lbrace\varrho_{\mathsf{S}},\sigma_{+}\sigma_{-}\rbrace \Big),
\label{e20}
\end{align}
where $S(\beta,\omega)=2[n(\beta,\omega)+1] J(\omega)$. This yields the following dynamical equations for the populations:
\begin{align}
\dot{\varrho}_{\mathrm{gg}}(\tau)&=S(\beta, \omega_0)-[S(\beta,-\omega_0)+S(\beta,\omega_0)] \varrho_{\mathrm{gg}}(\tau),\nonumber\\
\dot{\varrho}_{\mathrm{ee}}(\tau)&=S(\beta,-\omega_0)-[S(\beta,-\omega_0)+S(\beta,\omega_0)] \varrho_{\mathrm{ee}}(\tau).
\end{align}
The solutions of these equations are given by
\begin{align}
\varrho_{\mathrm{gg}}(\tau)&=\frac{S(\beta,\omega_0)}{S(\beta,-\omega_0)+S(\beta,\omega_0)}\big[1-e^{-[S(\beta,-\omega_0)+S(\beta,\omega_0)]\tau}\big]+ \varrho_{\mathrm{gg}}(0)e^{-[S(\beta,-\omega_0)+S(\beta,\omega_0)]\tau},\\
\varrho_{\mathrm{ee}}(\tau)&=\frac{S(\beta,-\omega_0)}{S(\beta,-\omega_0)+S(\beta,\omega_0)}\big[1-e^{-[S(\beta,-\omega_0)+S(\beta,\omega_0)]\tau}\big]+\varrho_{\mathrm{ee}}(0)e^{-[S(\beta,-\omega_0)+S(\beta,\omega_0)]\tau}.
\end{align}
To compare the Redfield solution with the exact solution, we rewrite the Redfield equation in the $\omega_0=0$ limit, which becomes equal to the Markovian Lindblad equation. Using the relation $\lim_{\omega_0 \rightarrow 0} S(\beta,\omega_0)=2\eta / \beta$, Eq. \eqref{e20} reduces to
\begin{equation}
\dot{\varrho}_{\mathsf{S}}(\tau)=-i[H_{\mathsf{S}},\varrho_{\mathsf{S}}(\tau)]+2(\eta / \beta) \big(\sigma_{x} \varrho_{\mathsf{S}}(\tau)\sigma_{x}-\varrho_{\mathsf{S}}(\tau)\big).
\end{equation}
Thus, we can obtain the differential equations for the diagonal elements of the density matrix,
\begin{align}
\dot{\varrho}_{ll}(\tau)=&2(\eta/\beta)\big(1-2\varrho_{ll}(\tau)\big),
\end{align} 	
where $l\in\{\mathrm{g},\mathrm{e}\}$. The solutions of these Redfield equations are given by
\begin{align}
\varrho_{\mathrm{gg}}(\tau)=(1/2)\big(1-e^{-4 (\eta/ \beta )\tau} \big)+ \varrho_{\mathrm{gg}}(0)e^{-4 (\eta/ \beta )\tau},\nonumber\\
\varrho_{\mathrm{ee}}(\tau)=(1/2)\big(1-e^{-4 (\eta/ \beta )\tau} \big)+ \varrho_{\mathrm{ee}}(0)e^{-4 (\eta/ \beta )\tau}.
\end{align}

\subsection{Time-local ULL$2$ equation}

If in the time-local ULL$2$ equation (\ref{eq:TL-ULL2}) we calculate the averages in $\widetilde{ \bm{H}}_{\mathrm{I}}(\tau)$ with respect to the initial system-bath state $\bm{\varrho}_{\mathsf{SB}}(0)=|\mathrm{e}\rangle\langle \mathrm{e}|\otimes \varrho^{\beta}_{\mathsf{B}}$ and put $\langle \circ\rangle_{\mathsf{S}(\mathsf{B})}\approx \langle \circ\rangle_{\mathsf{S}_{0}(\mathsf{B}_{0})}$, we arrive at $\widetilde{ \bm{H}}_{\mathrm{I}}(\tau)\approx { \bm{H}}_{\mathrm{I}}(\tau)$, $\widetilde{ \bm{H}}_{\mathsf{S}}(\tau) \approx { \bm{H}}_{\mathsf{S}}(\tau)$, which reduces the time-local ULL$2$ equation to
\begin{equation}
\dot{\bm{\varrho}}_{\mathsf{S}}(\tau) =-\textstyle{\int_{0}^{\tau}} ds\, \mathrm{Tr}_{\mathsf{B}}\big[{\bm{H}}_{\mathrm{I}}(\tau),[{ \bm{H}}_{\mathrm{I}}(s),{\bm{\varrho}}_{\mathsf{S}}(\tau) \otimes {\bm{\varrho}}_{\mathsf{B}}(0)]\big].
\label{IMML2}
\end{equation}
This is identical to the TCL$2$ equation (\ref{eq:TCL2})---which is discussed in detail below.

\subsection{TCL$2$ equation}

Following the derivation of the TCL$2$ equation
(e.g., see Ref. [4] of the main text) we obtain
\begin{align}
\dot{\bm{\varrho}}_{\mathsf{S}}(\tau)=-\textstyle{\int_{0}^{\tau}} du\,   \big[ &\big(\bm{\sigma}_{x}(\tau) \bm{\sigma}_{x}(\tau-u) \bm{\varrho}_{\mathsf{S}}(\tau) - \bm{\sigma}_{x}(\tau-u) \bm{\varrho}_{\mathsf{S}}(\tau) \bm{\sigma}_{x}(\tau) \big) G(\tau,\tau-u)\nonumber\\
&+\big(\bm{\varrho}_{\mathsf{S}}(\tau) \bm{\sigma}_{x}(\tau-u) \bm{\sigma}_{x}(\tau)  - \bm{\sigma}_{x}(\tau) \bm{\varrho}_{\mathsf{S}}(\tau) \bm{\sigma}_{x}(\tau-u) \big) G(\tau-u,\tau) \big],
\end{align}
where boldface denotes the interaction picture, e.g., $\bm{\sigma}_{x}(s)=U^{\dag}_{0}(s)\sigma_{x} U_{0}(s)$, $G(\tau,\tau-u)=\ave{\bm{\mathcal{O}}_{\mathsf{B}}(\tau)\bm{\mathcal{O}}_{\mathsf{B}}(\tau-u)}_{{\mathsf{B}}_0}$, and $G(\tau-u,\tau)=\ave{\bm{\mathcal{O}}_{\mathsf{B}}(\tau-u)\bm{\mathcal{O}}_{\mathsf{B}}(\tau)}_{{\mathsf{B}}_0}$. Since $\omega_0=0$, the interaction picture and Schr\"{o}dinger picture are equivalent for the system operators; hence the above equation can be rewritten as
\begin{align}
\dot{\varrho}_{\mathsf{S}}(\tau)=-\textstyle{\int_{0}^{\tau}} du\, \big[ & \big(\varrho_{\mathsf{S}}(\tau) - \sigma_{x} \varrho_{\mathsf{S}}(\tau)\sigma_{x} \big) C(\tau,\tau-u) + \big(\varrho_{\mathsf{S}}(\tau) -\sigma_{x} \varrho_{\mathsf{S}}(\tau)\sigma_{x} \big) G(\tau-u,\tau) \big].
\end{align}
Noting that the bath initial state is a thermal state and using a spectral density $J(\omega)$, we have $G(\tau,\tau-u)=\int_{0}^{\infty}d\omega\,  J(\omega)\, [\big(n(\beta,\omega)+1\big)e^{-i\omega \tau}+n(\beta,\omega) e^{i\omega \tau}]$ and $G(\tau-u,\tau)=\int_{0}^{\infty} d\omega\, J(\omega) [\big(n(\beta,\omega)+1\big)e^{i\omega \tau}+n(\beta,\omega) e^{-i\omega \tau}]$, where $n(\beta,\omega)=1/(e^{\beta \omega}-1)$. Using these quantities the TCL$2$ dynamical equation is given by
\begin{align}
\dot{\varrho}_{\mathsf{S}}(\tau)=\gamma(\tau) \big(\sigma_{x} \varrho_{\mathsf{S}}(\tau)\sigma_{x}-\varrho_{\mathsf{S}}(\tau)\big),
\end{align}
where $\gamma(\tau)=2\textstyle{\int_{0}^{\tau}}d\tau\, \int_{0}^{\infty} d\omega\, J(\omega) \cos(\omega \tau)[2 n(\beta,\omega)+1] $.

\subsection{Short-time dynamics}

Since $\ave{H_{\mathrm{I}}}_{\mathsf{B}_{0}}=0$ we have $\widetilde{H}_{\mathsf{S}}=H_{\mathsf{S}}$. Thus, the initial state of the atom $\varrho_{\mathsf{S}}(0)=|\mathrm{e}\rangle\langle \mathrm{e}|$ commutes with $\widetilde{H}_{\mathsf{S}}=H_{\mathsf{S}}=\omega_0 \sigma_{+} \sigma_{-}$. Hence from Eq. (20) of the main text, the short-time dynamics of the atom is obtained as
\begin{align}
\varrho_{\mathsf{S}}(\tau)=\varrho_{\mathsf{S}}(0)+\tau^{2}\, \mathrm{Cov}_{\mathsf{B}_0}(\mathpzc{O}_{\mathsf{B}},\mathpzc{O}_{\mathsf{B}})\big(\sigma_{x} \varrho_{\mathsf{S}}(0)\sigma_{x}- \varrho_{\mathsf{S}}(0)\big)+O(\tau^3),
\end{align}
from which it is seen that $\varrho_{\mathrm{ee}}(\tau)=\varrho_{\mathrm{ee}}(0)+\tau^{2}\, \mathrm{Cov}_{\mathsf{B}_0}(\mathpzc{O}_{\mathsf{B}},\mathpzc{O}_{\mathsf{B}})\big(\varrho_{\mathrm{gg}}(0) -\varrho_{\mathrm{ee}}(0)\big)+O(\tau^3)$. Since the atom is assumed to be initially in the excited state $\varrho_{\mathrm{ee}}(0)=1$, thus
\begin{align}
\varrho_{\mathrm{ee}}(\tau)=1-\tau^{2}\, \mathrm{Cov}_{\mathsf{B}_0}(\mathpzc{O}_{\mathsf{B}},\mathpzc{O}_{\mathsf{B}})+O(\tau^3),
\label{exp2-short-time}
\end{align}
which is in accordance with the short-time expansion of Eq. (21) of the main text. To compare Eq. \eqref{exp2-short-time} with the short-time expansion of the exact solution, we note that based on Eq. (9) of Ref. [22] of the main text 
and for the case of a single-mode bath $\langle \pm|\varrho_{\mathsf{S}}(\tau)|\mp\rangle = e^{-4 f(\tau)}\langle \pm|\varrho_{\mathsf{S}}(0)|\mp\rangle$. Using the identity $\varrho_{\mathrm{ee}}(\tau)=\frac{1}{2}\big(1-\langle +|\varrho_{\mathsf{S}}(\tau)|-\rangle-\langle -|\varrho_{\mathsf{S}}(\tau)|+\rangle \big)$, which is obtained with a simple basis transformation using the relations $|+\rangle=(|\mathrm{g}\rangle+|\mathrm{e}\rangle)/\sqrt{2}$ and $|-\rangle=(|\mathrm{g}\rangle-|\mathrm{e}\rangle)/\sqrt{2}$, we conclude that $\varrho_{\mathrm{ee}}(\tau)=\frac{1}{2}(1+e^{-4 f(\tau)})$, in which $f(\tau)=\tau^2 \ave{\mathpzc{O}_{\mathsf{B}}^2}_{\mathsf{B}_0}/2$. Thus, the short-time behavior from the exact solution becomes $1-\tau^2\ave{\mathpzc{O}_{\mathsf{B}}^2}_{\mathsf{B}_0}+O(\tau^3)$, which---considering $\ave{\mathpzc{O}_{\mathsf{B}}}_{\mathsf{B}_0}=0$---coincides with Eq. \eqref{exp2-short-time}.

\subsection{Comparison with the CG equation}

In Fig. \ref{fig-ex2-RF-CG-Ex-TCL2-MLL}, we compare our data with the particular CG method delineated in Ref. \cite{Schaller-CG} (Sec. III C 3 therein) with the MLL, ULL$2$, and other methods. We observe that here the MLL, ULL$2$, and TCL$2$ solutions outperform the CG and the Redfield solutions.

\begin{figure}[tp]
\includegraphics[width=0.4\linewidth]{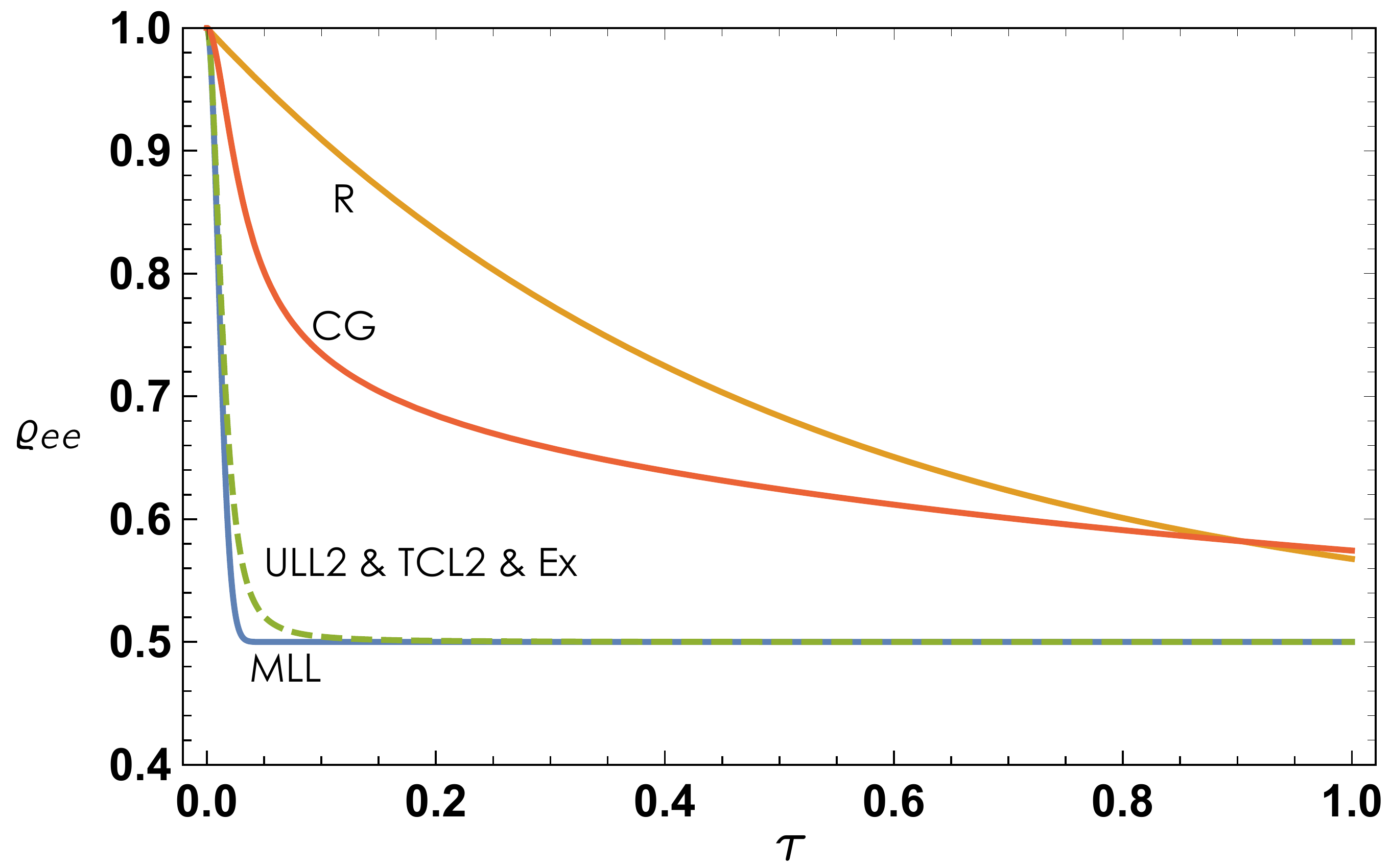}
\caption{Comparison of different methods for example II. Orange is the solution of the Redfield (R) equation, red is the solution of the coarse-grained (CG) dynamical equation of Ref. \cite{Schaller-CG}, dashed green is the exact (Ex) solution (which is equivalent to the TCL$2$ as well as the time-local ULL$2$ equations), and blue is the solution of the MLL equation. Here $\eta=0.5$, $\beta=1$, and $\omega_{c}=100$ in natural units.}
\label{fig-ex2-RF-CG-Ex-TCL2-MLL}
\end{figure}
\section{Details of example III: Harmonic oscillator within a bath of oscillators}
\label{sec:exIII}

Here we provide the details related to example III introduced in Sec. VI B of the main text.

\subsection{TCL$2$ equation}

We recall the TCL$2$ equation (\ref{eq:TCL2}).
The interaction Hamiltonian for a harmonic oscillator interacting with a bath of similar oscillators is given by
\begin{equation}
\bm{H}_{\mathrm{I}}(\tau)= \textstyle{\sum_{k=1}^M} g_{k}\big[\hat{\bm{a}}^{\dagger} (\tau)\,\hat{\bm{b}}_{k}(\tau) + \hat{\bm{a}}(\tau) \,\hat{\bm{b}}_{k}^{\dagger} (\tau)\big],
\end{equation}
where $ \hat{\bm{a}}(\tau)=\hat{a}e^{-i\omega_0\tau}$ and $\hat{\bm{b}}_{k}(\tau)=\hat{b}_{k}e^{-i\omega_k \tau}$ (see Ref. [39] of the main text for derivation). Now we can evaluate each term in the TCL$2$ equation. We have
\begin{align}
\mathrm{Tr}_{\mathsf{B}}[\bm{H}_{\mathrm{I}}(\tau) \bm{H}_{\mathrm{I}}(s) \bm{\varrho}_{\mathsf{S}}(\tau) \otimes \bm{\varrho}_{\mathsf{B}}(0)]=&  \textstyle{\sum_{k=1}^M} \textstyle{\sum_{k'=1}^M} g_{k} g_{k'}\Big( \hat{\bm{a}}^{\dag}(\tau) \hat{\bm{a}}^{\dag}(s) \bm{\varrho}_{\mathsf{S}}(\tau) \mathrm{Tr}_{\mathsf{B}}[\hat{\bm{b}}_{k}(\tau) \hat{\bm{b}}_{k'}(s) \bm{\varrho}_{\mathsf{B}}(0)]\nonumber\\
&+ \hat{\bm{a}}(\tau) \hat{\bm{a}}(s) \bm{\varrho}_{\mathsf{S}}(\tau) \mathrm{Tr}_{\mathsf{B}}[\hat{\bm{b}}_{k}^{\dag}(\tau) \hat{\bm{b}}_{k'}^{\dag}(s) \bm{\varrho}_{\mathsf{B}}(0)] \nonumber \\
&+ \hat{\bm{a}}^{\dag}(\tau) \hat{\bm{a}}(s) \bm{\varrho}_{\mathsf{S}}(\tau) \mathrm{Tr}_{\mathsf{B}}[\hat{\bm{b}}_{k}(\tau) \hat{\bm{b}}_{k'}^{\dag}(s) \bm{\varrho}_{\mathsf{B}}(0)] \nonumber\\
&+ \hat{\bm{a}}(\tau) \hat{\bm{a}}^{\dag}(s) \bm{\varrho}_{\mathsf{S}}(\tau) \mathrm{Tr}_{\mathsf{B}}[\hat{\bm{b}}_{k}^{\dag}(\tau) \hat{\bm{b}}_{k'}(s) \bm{\varrho}_{\mathsf{B}}(0)]\Big) \nonumber\\
 =& \textstyle{\sum_{k=1}^M}\textstyle{\sum_{k'=1}^M} g_{k} g_{k'}\Big( \hat{a}^{\dag} \hat{a}^{\dag} e^{i\omega_0(\tau+s)} \bm{\varrho}_{\mathsf{S}}(\tau) \mathrm{Tr}_{\mathsf{B}}[\hat{b}_k \hat{b}_{k'} \bm{\varrho}_{\mathsf{B}}(0)] e^{-i(\omega_{k}\tau+\omega_{k'}s)} \nonumber\\
 &+ \hat{a} \hat{a} e^{-i\omega_0(\tau+s)} \bm{\varrho}_{\mathsf{S}}(\tau) \mathrm{Tr}_{\mathsf{B}}[\hat{b}_{k}^{\dag} \hat{b}_{k'}^{\dag} \bm{\varrho}_{\mathsf{B}}(0)] e^{i(\omega_k\tau+\omega_{k'}s)} \nonumber\\
 &+ \hat{a}^{\dag} \hat{a} e^{i\omega_0(\tau-s)} \bm{\varrho}_{\mathsf{S}}(\tau) \mathrm{Tr}_{\mathsf{B}}[\hat{b}_{k} \hat{b}_{k'}^{\dag} \bm{\varrho}_{\mathsf{B}}(0)] e^{-i(\omega_{k}\tau-\omega_{k'}s)} \nonumber\\
 & + \hat{a} \hat{a}^{\dag} e^{-i\omega_{0}(\tau-s)} \bm{\varrho}_{\mathsf{S}}(\tau) \mathrm{Tr}_{\mathsf{B}}[\hat{b}_{k}^{\dag} \hat{b}_{k'}\bm{\varrho}_{\mathsf{B}}(0)]e^{i\omega_{k}(\tau-s)}\Big).
\label{eq1}
\end{align}
We assume that the bath is at zero temperature and simplify Eq. (\ref{eq1}) further (using the relations $\mathrm{Tr}_{\mathsf{B}}[\hat{b}_k \hat{b}_{k'}^{\dag} \bm{\varrho}_{\mathsf{B}}(0)]=\delta_{kk'}$, $\mathrm{Tr}_{\mathsf{B}}[ \hat{b}_{k}^{\dag} \hat{b}_{k'} \bm{\varrho}_{\mathsf{B}}(0)]=0$, $\mathrm{Tr}_{\mathsf{B}}[\hat{b}_k \hat{b}_{k'} \bm{\varrho}_{\mathsf{B}}(0)]=0$, and $\mathrm{Tr}_{\mathsf{B}}[\hat{b}_{k}^{\dag} \hat{b}_{k'}^{\dag}\bm{\varrho}_{\mathsf{B}}(0)]=0$),
\begin{equation}
\begin{split}
\mathrm{Tr}_{\mathsf{B}}[\bm{H}_{\mathrm{I}}(\tau) \bm{H}_{\mathrm{I}}(s) \bm{\varrho}_{\mathsf{S}}(\tau) \otimes \bm{\varrho}_{\mathsf{B}}(0)] = \textstyle{\sum_{k=1}^M} g_{k}^2\, \hat{a}^{\dag} \hat{a} \bm{\varrho}_{\mathsf{S}}(\tau) e^{i(\omega_0-\omega_k)(\tau-s)}.
\end{split}
\label{eq2}
\end{equation}
Thus, the TCL$2$ equation reads as
\begin{equation}
\begin{split}
\dot{\bm{\varrho}}_{\mathsf{S}}(\tau)=-\textstyle{\int_{0}^{\tau}} ds\, \textstyle{\sum_{k=1}^M} g_{k}^2\Big(& \hat{a}^{\dag} \hat{a} \bm{\varrho}_{\mathsf{S}}(\tau) e^{i(\omega_0-\omega_k)(\tau-s)} - \hat{a} \bm{\varrho}_{\mathsf{S}}(\tau) \hat{a}^{\dag}  e^{-i(\omega_0-\omega_k)(\tau-s)}- \hat{a} \bm{\varrho}_{\mathsf{S}}(\tau) \hat{a}^{\dag} e^{i(\omega_0-\omega_k)(\tau-s)}\\&+ \bm{\varrho}_{\mathsf{S}}(\tau) \hat{a}^{\dag} \hat{a} e^{-i(\omega_0-\omega_k)(\tau-s)}\Big).
\end{split}
\label{eq3}
\end{equation}
If we set $x =\tau-s$ we obtain
\begin{equation}
\begin{split}
\dot{\bm{\varrho}}_{\mathsf{S}}(\tau)=- \textstyle{\sum_k} g_{k}^2 \textstyle{\int_{0}^{\tau}} dx\, \Big(&\hat{a}^{\dag} \hat{a} \bm{\varrho}_{\mathsf{S}}(\tau) e^{i\Delta_{k} x} - \hat{a} \bm{\varrho}_{\mathsf{S}}(\tau) \hat{a}^{\dag}  e^{-i\Delta_{k} x}- \hat{a} \bm{\varrho}_{\mathsf{S}}(\tau) \hat{a}^{\dag}  e^{i\Delta_{k} x}+ \bm{\varrho}_{\mathsf{S}}(\tau) \hat{a}^{\dag} \hat{a} e^{-i\Delta_{k} x}\Big),
\end{split}
\label{eq3}
\end{equation}
where $\Delta_{k}=\omega_0-\omega_{k}$. Using the integral relation
\begin{equation}
{\textstyle{\int_{0}^{\tau}}} dx\, e^{\pm i\Delta_{k} x}=\pm i \frac{2\sin^2(\Delta_{k} \tau/2)}{\Delta_{k} }+\frac{\sin(\Delta_{k} \tau)}{\Delta_{k}}
\end{equation}
yields
\begin{equation}
\dot{\bm{\varrho}}_{\mathsf{S}}(\tau) =-i\kappa(\tau)[\hat{a}^{\dag} \hat{a},\bm{\varrho}_{\mathsf{S}}(\tau) ]+\gamma(\tau)\big(2 \hat{a} \bm{\varrho}_{\mathsf{S}}(\tau) \hat{a}^{\dag}-\{\hat{a}^{\dag} \hat{a},\bm{\varrho}_{\mathsf{S}}(\tau)\}\big),
\label{eq4}
\end{equation}
where $\gamma(\tau)=\sum_{k}g_{k}^2\sin(\Delta_{k} \tau )/\Delta_{k}$ and $\kappa(\tau) =\sum_{k} 2 g_{k}^2 \sin^2(\Delta_{k} \tau /2)/\Delta_{k}$. After transforming back to the Schr\"{o}dinger picture, we obtain the following TCL$2$ equation:
\begin{equation}
\dot{\varrho}_{\mathsf{S}}(\tau)=-i\big(\omega_0+\kappa(\tau)\big) [\hat{a}^{\dag} \hat{a},\varrho_{\mathsf{S}}(\tau) ]+\gamma(\tau)\big(2 \hat{a} \varrho_{\mathsf{S}}(\tau) \hat{a}^{\dag}-\{\hat{a}^{\dag} \hat{a},\varrho_{\mathsf{S}}(\tau)\}\big).
\label{eq5}
\end{equation}

As a remark, note that $\gamma(\tau)\approx G \tau$ for small $\tau$'s and $\lim_{\tau \rightarrow 0}\kappa(\tau) =0$, hence in this limit this TCL$2$ equation reduces to the MLL equation (23) of the main text.

\begin{figure*}[tp]
\includegraphics[width=8cm]{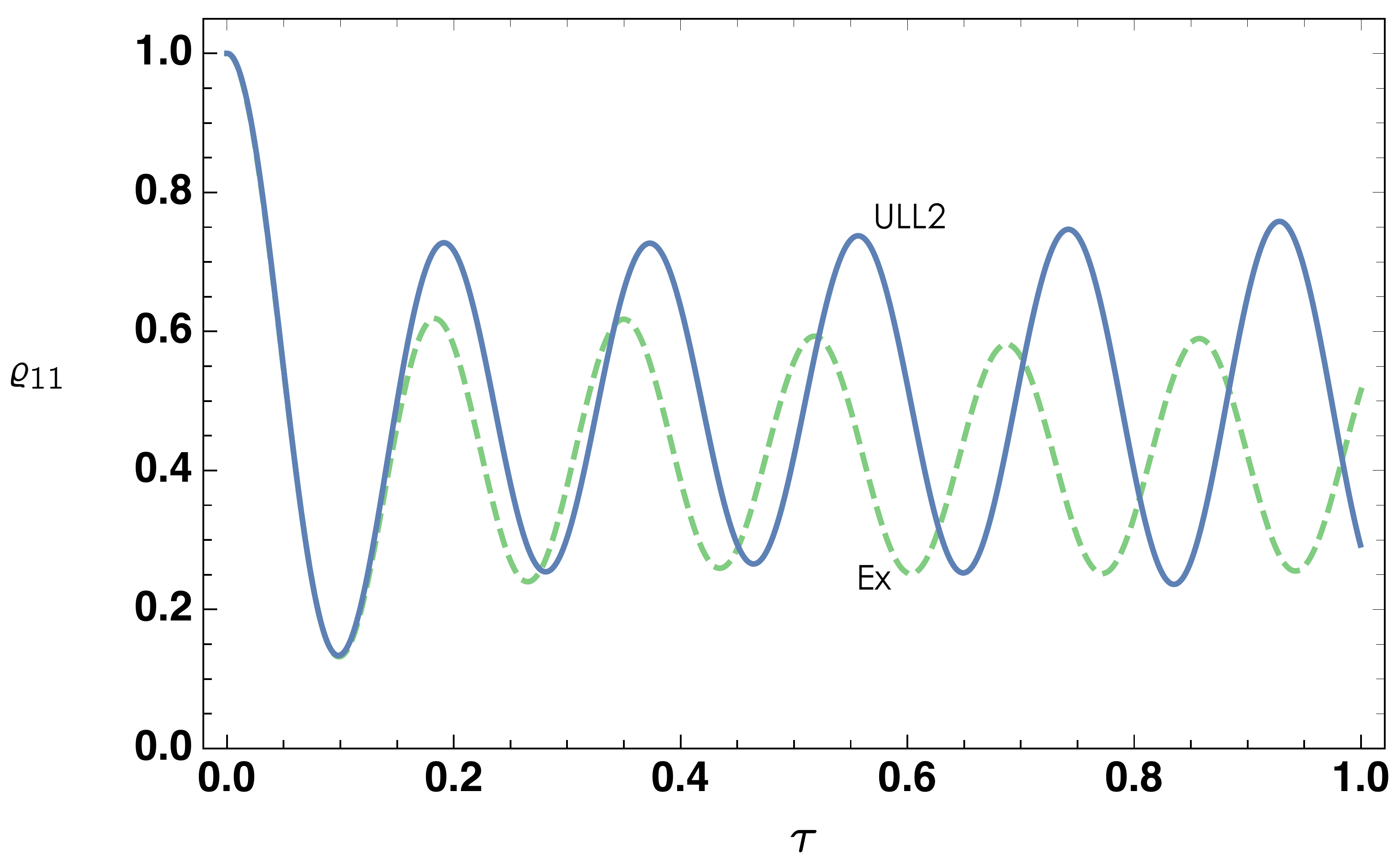}
\caption{Damped harmonic oscillator. Comparing the ULL$2$ and the exact dynamics when $\omega_c=10$.}
\label{fig:eMLL-cutoff10}
\end{figure*}
\subsection{CG equation}

Here we use the CG method as per Ref. \cite{Majenz} and obtain the following CG master equation for a harmonic oscillator within a bath of oscillators:
\begin{equation}
 \dot{\varrho}_{\mathsf S}(\tau) =-K_{\omega\omega}(\Delta t) [\hat{a}^\dagger \hat{a},\varrho_{\mathsf S}(\tau) ]+\gamma_{\omega\omega}(\Delta t)\big(2 \hat{a} \varrho_{\mathsf S}(\tau)\hat{a}^\dagger-\{\hat{a}^\dagger \hat{a},\varrho_{\mathsf S}(\tau)\}\big),
 \label{CG-SME1}
\end{equation}
where
\begin{align}
\gamma_{\omega\omega}(\Delta t)=& \frac{e^{-\frac{\omega}{\omega_c}}}{\pi \Delta t} \Big[(1-\frac{\omega}{\omega_c}-i\omega \Delta t){\rm Ei}(\frac{\omega}{\omega_c}+i\omega \Delta t)+(1-\frac{\omega}{\omega_c}+i\omega \Delta t){\rm Ei}(\frac{\omega}{\omega_c}-i\omega \Delta t)+2(\frac{\omega}{\omega_c}-1){\rm Ei}(\frac{\omega}{\omega_c})\Big] \nonumber\\
&+\frac{2[1-\cos(\omega \Delta t)]}{\pi},\\
K_{\omega\omega}(\Delta t)=&\frac{\Delta t}{\pi}\Big[\omega_c- \omega e^{-\frac{\omega}{\omega_c}} {\rm Ei}(\frac{\omega}{\omega_c})\Big]+\frac{e^{-\frac{\omega}{\omega_c}}}{2i\pi } \Big[(1-2\frac{\omega}{\omega_c}+2i\omega \Delta t){\rm Ei}(\frac{\omega}{\omega_c}+i\omega \Delta t)-(1-2\frac{\omega}{\omega_c}-2i\omega \Delta t){\rm Ei}(\frac{\omega}{\omega_c}-i\omega \Delta t)\Big] \nonumber\\
&+\frac{\sin(\omega \Delta t)}{\pi},\\
\mathrm{Ei}(z)=&-\textstyle{\int_{-z}^{\infty}dx}\, e^{-x}\mathbbmss{P}[1/x].
\end{align}
We have used the Ohmic spectral density  $J(\omega)=(\omega/\pi)e^{-\omega/\omega_c}$, where $\omega_c$ is the cutoff frequency. The optimal value of the free parameter $\Delta t$ can be fixed by comparing with the exact solution.

\subsection{ULL$2$ equation}

We use the coupled ULL$2$ dynamical equations (\ref{rhoS-1}) and (\ref{eMLL-B}).
The results of the numerical simulations of the the system dynamical equation have been shown in Fig. 4 of the main text. In Fig. \ref{fig:eMLL-cutoff10} of this note we have depicted the population of the first excited state for example III for both ULL$2$ and exact solutions, when $\omega_c=10$ and the rest of the parameters have values equal to those used in Fig. 4 of the main text. Again, the ULL$2$ equation outperforms the other methods (see also the inset of Fig. 4 of the main text).


\twocolumngrid

\end{widetext}
\end{document}